\documentclass[preprint,12pt,3p]{elsarticle}

\usepackage{graphicx}

\usepackage[utf8]{inputenc}
\usepackage[usenames,dvipsnames]{xcolor}
\usepackage{amsmath}
\usepackage{mathrsfs}
\usepackage{amsfonts}
\usepackage{amssymb}
\usepackage{hyperref}
\usepackage{amsthm}
\usepackage{algorithm}
\usepackage{algorithmicx}
\usepackage{algpseudocode}

\usepackage[normalsize,tight]{subfigure}

\biboptions{authoryear}
\renewcommand{\cite}[1]{\citep{#1}}

\definecolor{dorange}{rgb}{0.6,0.1,0.1}
\definecolor{dgreen}{rgb}{0.0,0.4,0.11}
\definecolor{barva}{rgb}{0.5,0.4,0.1}
\definecolor{barvaE}{rgb}{0.3,0.4,0.1}
\definecolor{lblue}{rgb}{0.01,0.5,0.1}

\usepackage{macros-rohan}

\newcommand{\newtext}[1]{{#1}}

\DeclareMathOperator{\eigs}{eig}
\newcommand{\bs}[1]{\boldsymbol{#1}}

\newcommand{\per}{\#}

\newcommand{\mean}[1]{\left\langle#1\right\rangle}

\newcommand{\set}[1]{\mathbb{#1}}

\newcommand{\xR}{\ensuremath{\mathbb{R}}}

\newcommand{\alp}{\ensuremath{\alpha}}

\newcommand{\del}{\ensuremath{\delta}}

\newcommand{\V}[1]{\bs{#1}}

\newcommand{\Valp}{\ensuremath{\V{\alp}}}

\newcommand{\Ve}{\ensuremath{\V{e}}}

\newcommand{\dV}{\mathcal{V}}

\newcommand{\Bsc}{\ol{B}}

\def\rr{r}
\def\ss{s}
\def\kk{k}

\def\mater{{\rm{phys}}}
\def\unit#1{{\rm #1}}

\newcommand\ApxA{\ref{apx-A}}
\newcommand\ApxB{\ref{apx-B}}

\newcommand\lamsr[1]{\sqrt{\lambda^{#1}}}
\newcommand\Lam{\Lambda}

\newcommand\ie{{\it{i.e.~}}}
\newcommand\eg{{\it{e.g.~}}}
\newcommand\cf{{{cf.~}}}
\newcommand\lhs{{{\rm l.h.s.~}}}

\newcommand{\ER}[1]{{#1}}
\newcommand{\newER}[1]{{#1}}

\def\pathHeczko{.}
\def\pathOpt{.}
\def\pathOptNew{.}
\def\pathScale{.}

\newcounter{remark}
\newenvironment{myremark}[1]%
{\vspace{7pt} \noindent{\bf Remark 
  \refstepcounter{remark}\theremark. \protect\label{#1} \hspace{-2mm}}\rm }%
{\vspace*{-7pt} \flushright $\triangle$\\ } 

{\vspace{7pt} \noindent{{\bf Proof} of \protect{#1}. \hspace{-2mm}}\rm}%
{\flushright $\square$ \\} 

\newcounter{proposition}
{\vspace{7pt} \noindent{\bf Proposition
  \refstepcounter{proposition}\theproposition. \protect\label{#1} 
\hspace{-2mm}}\it}%
{\vspace*{7pt}}

\newcounter{definition}
{\vspace{7pt} \noindent{\bf Definition 
  \refstepcounter{definition}\thedefinition. \protect\label{#1} 
{\it({#2})}
\hspace{-2mm}}\rm}%
{} 

\begin{document}

\begin{frontmatter}

\title{Shape optimization of phononic band gap structures using the homogenization approach}



\author[1]{Jaroslav Vond\v{r}ejc
}
\ead{vondrejc@gmail.com}

\author[2]{Eduard Rohan\corref{cor1}}
\ead{rohan@kme.zcu.cz}

\author[2]{Jan Heczko}
\ead{jheczko@ntis.zcu.cz}

\address[1]{Institute of Scientific Computing, Technische Universit\"{a}t Braunschweig, 
Mühlenpfordstrasse~23, 38106 Braunschweig, Germany}

\address[2]{European Centre of Excellence, NTIS -- New Technologies for
Information Society Faculty of Applied Sciences, University of West Bohemia,
Univerzitn\'{\i} 22, 30614 Pilsen, Czech Republic}

\cortext[cor1]{Corresponding author}


\begin{abstract}
 The paper deals with optimization of the acoustic band gaps computed using the homogenized model of strongly heterogeneous elastic composite which is constituted by soft inclusions periodically distributed in stiff elastic matrix.
 We employ the homogenized model of such medium to compute intervals --- band gaps --- of the incident wave frequencies for which acoustic waves cannot propagate.  
 It was demonstrated that the band gaps distribution can be influenced by changing the shape of inclusions.
 Therefore, we deal with the shape optimization problem to maximize low-frequency band gaps; their bounds are determined by analysing the effective mass tensor of the homogenized medium.
 Analytic transformation formulas are derived which describe dispersion effects of resizing the inclusions. The core of the problem lies in sensitivity of the eigenvalue problem associated with the microstructure.
 Computational sensitivity analysis is developed, which allows for efficient usage of the gradient based optimization methods.
 Numerical examples with 2D structures are reported to illustrate the effects of optimization  with stiffness constraint.
 This study is aimed to develop modelling tools which can be used in optimal design of new acoustic devices for ``smart systems''.
\end{abstract}

\begin{keyword}
phononic materials \sep acoustic waves \sep band gap structures \sep optimization \sep sensitivity analysis \sep homogenization
\end{keyword}

\end{frontmatter}

\section{Introduction}
\label{sec:introduction}

Phononic materials, often called phononic crystals \cite{Laude2015book}, are elastic composites with periodic structure and with large contrasts in elasticity of the constituents.
For certain frequency ranges called \emph{band gaps}, such elastic structures can effectively attenuate propagation of incident acoustic waves, see \eg \cite{Henderson2001,Vasseur1998} where this property has been studied experimentally.
Due to this behaviour, the phononic materials may be used in modern technologies to generate frequency filters, as beam splitters, sound or vibration protection devices (for noise reduction), or they may serve as waveguides. {In this context, optimization of such materials is of a great importance.
To control acoustic waves propagation, also a mechanical tuning of phononic band gaps by deformation can be considered \cite{Wang2012}.}

It has been noted as early as in eighties of the last
century, cf. \cite{Auriault1985}, see the reviews
\cite{Milton2007,Auriault2012}, that constitutive relationships of the
elastodynamics are non-local in space and time and that the
\emph{effective inertia mass} presents, in general, an anisotropic
second order tensor which depends on the frequency of imposed
oscillations, or incident waves. 
Due to this property, the wave dispersion may become very strong for frequencies
in whole bands -- \emph{band gaps} -- where the effective (homogenized) inertia mass
becomes indefinite, or even negative definite, such that the elastic composites may
\emph{loose their potential capability to behave as harmonic
  oscillators}\footnote{The notion of
band gaps is introduced as the range of frequencies for which the
structure is blocked from free vibrations. However, as we shall see
  below, the definition of the band gaps requires more complexity in order to respect the
  material anisotropy.}.

\subsection{Modelling of the phononic materials and band gap analysis}

The classical continuum approach to the band gap analysis, also frequently reported in the literature in the context of the material optimization, \eg \cite{Vatanabe2014,Liu2014,Men2010photonic,Dong2014PLA,Jensen2006,Qian2011}, is based on the Bloch-Floquet theory for waves in an unbounded medium. This treatment requires that the band gaps are searched for specified wave directions on the boundary of the first Brillouin zone.
{The transfer matrix method \cite{Fomenko2014} can be used to study layered functionally graded phononic structures subject to plane wave propagation, so that the band gaps are obtained upon analysing eigenvalues of the transfer matrices.}

An alternative and effective way of modelling the phononic materials, which is also adhered to in this paper,
is based on the asymptotic homogenization method applied to the strongly
heterogeneous elastic \cite{Auriault1985,Avila2008MMS_BG}, or piezoelectric medium \cite{pzph-roma06,Rohan2009-WCSMO,acm-phono2010}.  
This approach was first developed  to study effective behaviour of the \emph{photonic crystals} used in optical devices, cf. \cite{Yablon}. 
In \cite{Bouchitte2004}, the scaling ansatz for periodically oscillating coefficients of the scalar Helmholtz equation governing the electromagnetic wave propagation was proposed, such that the band gaps could be analysed according to the sign of the effective coefficient. The idea was extended to elasticity due to similarity of the mathematical description, although the band gap analysis becomes more complex with respect to the wave polarization \cite{Avila2008MMS_BG}, cf. \cite{RohanSeifrt2009}. The same mathematical tools were used in \cite{Smyshlyaev2009} to treat strongly anisotropic elastic structures, so that effective band gaps are relative to the wave polarization and the direction of propagation.

Very recently, using the Bloch-Floquet theory and the Hill-Mandel lemma,  an elastodynamic homogenization theory has bee proposed in \cite{Nassar2016IJSS} which generalizes the contribution due to \cite{Milton2007}, cf. \cite{RohanSeifrt2009}.
It is worth to note, that higher-order homogenization theory is now being developed to 
respect transition effects which are not captured by the first order theory, see \eg \cite{Pham2013}. {Beyond the classical continuum theory, the relaxed micromorphic continua \cite{Madeo2015bg} serve as a good basis for studying acoustic wave dispersion and band gaps.}

We consider periodically heterogeneous elastic composites 
with the characteristic size of the heterogeneities proportional to $\veps$; this small parameter also determines the size of the so-called \emph{representative periodic cell} (RPC).
Although, in general, the material properties
may vary arbitrarily in the RPC,
we confine to materials where the RPC is formed by an elastic matrix and a very soft inclusion with the shape, which influences the effective medium properties.
The medium is subject to a periodic
excitation. The limit homogenized model, obtained by the asymptotic analysis with $\veps\rightarrow 0$, describes acoustic wave
propagation in the effective elastic medium and is characterized by the
homogenized elastic fourth order tensor and by
the homogenized second order mass tensor.  By designing the shape of the "soft"
inclusions in the RPC, we can manipulate the range of
frequencies for which this homogenized \emph{mass tensor becomes
  negative} \wrt waves of some polarizations so that, as the consequence, those waves cannot propagate.
It has been shown in \cite{RohanSeifrt2009} that analysis of the eigenvalues of the mass tensor is sufficient to identify the band gaps independently of the direction of the wave propagation. Therefore, the band gaps can be optimized without any need to evaluate the Brillouin zone. \ER{However, the model based on the asymptotic analysis provides a good approximation of the heterogeneous structure response for waves of low wave numbers only, such that it cannot describe the whole Brillouin zone, see Remark~\ref{rem-10}.}

Although in this paper we focus on the elastic composites, an extension for \emph{piezoelectric composite materials} is possible following the same approach, see \eg \cite{pzph-roma06} where the associated modifications concerning the homogenized model and the sensitivity analysis were discussed.
In these materials, besides the passively controlled band gaps by designing microstructure-related geometrical parameters, see \cite{Rohan2006sa_piezoel,MiaraRohan2005piezo}, alternatively the \emph{piezo-phononic} structures can also be controlled ``actively'' by external electrical fields, due to the electro-mechanical coupling.

\subsection{Optimal design of the band gap structures}
Optimization of \emph{band gap structures}, or phononic crystals enjoys increasing attention over the last decade.
The topic is challenging owing to the wide applicability of the metamaterials featured by the band gap property.
Moreover, new technologies like 3D printing show new perspectives in the optimal metamaterial design based on computational analysis. 
There is a vast body of literature devoted to the phononic material optimization, see \eg
\cite{Sigmund2003bandgaps,Dong2014SMO,Krushynska2014,Farhat2009,Yuksel2015,Wautier2015,Park2014a}.
Typically, the optimization merit is to
modify widths of the band gap intervals. The design parameters are related to the shape of one or more inclusions situated in the RPC, as in our case, or to the microstructure topology. The topology optimization of the phononic crystals within the classical Bloch-Floquet theory was first treated by Sigmund and Jensen in \cite{Sigmund2003bandgaps}, where also the global problem in a finite domain was considered. Recently, using the same framework,  the topology optimization was extended for the piezo-phononic structures in \cite{Vatanabe2014}. 

Although, in general, the topology optimization, e.g. \cite{Schury2012}, leads to better designs than the shape optimization with a fixed topology, the latter approach is of pertaining interest.
From the practical point of view, it leads to metamaterial designs which can be manufactured relatively easier than quite general designs arising from the topology optimization.
Moreover, it was pointed out by Milton and Willis \cite{Milton2007}, that the band gap property depends on the size and the shape of inclusions, however, it is 
insensitive to deviations form their exactly periodic distribution. This observation is coherent with the homogenization result and the related band gap definition employed in this paper, cf. \cite{RohanSeifrt2009}. In this context, responses of locally resonant acoustic metamaterials designed as rubber-coated hard particles in a compliant matrix (cf. \cite{acm-phono2009}) were studied in  \cite{Krushynska2014}, where the same phenomenon of insensitivity of the band gaps \wrt  periodic placements of the particles was shown.
From the theoretical point of view, it is not yet understood, how the inclusion shape can contribute to the band gap widening while design constraints are imposed to control the material stiffness. The present paper should give answers related to this issue.


In this paper, we rely on the homogenization approach \cite{Avila2008MMS_BG,RohanSeifrt2009}
of modelling phononic materials in 2D under the plane {strain condition; alternatively the plane stress condition can be prescribed, so that the 2D problem
can describe} the in-plane waves in plates (the homogenized phononic plates were introduced in \cite{rohan-miara-ZAMM2015,rohan-miara-CRAS2011}, and elaborated further in \cite{Rohan2015bg-plates}). The optimal shape problem is formulated for a single ``soft'' elastic inclusion embedded in the matrix of the RPC. The shape is described by the circular B-spline which ensures enough regularity independently of the numerical discretisation. The constraints are related to the effective stiffness of the composite. Besides solving the optimal design problem, we explore dependence of the band gap distribution on the volume fraction of the soft phase, \ie on size of the inclusion.



The plan of the paper is as follows. In Section~\ref{sec:pbsetting} we introduce the mathematical model of the homogenized phononic material and explain the band gap analysis based on the effective mass tensor.
The effects related to the change of RPC size and the inclusion size are discussed in Section~\ref{sec:rescaling}, where rescaling formulas are given; it enables to transform the band gaps distribution accordingly.
The optimal shape problem is formulated in Section~\ref{sec:opt}, followed by the sensitivity analysis in  Section~\ref{sec:sa}, where the computation of the total gradients of the band gap bounds \wrt the design variables are introduced. 
Numerical algorithms are summarized in Section~\ref{sec:numerical-examples}, where also some selected examples of solved problems are reported.
As initial designs, we consider the square-shaped and L-shaped inclusions;
it is shown, how maximization of the first and the second band gap leads to different optimal shapes.
In the conclusion, we summarize the particular amendments of the paper, and comment on some particular issues which will be handled in a future work.
Some technical contributions of the paper are postponed in the Appendix.

\section{Problem setting}\label{sec:pbsetting}
An elastic composite with a periodic structure is featured by the
characteristic size\footnote{The so-called characteristic size of the
  ``microstructure''.} $\veps = \ell/L$. This is the ratio between the
diameter $\ell$ of a generating representative cell and a ``macroscopic'' length $L$
corresponding to the wavelength of propagating waves, or to the size
of an open bounded domain $\Om\subset \RR^d$, $d = 2,3$ occupied by the
composite. It is constituted by two different materials: the one
occupying the domain $\Om_1^\veps$ called \textit{the matrix} and the
other one situated in the domain $\Om_2^\veps$ called \textit{the
  inclusions}. 
Importantly, we shall assume that domain $\Om_1^\veps$ is connected. 
The material constituents vary periodically with the
local position. 

 Throughout the text all the quantities varying with this
microstructural periodicity are labelled with superscript $^\veps$.
We use the usual boldface notation to denote the vectors $\ab = (a_i)$ or the 2nd order tensors $\Bb = (B_{ij})$ for $i,j = 1,\dotsc,d$.

The following functional
spaces are used: by $L^2(D)$ we refer to square integrable functions defined in
an open bounded domain $D$; by $H^1(D)$ we mean the Sobolev space $W^{1,2}(D) \subset L^2(D)$
composed of square integrable functions including their first generalized
derivatives; space $H_0^1(D) \subset H^1(D)$ is constituted by functions with
zero trace on $\pd D$. Bold notation is used to denote spaces of vector-valued
functions, \eg $\Hdb(D)$. We shall consider periodic functions defined in special domains $Y = ]0,1[^d \subset \RR^d$, where $d = 2,3$; we shall need spaces of   $Y$-periodic functions, labelled by subscript $_\#$, thus, $\Hpdb(Y)$ contains  functions $\fb \in \Hdb(Y)$ extended by Y-periodicity from $Y$ to $\RR^d$ using $\fb(y') = \fb(y)$ with $y' = y + k$ for any $k \in \ZZ^d$.

\subsection{Heterogeneous elastic structure}
The material properties are associated to the periodic geometrical
decomposition which is now introduced.  We consider 
the  \emph{reference (unit) cell} $Y = ]0,1[^d$ which generates the structure as the periodic lattice. The cell consists of
an  inclusion $\ol{Y_2} \subset Y$, embedded in the matrix
part $Y_1 = Y \setminus \ol{Y_2}$ so that the interface $\Gamma = \pd Y_2$ is contained in $Y$. In general, cell $Y$ may be
defined as a parallelepiped, the particular choice of the unit cube in $\RR^d$ is used
just for an ease of explanation; we shall return to this issue
in Remark~\ref{rem-4}. 
Using the reference cell we generate
the decomposition of $\Om$ as the union of all inclusions (which
should not penetrate boundary $\pd \Om$), having the size $\approx \veps$,
\begin{equation}\label{eq-1}
\begin{split}
\Om^\veps_2 & = \bigcup_{k \in \KK^\veps} \veps( Y_2 + k)\;, \mbox{
where }\KK^\veps = \{k \in \ZZ|\; \veps(k+ \ol{Y_2}) \subset
\Om\}\;,
\end{split}
\end{equation}
whereas the perforated matrix is $\Om^\veps_1  = \Om \setminus \Om^\veps_2$. Also we introduce the interface $\Gamma^\veps =
\ol{\Om_1^\veps}\cap\ol{\Om_2^\veps}$, 
so that $\Om = \Om_1^\veps \cup \Om_2^\veps \cup \Gamma^\veps$.

Properties of a three dimensional body made of the elastic
material are described by the elasticity tensor $\Cop^\veps = (C_{ijkl})^\veps$, where $i,j,k = 1,2,\dots,d$. 
As usually, we assume both major and minor symmetries of $C_{ijkl}^\veps$ (\ie $C_{ijkl}^\veps = C_{klij}^\veps = C_{jikl}^\veps$), and the ellipticity,  

We assume that inclusions are occupied by a ``{very soft material}'' in such a sense that
coefficients of \emph{the elasticity tensor in the inclusions} are significantly smaller than those of
the matrix compartment, however \emph{the material density} is comparable in
both the compartments. The physical aspects of such an arrangement of the periodic structure has been discussed \eg in \cite{RohanSeifrt2009}; the soft inclusions behave as passive dumpers inducing an anti-resonance effect for certain frequencies, as explained in Section~\ref{sec-bandgap}.

The  material parameters are defined with respect to the decomposition of generating cell $Y$, as follows:
\begin{equation}\label{eq-10}
\begin{split}
\rho^\veps(x) = \left \{
\begin{array}{ll}
\rho^1(y) & \mbox{ for } y\in Y_1,\\
\rho^2(y) & \mbox{ for } y\in Y_2,
\end{array} \right .
 & \quad \quad 
C_{ijkl}^\veps(x) = \left \{
\begin{array}{ll}
C_{ijkl}^1(y) & \mbox{ for }  y\in Y_1,\\C_{ijkl}^{2,\veps}(y) = 
\veps^2\bar C_{ijkl}^2(y) & \mbox{ for } y\in Y_2,
\end{array} \right .
\end{split}
\end{equation}
where, by virtue of the decomposition introduced in \eq{eq-1}, see e.g. \cite{Cioranescu2008unfolding}, $y := \{\frac{x}{\veps}\}_Y$ is the local ``microscopic'' coordinate associated with the global ``macroscopic'' position $x \in \Om$. 
As an important feature of the modelling based on the asymptotic analysis, the
\emph{strong heterogeneity} in the elastic coefficients is related to the geometrical scale $\veps$ of the
underlying microstructure by the coefficient $\veps^2$.
Due to this scaling an internal scale of the structure is retained when passing to the limit $\veps\rightarrow 0$ and the
homogenized model exhibits dispersive behaviour; this phenomenon
cannot be observed when standard two-scale homogenization procedure is
applied to a medium with scale-independent material parameters, 
cf. \cite{Auriault1985}.

\begin{myremark}{rem-5}
The limit homogenized model of the strongly heterogeneous medium can be used to describe a real situation only for a given scale, \ie for a given $\veps>0$,
such that $\veps Y = ]0,\veps[^2$ is the actual \emph{real-sized reference cell}. 
In fact, the elasticity $\veps^2 \bar\Cop^2$ corresponds to the physical values of the elasticity tensor, so that, due to \eq{eq-10}, the elasticity $\Cop^\mater$ of a given real material situated in the inclusions $\Om_2^\veps$ is related to the model coefficients $\bar\Cop^2$ by the following relationship: $\Cop^\mater(x) \equiv \Cop^\veps(x) =\veps^2 \bar\Cop^2(y)$ for $y = \{x/\veps\}_Y \in Y_2$.
Te derive the limit model using the asymptotic analysis for $\veps\rightarrow 0$, $\bar \Cop^2$ is treated as a fixed parameter.
\end{myremark}


\subsection{Modelling the stationary waves}
We consider stationary wave propagation in the medium introduced
above. Although the problem can be treated for a general case of
boundary conditions, for simplicity we restrict the model to the
description of clamped structures loaded by volume forces.
Assuming a harmonic single-frequency volume forces,
\begin{equation*}
\begin{split}
\Fb(x,t) = \fb(x)e^{\imu\om t}\;, 
\end{split}
\end{equation*}
where $\fb = (f_i),i=1,{\dotsc,d}$ is its local amplitude and $\om$ is the
frequency, $\imu^2 = -1$. We consider a dispersive displacement field with the local
magnitude $\ub^\veps$
\begin{equation*}
\begin{split}
\Ub^\veps(x,\om,t) = \ub^\veps(x,\om)e^{\imu\om t}\;,
\end{split}
\end{equation*}
This allows us to study the steady
periodic response of the medium, as characterized by displacement field $\ub^\veps$ which satisfies the following
boundary value problem:

\begin{equation}\label{eq-11}
\begin{split}
-\om^2\rho^\veps \ub^\veps - \dvg \sigmabf^\veps  & = \fb  \quad \mbox{ in } \Om, \quad 
\ub^\veps  = 0  \quad \mbox{ on } \pd\Om, 
\end{split}
\end{equation}
where the stress tensor $\sigmabf^\veps = (\sigma_{ij}^\veps)$ is expressed by the Hooke's law 
$\sigmabf^\veps = \Cop\eeb^\veps$ involving the linearised strain tensor
$\eeb^\veps = (e_{ij}^\veps)$ which for a displacement field $\vb = (v_i)$ is given by  $e_{ij}^\veps(\vb) = 1/2(\pd_jv_i + \pd_iv_j)$. 
The problem \eq{eq-11} can be  formulated in a weak form as follows: Find
$\ub^\veps \in \HOdb(\Om)$, such that
\begin{equation*}
\begin{split}
-\om^2 \int_\Om \rho^\veps \ub^\veps \cdot \vb + \int_\Om [\Cop^\veps \eeb(\ub^\veps)] : \eeb(\vb)
& = \int_\Om \fb\cdot\vb \quad \mbox{ for all } \vb \in \HOdb(\Om)
 \;,
\end{split}
\end{equation*}
where $\HOdb(\Om)$ is the standard Sobolev space of vectorial functions with square integrable 
generalized derivatives and with vanishing traces on $\pd \Om$, as required by \eq{eq-11}$_2$.

\subsection{Homogenized model}\label{sec-homogeq}
The periodic homogenization method is widely accepted as a modelling tool which
allows to associate a medium characterized by some heterogeneous microstructure
with a homogeneous model relevant to the macroscopic scale.   In \cite{Avila2008MMS_BG},
the unfolding operator method of homogenization \cite{Cioranescu2008unfolding}
was applied with the strong heterogeneity assumption \eq{eq-10}; it
can be shown that there is a limit macroscopic displacement field
$\overline{\ub}$ which satisfies the equation of the homogenized
medium; in fact for $\veps \rightarrow 0$, $\ub^\veps$  converges weakly to
the macroscopic displacement field $\ub$ 
with perturbations
localized in the inclusions. 
These local perturbations are related to $\ub$ 
 by characteristic responses introduced below.
They give rise to a correction term of the ``effective'' mass coefficients which is responsible for the wave dispersion and the band gap effect, in particular. 
We shall now record the resulting system of coupled equations, as derived in \cite{Avila2008MMS_BG}, which
describe the structure behaviour at two scales, the \emph{macroscopic}
one and the \emph{microscopic} one.

First we define the \emph{homogenized coefficients}\footnote{Often called the effective material parameters.} involved in the homogenized model of wave propagation.
The ``frequency--dependent coefficients'' are determined just by
material properties of the inclusion and by the 
material density $\rho^1$, whereas the elasticity coefficients are related
exclusively to the matrix material occupying the perforated domain.

\emph{Frequency--dependent homogenized coefficients} 
 involved in the
macroscopic momentum equation are expressed in terms of eigenelements
$(\lam^r,\phibf^r) \in \RR \times \HOdb(Y_2)$, $r = 1,2,\dots$ of the
elastic spectral problem which is imposed in inclusion $Y_2$ with $\phibf^r = 0$ on $\pd Y_2$:
\begin{equation}\label{eq-36a}
\begin{split}
\int_{Y_2} [\bar\Cop^2 \eeb^y(\phibf^r)]:\eeb^y(\vb) = \lam^r \int_{Y_2} \rho^2 \phibf^r \cdot \vb
\quad \forall \vb \in \HOdb(Y_2)\;, \quad \int_{Y_2} \rho^2 \phibf^r\cdot\phibf^s = \delta_{rs}\;.
\end{split}
\end{equation}

To simplify the notation, the \emph{eigenmomentum}
$\mb^r = (m_i^r)$, and the averaged density $\aver{\rho}$ are introduced:
\begin{equation}\label{eq-36}
\begin{split}
\mb^r = \int_{Y_2} \rho^2 \phibf^r\;, \quad \aver{\rho} = \sum_{k=1,2}\int_{Y_k} \rho^k\;.
\end{split}
\end{equation}
Due to the above definition of $\mb^r$, in what follows we confine spectrum with the following index set:
\begin{equation}\label{eq-evS1}
\Rcal = \{r \in \NN|\;(\lam^r,\phibf^r) \mbox{ solve }\eq{eq-36a}\;, \; |\int_{Y_2}\rho^2\phibf^r | > 0\}.
\end{equation}
The macroscopic model of the acoustic wave propagation in the homogenized medium is introduced in terms of the following tensors, all depending on $\om^2$:

\begin{itemize}
\item \emph{Mass tensor} $\Mb^{} = (M_{ij}^{})$
\begin{equation}\label{eq-37}
\begin{split}
M_{ij}^{}(\om) = \frac{1}{|Y|}\int_{Y}\rho \delta_{ij} - \frac{1}{|Y|}\sum_{r \in \Rcal} \frac{\om^2}{\om^2 - \lam^r} m_i^r m_j^r\;;
\end{split}
\end{equation}

\item Applied \emph{load tensor} $\Bb^{} = (B_{ij}^{})$
\begin{equation*}
\begin{split}
B_{ij}^{}(\om) = \delta_{ij} - \frac{1}{|Y|}\sum_{r \in \Rcal} \frac{\om^2}{\om^2 - \lam^r} m_i^r \int_{Y_2} \vphi_j^r\;.
\end{split}
\end{equation*}

\end{itemize}

\emph{The elasticity coefficients}
$\Dop=(D_{ijkl})$ are computed using the same
formula as the one obtained for the homogenized perforated
domain, thus being independent of
the  material in inclusions:
\begin{subequations}
\label{eq:homog_elas}
\begin{equation}\label{eq-C*}
\begin{split}
D_{ijkl}^{} & = \frac{1}{|Y|} \int_{Y_1} [\Cop^1 \eeb^y(\wb^{kl}+\Pibf^{kl})]:\eeb^y(\wb^{ij}+\Pibf^{ij})\;,
\end{split}
\end{equation}
where $\Pibf^{kl} = (\Pi_i^{kl}) = (y_l \delta_{ik})$ and $\wb^{kl} \in \Hpdb(Y_1)$ are the corrector functions satisfying
\begin{equation}\label{eq-C*a}
\begin{split}
\int_{Y_1} [\Cop^1 \eeb^y(\wb^{kl}+\Pibf^{kl})]:\eeb^y(\vb) = 0\quad \forall \vb \in \Hpdb(Y_1)\;.
\end{split}
\end{equation}
\end{subequations}
Above $\Hpdb(Y_1)$ is the restriction of $\Hdb(Y_1)$ to the Y-periodic functions (periodicity w.r.t. the homologous points on the opposite edges of $\pd Y$).

The wave propagation in the \emph{homogenized phononic material} is described by the macroscopic problem which involves 
the homogenized coefficients introduced above. We find $\ub \in \HOdb(\Om)$ such that

\begin{equation}\label{eq-45} 
\begin{split}
 -\om^2 \int_\Om [\Mb^{}(\om)\ub]\cdot \vb 
 + \int_\Om [\Dop^{} \eeb(\ub)]:\eeb(\vb) 
= & \int_\Om [\Bb^{}(\om)\fb]\cdot \vb 
 \quad \forall \vb \in \HOdb(\Om)\,.
\end{split}
\end{equation}

\begin{myremark}{rem-homproc}
\ER{The model of the \emph{homogenized phononic material} has been derived rigorously in \cite{Avila2008MMS_BG} using the unfolding method of the periodic homogenization.
For the readers convenience we summarize main steps of the procedure.
First a~priory estimates are obtained for the displacement field $\ub^\veps$ and the associated strains.
For this, the solution is decomposed in two parts, \ie $\ub^\veps = \ub^{1,\veps} + \ub^{2,\veps}$, whereby $\ub^{1,\veps}$ is the bulk response, while $\ub^{2,\veps}$ is a ``bubble function'' supported in the inclusions $\Om_2^\veps$ only.
Whereas strains $\eeb(\ub^{1,\veps})$ are bounded in $L^2(\Om)$ for each $\varepsilon$, only the rescaled strains $\veps\eeb(\ub^{1,\veps})$ are bounded uniformly in $L^2(\Om)$, as the consequence of the ``weak inclusion'' ansatz introduced by the $\veps^2$  scaling of the elasticity coefficients in \eq{eq-10}.
This leads to qualitatively different convergence of the two fields, when passing to the limit $\veps\rightarrow 0$;
the bulk field converges to $\ub(x)$, whereby the strains in the matrix part are corrected by virtue of the standard two-scale fluctuations $\ub^1(x,y)$.
In the inclusions, the bubble field converges to a two-scale function $\ub^2(x,y)$. 
The limit two-scale wave equation can be split into local problems solved in the representative cell $Y$ and to the global problem.
The local problems in the matrix part $Y_m$ involve $\ub^1$ which can be expressed in terms of corrector fields $\wb^{ij}$ satisfying \eq{eq-C*a}, so that $\ub^1(x,y) = \wb^{ij}(y)e_{ij}(\ub(x))$.
In the inclusions, $\ub^2(x,y)$ is expressed in terms of the eigenmodes defined by \eq{eq-36a} which (by virtue of the Fourier method) are also used to express contributions of the macroscopic variables $\ub^1(x)$ and of the external loads. 
Upon substituting in the global equation the two-scale functions $\ub^1$ and $\ub^2$ using the local autonomous responses $\wb^{ij}$ and the eigenmodes $(\lam^r,\phibf^r)$, the homogenized coefficients can be introduced, which leads to the macroscopic model  \eq{eq-45}. }
\end{myremark}

\subsection{Band gap prediction}
\label{sec-bandgap}

Heterogeneous periodic structures with finite scale of heterogeneities exhibit
the frequency \emph{band gaps} -- for certain frequency bands the wave propagation is disabled, or restricted in the polarization, cf. \cite{Smyshlyaev2009}. In the
\emph{homogenized medium}, the wave propagation depends on the
positivity of mass tensor $\Mb^{}(\om)$.  Thus, as the main advantage of
the homogenized model derived in \cite{Avila2008MMS_BG}, {by analysing the dependence $\om \mapsto \Mb^{}(\om)$, one can determine bounds of the band gaps} with
significantly less effort than in case of computing  them in the
standard way, see e.g. \cite{Sigmund2003bandgaps}.

\begin{figure}
\centering
\includegraphics[width=.5\textwidth]{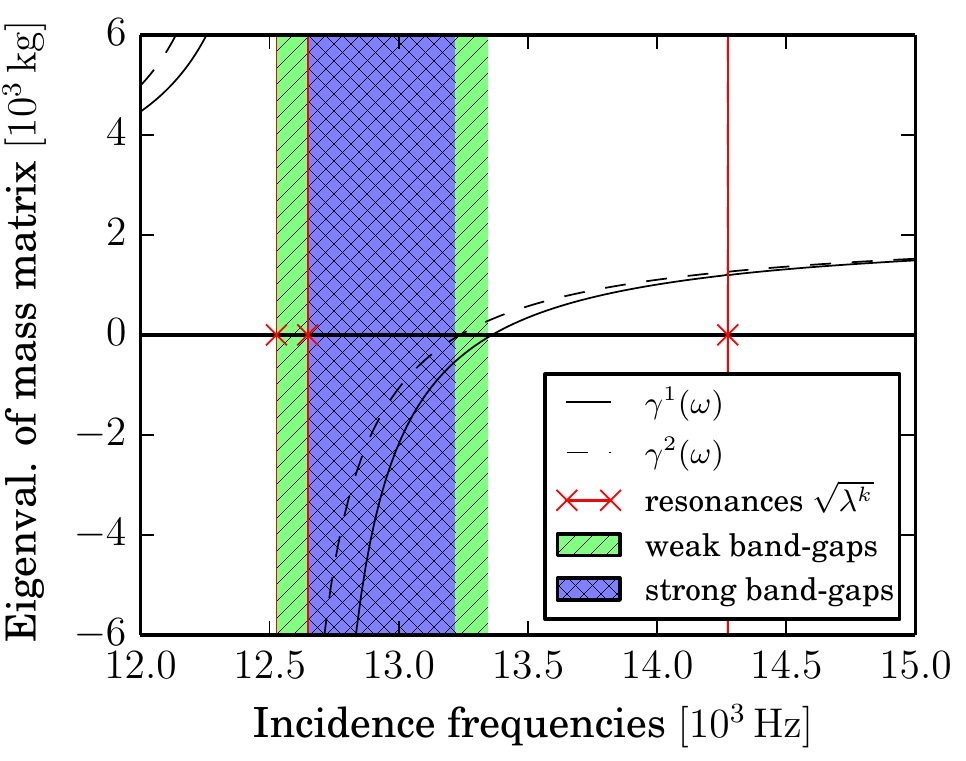}
\caption{The distribution of band gaps, resonances, and eigenvalues of mass matrix for a periodic cell with an oval inclusion which is obtained as a perturbation of the circular inclusion; all types of frequency bands (weak band gap, strong band gap, and propagation zone) can be observed.}
\label{fig:bg_types}
\end{figure}

The band gaps can be classified \wrt the polarization of waves which cannot
propagate. We now summarize the results of \cite{Avila2008MMS_BG} and \cite{RohanSeifrt2009} which are important for the optimization problem formulation.
There are three principal modes of the wave propagation which are
distinguished below according to the signs of eigenvalues $\gamma^r(\om)$, $ r =
1,\dots,d$ of the mass
tensor $\Mb(\om)$ analysed for a given frequency $\omega$; more precisely, we consider 
\begin{equation}\label{eq-evM} 
\begin{split}
\gamma^r:\set{R}^+\setminus\{\lam^k\}_{k\in\set{N}}\rightarrow\set{R}
\text{ such that }
(\Mb^{}(\om)- \gamma^r(\om)\Ib)\vb^r = \bmi{0}\;,\quad|\vb^r| = 1\;.
\end{split}
\end{equation}
Further we consider the ordering $\gamma^1(\om) \leq
\gamma^2(\om)\dots \leq\gamma^d(\om)$ for any ``non-resonant'' $\om
\not = \lamsr{k}$, $k = 1,2,\dots$.  In general, $\gamma^1 \rightarrow
-\infty$ for $\om \searrow \lamsr{k}$ and $\gamma^d \rightarrow \infty$
for $\om \nearrow \lamsr{k+1}$, see Figure~\ref{fig:bg_types}.  

Let us denote by $\omega_k^r$ the root
of the eigenvalue $\gamma^r(\omega)$ within a particular interval
between adjacent eigenvalues of problem \eq{eq-36a}, thus
\begin{align}\label{eq-evroot}
\omega_k^r \in ]\lamsr{k},\lamsr{k+1}[ \text{ such that }\gamma^r(\omega_k^r)=0.
\end{align}
Such a root of $\gamma^r(\omega)$ may not exist for any $r \in \{1,\dots,d\}$, however.  

In Section~\ref{sec:opt}, when dealing with the band gap optimization
problem, we assume that the root $\omega_k^1$ exists\footnote{Note
that $\om_k^r\leq\om_k^{r-1}$ for any two existing roots.}, see Remark~\ref{rem-1}$_{(v)}$.
Under this assumption, the wave propagation for $\om$ in the interval
$]\lamsr{k},\omega_k^1[$ is restricted to the waves with the local
polarizations $\ub(x)$ orthogonal to eigenvector $\vb^1$ and also to all other $\vb^s$, $s \in \{2,\dots,d\}$ for which 
$\gamma^s(\om)< 0$. For more details we refer to \cite{RohanSeifrt2009} and  \cite{Smyshlyaev2009}, where this topic was discussed in detail.
Depending on the frequency considered, $\om \in ]\lamsr{k},\omega_k^1[$
and on the existence of other roots $\omega_k^r$, $r \in
\{2,\dots,d\}$, the following 3 cases of the wave propagation can be distinguished:
\begin{enumerate}
\item Non-restricted propagation: $d$ waves governed by the homogenized model  \eq{eq-45} can propagate without any restriction of the wave polarization;
this can only happen for $\om \in P_k:=]\omega_k^1,\lamsr{k+1}[$. 
\item No wave can propagate, if all eigenvalues of $\gamma^r(\om)$, $r = 1,\dots,d$  are negative, so that 
the mass tensor $\Mb(\om)$ is negative definite. This only can happen, if the root $\om_k^d$ exists, for frequencies $\omega\in G_k^S := ]\lamsr{k},\omega_k^d[$.
\item Propagation is possible only for waves polarized in a manifold determined by
eigenvectors $\vb^r$ associated with positive eigenvalues $\gamma^r(\om) >0$. Tensor $\Mb(\omega)$ is indefinite for a such given $\omega$, i.e. there is at least
one negative and one positive eigenvalue, $\gamma^1(\om)<0< \gamma^d(\om)$.
\end{enumerate}   

In general, any interval $\Lam_k := ]\lamsr{k},\lamsr{k+1}[$ splits into three bands characterized by the above three cases and the three corresponding sets $P_k$ and $G_k^S$, and $G_k^W$, whereby $G_k^W :=  \Lam_k \setminus \ol{(G_k^S \cup P_k)}$,

\begin{enumerate}
\item {\bf strong band gap} $G_k^S := \{\om \in \Lam_k| \gamma^s(\om) <0, \forall s = 1,\dots,d\}$.
\item {\bf weak band gap} $G_k^W :=  \{\om \in \Lam_k| \gamma^1(\om) <0 \text{ and }\gamma^d(\om) > 0\}$.
\item {\bf propagation zone} $P_k := \{\om \in \Lam_k| \gamma^s(\om) >0, \forall s = 1,\dots,d\}$.
\end{enumerate}
For illustration, in Figure~\ref{fig:bg_types}, the three sets are depicted and the curves $\gamma^r(\om)$ are displayed.
We conclude this section by following comments: 

\begin{myremark}{rem-1}

\begin{list}{}{}
\item (i)
The above definition of the weak band gap $G_k^W$ is not explicit in the sense that it depends on the existence of $P_k$ and $G_k^S$. However, none of these can be guaranteed. 

\item (ii) Functions $\gamma^r(\om)$ are monotone in $\Lam_k$.

\item (iii) It may happen that ${\gamma^1}(\om) < 0$ for all $\om \in \Lam_k$, which would mean that $P_k = \emptyset$.

\item (iv)
If inclusions (considered in 2D) are symmetric \wrt more than 1 axis of symmetry, then only
strong band gaps exist; the band gap properties and
their relationship to the dispersion of guided waves were discussed in
\cite{RohanSeifrt2009}.

\item (v) In what follows we shall consider only such situations where $P_k \not = \emptyset$ and $G_k^W\not = \emptyset$. Therefore, we assume existence of the root $\om_k^1 \in \Lam_k$, so that  $G_k^W = ]\lamsr{k},\om_k^1[$.
Note that, if such a $\om_k^1$ does not exist within $\Lam_k$ for a particular $k$, the weak band gap spans over the whole interval $\Lam_k$, hence $P_k = \emptyset$. For example, this situation arises when a symmetric shape $Y_2$ with two axes of symmetry is slightly perturbed, so that a resonance value $\lam^k$ with multiplicity 2 splits into two closed values $\lam^k$ and $\lam^{k+1}$, as reported in Figure~\ref{fig:bg_types}.

\end{list}

\end{myremark}

\begin{myremark}{rem-10}
\ER{The model employed in this paper has been validated in \cite{RohanSeifrt2009}(see Section~6.3 therein) using the well accepted Bloch-Floquet theory of guided wave propagation in periodically heterogeneous media. In accordance with the modelling assumptions, 
the model captures dispersion properties of the medium for low wave numbers, if the material elasticity contrast is sufficiently large, \ie for $r :=  |\Cop^\mater| / |\Cop^1| \ll 1$. To give an example, with a contrast $r \approx 0.1$,  the first two band gaps are well approximated by the model for structures characterized by wave lengths longer than 5 material periods. Another aspect explored in the cited reference is the polarization stability of waves in the weak (partial) band gaps; this stability is good for stationary waves, but deteriorates with increasing wave numbers.}
\end{myremark}

\section{Rescaling and interpretation of the results}\label{sec:rescaling}
As pointed out above, see Remark~\ref{rem-5},
for interpretation of the limit model, it is necessary to consider a  given ``fixed scale'' $\veps = \veps_0$,
which yields the 
elasticity tensor  $\bar\Cop^2:= \Cop^\mater/\veps_0^2$.
In this section we discus how the band gap distribution depends on the real microstructure size of the heterogeneity and on the material contrast, cf. \cite{RohanSeifrt2009}.

For elasticity tensor $\Cop^{2,\veps} = \Cop^\mater$ of a given existing material and for a given microstructure represented by the inclusion $Y_2$ placed in the unit periodic cell $Y$, simple rescaling of the eigenvalue problem  \eq{eq-36a} yields the following equivalent problem imposed in the ``real-sized'' inclusion $\veps Y_2$,
\begin{equation}\label{eq-R1}
\begin{split}
\int_{\veps Y_2} \Cop^{2,\veps}\eeb^x(\phibf^\veps)]:\eeb^x(\vb) = \lam^\veps \int_{\veps Y_2} \rho^2 \phibf^\veps \cdot \vb
\quad \forall \vb \in \HOdb(Y_2)\;, \quad \int_{Y_2} \rho^2 \phibf^r\cdot\phibf^s = \delta_{rs}\;.
\end{split}
\end{equation}
whereby, the eigenvalues $\lam^\veps$ are equivalent to the ones computed in \eq{eq-36a}, \ie $\lam = \lam^\veps$ and also $\phibf = \phibf^\veps$.

As the consequence, for describing the band gaps by analysing homogenized mass tensor $\Mb(\om)$, the eigenvalue problem has to be solved either in the reference domain $Y_2$ with the rescaled coefficients $\bar\Cop^2$, or equivalently using \eq{eq-R1} with ``true'' coefficients $\Cop^\mater$ (in the sense of Remark~\ref{rem-5}) and the real-sized inclusion  $\veps Y_2$.

\begin{myremark}{rem-3}(\emph{Two-phase piecewise constant material.})
When dealing with an optimal material design problem, 
it is quite natural to consider a composite material formed by two homogeneous components. Therefore, in what follows, with reference to the definition \eq{eq-10}, we shall consider 
$\Cop^1$, $\bar\Cop^2$ and $\rho^\beta$, $\beta=1,2$ as constants independent of the position $y\in Y$. With this assumption in hand much of the subsequent development  could be simplified.  
\end{myremark}

\subsection{Influence of the size of the microstructure}\label{sec-RS1}

As explained above, the spectrum given equivalently by \eq{eq-36a}, or \eq{eq-R1}, depends on $\veps$.
In this paragraph we explore how the band gap distribution changes with the characteristic size of the microstructure, whereby
 the material properties of the composite are fixed.

\begin{table}[th]
\centering
\begin{tabular}{c|c|c|ccc|c}
problem & $\veps$ & $\bar E_2$ $[\unit{Pa}]$ & $\sqrt{\lam^2}$ & ${\om_2^1}$ & $\sqrt{\lam^3}$ & BG type \\
\hline
& 1 & $3.772\cdot 10^9$ & $1.2511 \cdot 10^4$ & $1.3182 \cdot 10^4$ & $1.4005 \cdot 10^4$ & \\
circle & 0.1 & $3.772\cdot 10^7$ & $1.2511 \cdot 10^5$ & $1.3182 \cdot 10^5$ & $1.4005 \cdot 10^5$ & strong \\
& 0.01 & $3.772\cdot 10^5$ & $1.2511 \cdot 10^6$ & $1.3182 \cdot 10^6$ & $1.4005 \cdot 10^6$ & \\
\hline
& 1 & $3.772\cdot 10^9$ & $1.3234 \cdot 10^4$ & $1.3898 \cdot 10^4$ & $1.5564 \cdot 10^4$ & \\
diamond & 0.1 & $3.772\cdot 10^7$ & $1.3234 \cdot 10^5$ & $1.3898 \cdot 10^5$ & $1.5564 \cdot 10^5$ & strong \\
& 0.01 & $3.772\cdot 10^5$ & $1.3234 \cdot 10^6$ & $1.3898 \cdot 10^6$ & $1.5564 \cdot 10^6$ & \\
\hline
& 1 & $3.772\cdot 10^9$ & $1.6596 \cdot 10^4$ & $1.7167 \cdot 10^4$ & $2.0934 \cdot 10^4$ & \\
L-shape & 0.1 & $3.772\cdot 10^7$ & $1.6596 \cdot 10^5$ & $1.7167 \cdot 10^5$ & $2.0934 \cdot 10^5$ & weak \\
& 0.01 & $3.772\cdot 10^5$ & $1.6596 \cdot 10^6$ & $1.7167 \cdot 10^6$ & $2.0934 \cdot 10^6$ & \\
\hline
\end{tabular}
\caption{Illustration of the spectral properties and the 2nd weak band gap dependence  on the microstructure size. The material is the aluminium-epoxy composite.
The Young's modulus and the Poisson's ratios for the epoxy resin inclusion are $ E_2 = 3.772 \; \unit{GPa}$ and $\nu_2 = 0.2743$, respectively; densities $\rho_1$ and $\rho_2$ are introduced in Table~\ref{tab:material}.
The frequencies ($\lamsr{1},\lamsr{2},{\om_2^1}$) are in [Hz]. The rescaled modulus is given by $\bar E_2 = \veps^{-2}E_2$.}
\label{tab:size_of_microstructure}
\end{table}

\begin{figure}[h]
\centering
\includegraphics[width=0.3\textwidth]{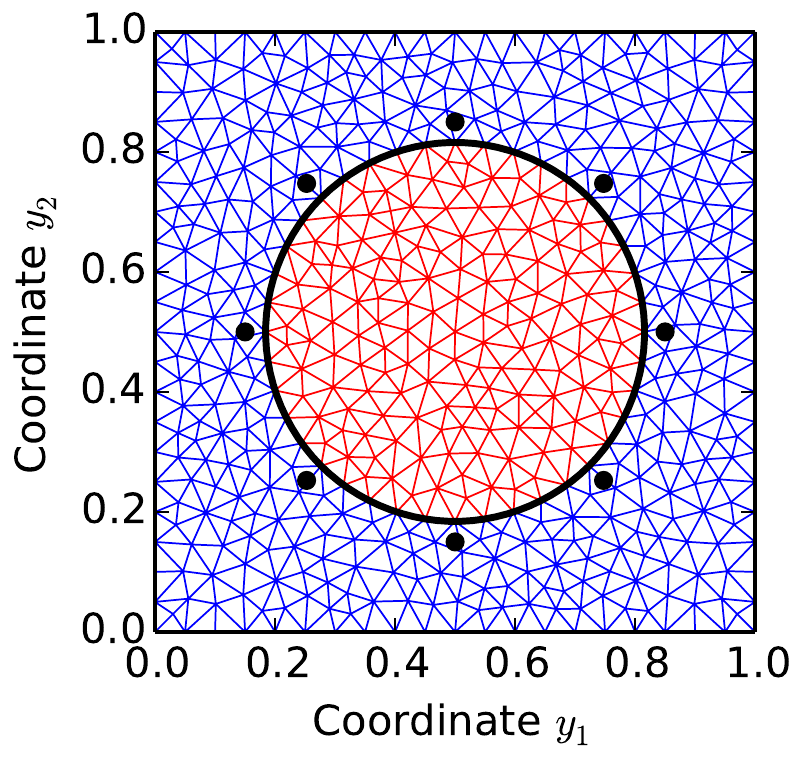}
\includegraphics[width=0.3\textwidth]{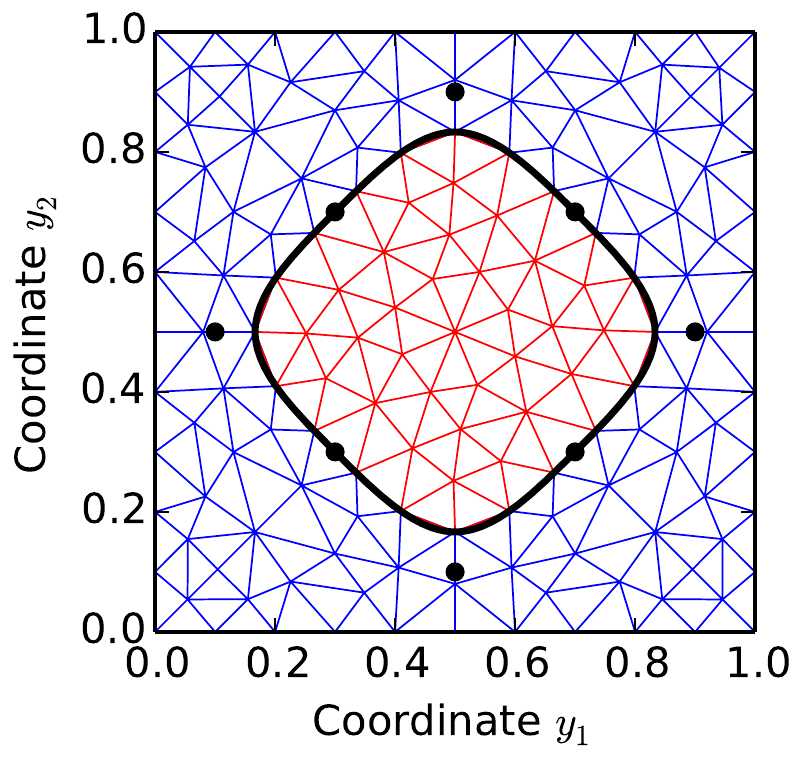}
\includegraphics[width=0.3\textwidth]{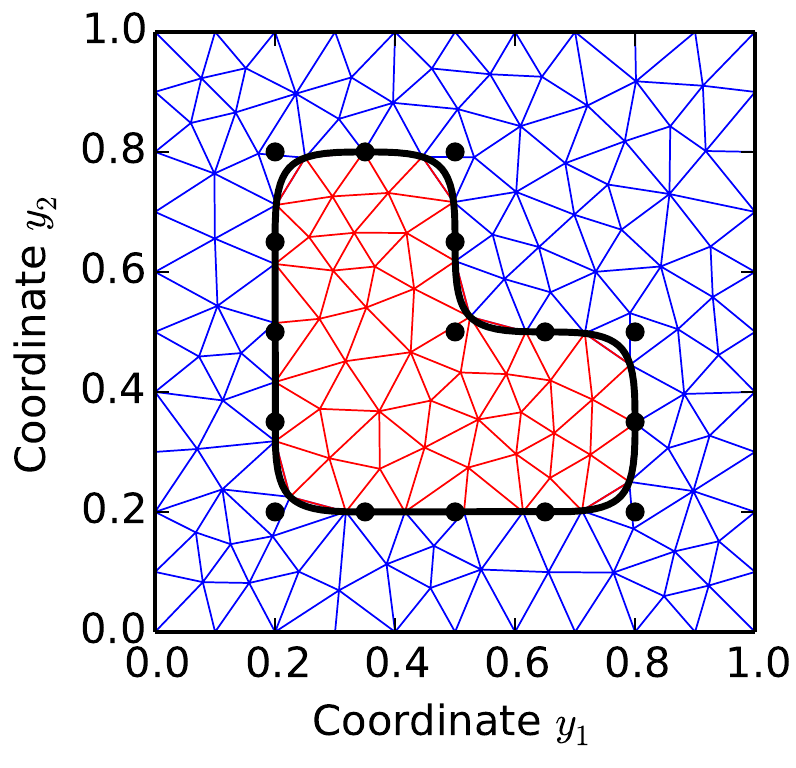}
\caption{Shapes of inclusions used in Tab.~\ref{tab:size_of_microstructure} (left to right): circle, diamond, L-shape.}\label{fig-HH}
\end{figure}

To illustrate the size effect, we may consider two structures characterized by $\veps_0$ and $\veps_1$, respectively, but both composed of the same two materials, the aluminium situated in $Y_1$, and epoxy resin situated in $Y_2$, with a given fixed volume fraction. The transformation formulas are derived in~\ApxA.
In Tab.~\ref{tab:size_of_microstructure}, for different $a = \veps_1/\veps_0$ and different shapes of inclusion $Y_2$ depicted in Fig.~\ref{fig-HH}, 
 the spectral responses of three the phononic structures are compared 
in terms of the 
two resonant frequencies, ${\sqrt{\lam^{2}},\sqrt{\lam^{3}}}$ and the {second} band gap bounds, such that 
${G_2^W=]\sqrt{\lam^2}, \omega_2^1]}$.
As the rule of the scaling, by getting the structure smaller with the factor $a <1$, any frequency band is multiplied by factor $1/a$; as the consequence, the band gaps become wider, being shifted to higher frequency ranges.

\begin{figure}[!htpb]
\centering
\subfigure[Periodic cell, $\veps = 1$]{
\includegraphics[height=0.36\textwidth]{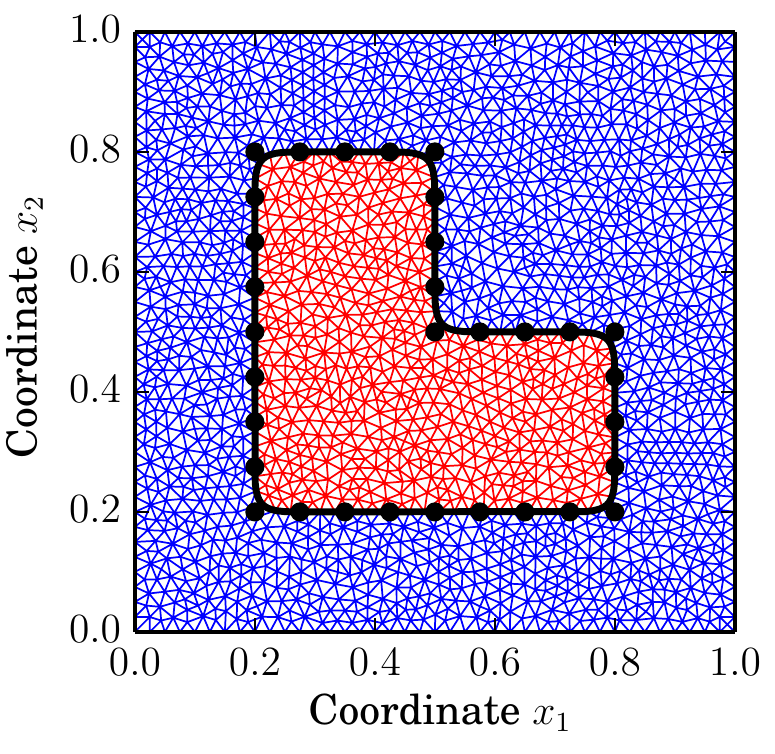}
}
\subfigure[band gaps and resonance frequencies]{
\includegraphics[height=0.36\textwidth]{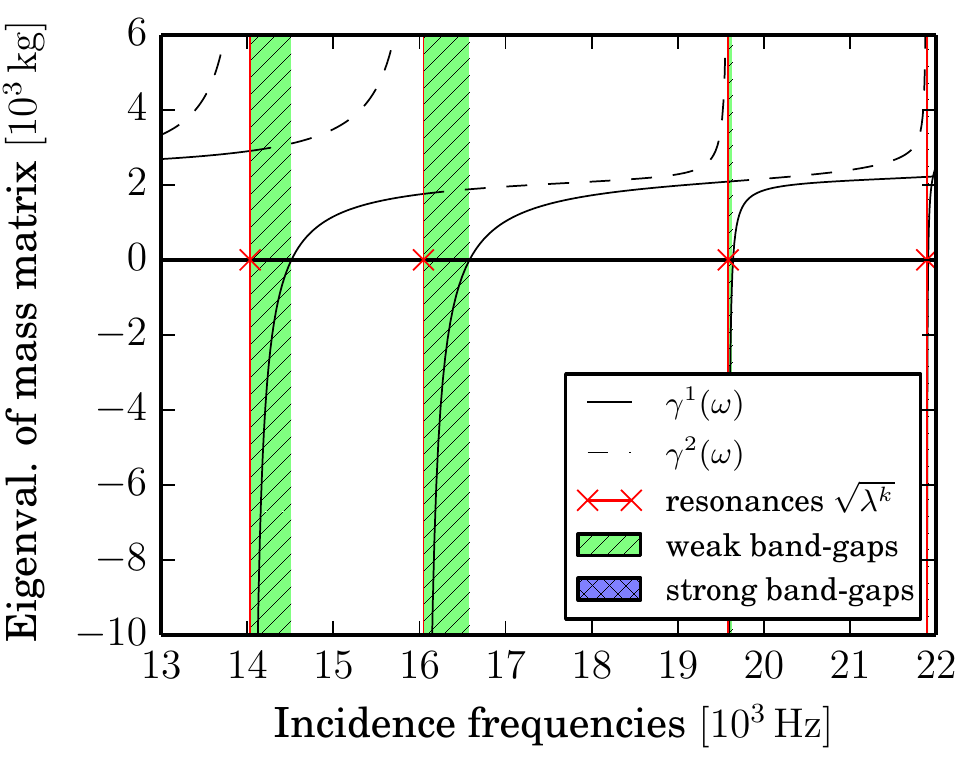}
}
\subfigure[Periodic cell, $\veps = 0.5$]{
\includegraphics[height=0.36\textwidth]{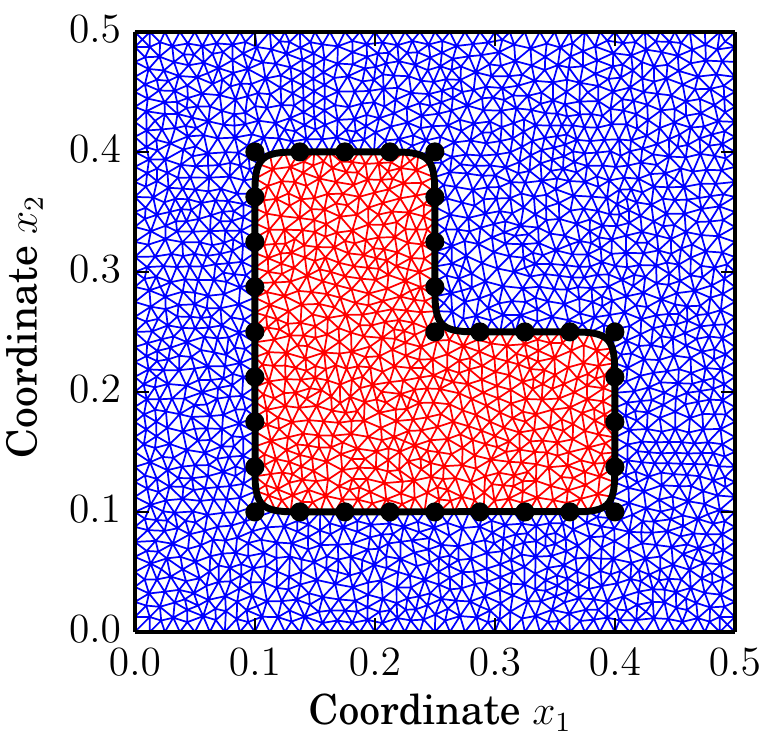}
}
\subfigure[band gaps and resonance frequencies]{
\includegraphics[height=0.36\textwidth]{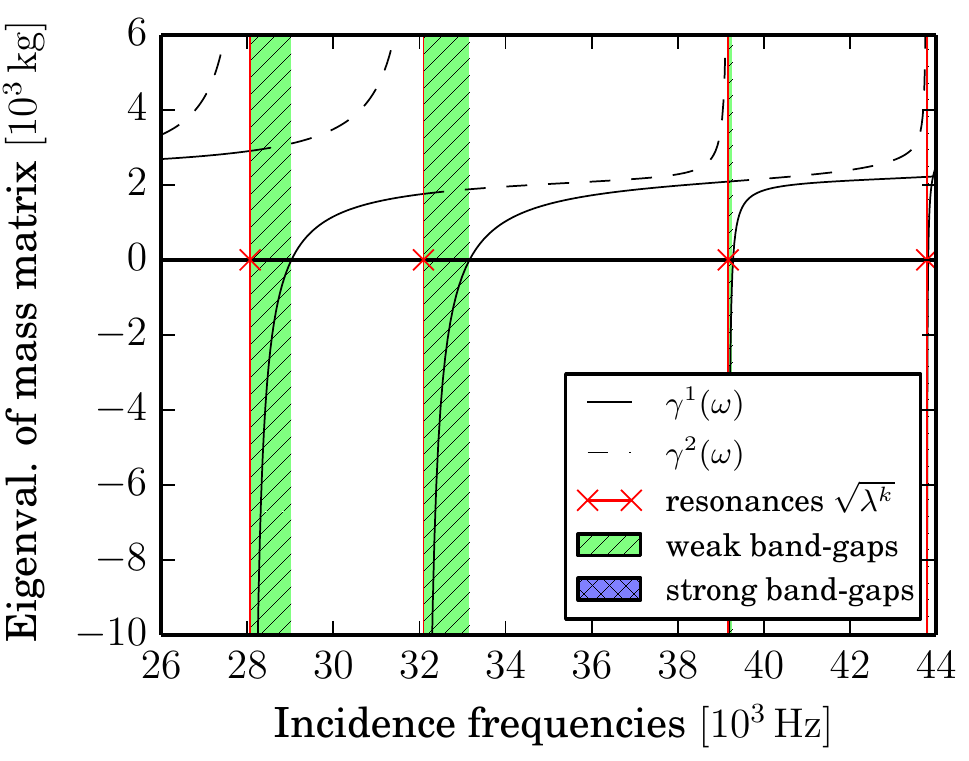}
}
\subfigure[Periodic cell, $\veps = 1$]{
\includegraphics[height=0.36\textwidth]{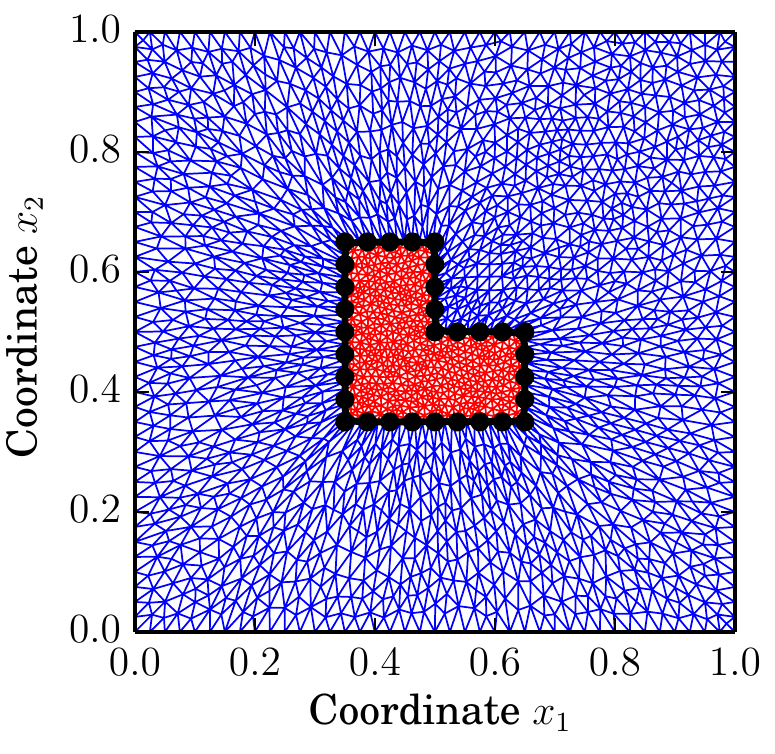}
}
\subfigure[band gaps and resonance frequencies]{
\includegraphics[height=0.354\textwidth]{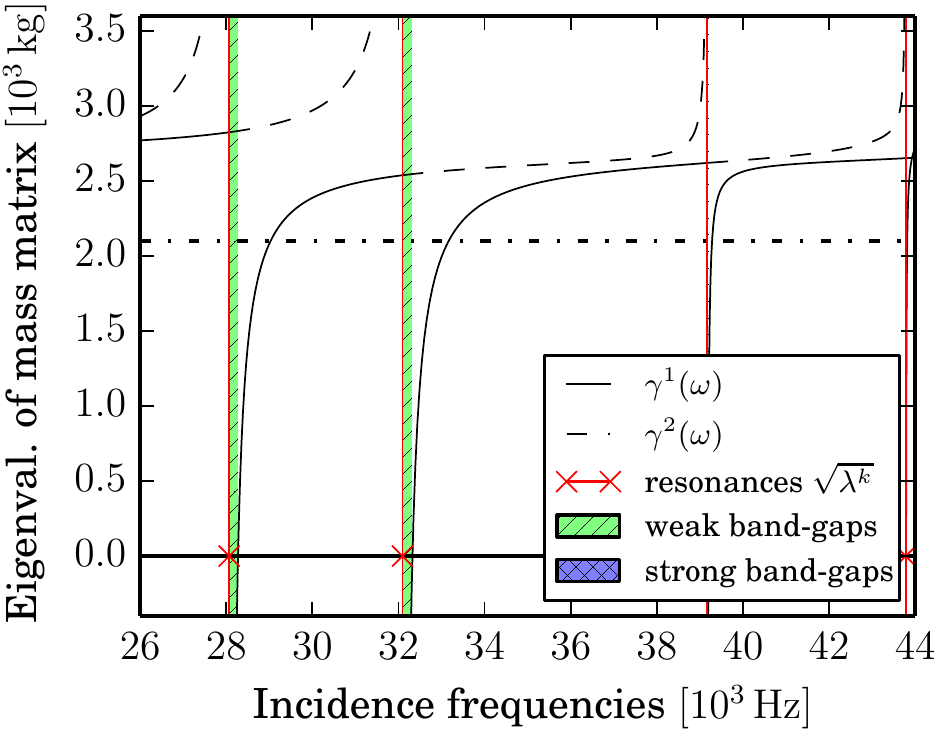}
}
\caption{The distribution of band gaps, resonances, and eigenvalues of mass matrix for different sizes of periodic cell and L-shaped inclusion;
in (f), the dot-and-dash line represents the initial $x$-axis of figures (b) and (d),
which is shifted by $\rho_1(1-\kappa^d)/\kappa^d=2.099\cdot10^3\,\mathrm{Hz}$,
{where $d=2$ and $\kappa=2$}.}
\label{fig:scale}
\end{figure}

We depict the size effect in Figures~\ref{fig:scale} (ab)
and (cd), where the band gap distribution, eigenvalues of mass matrix, and resonance frequencies are plotted for two sizes of periodic cell, $\veps=1$ and $\veps=0.5$, respectively.

It can be seen that decreasing the microstructure size by $1/2$ produces an increase of the frequencies $\{\lamsr{r}\}$, $r\in \Rcal$ of the spectrum by the corresponding factor of $2$;
thus, all the relevant properties are invariant under rescaling the frequency axis, see the difference between Figures~\ref{fig:scale}~(b) and (d).

\begin{figure}[h]
\centering
\includegraphics[height=0.5\textwidth]{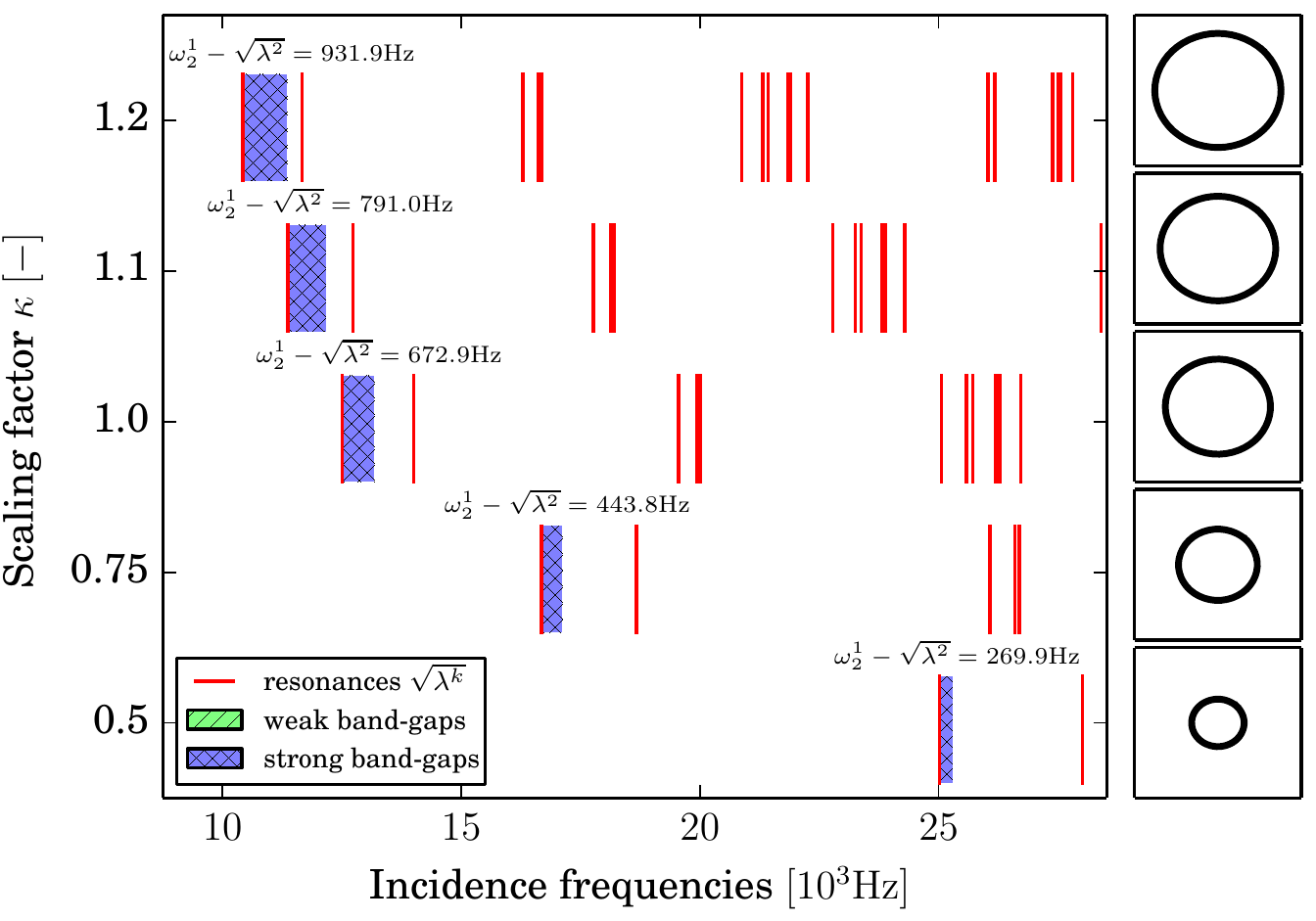}
\caption{The change of bang-gap size depending on a change in the volume fraction $\phi_2 = |Y_2|/|Y|$ by the scaling factor $\kappa$.}
\label{fig:study-fraction}
\end{figure}

\subsection{Effect of volume fraction $|Y_2|/|Y|$}
\label{sec-RS2}
Intuitively, by increasing the volume fraction $\phi_2 = |Y_2|/|Y|$, phononic structures with larger band gaps can be designed while retaining the same range of frequencies.
In the previous section, we have discussed the proportional scaling of the whole microstructure, which produces the corresponding scaling of frequencies.
In Figure~\ref{fig:study-fraction} and Table~\ref{tab:effect_of_volume_fraction}, we show more complicated change of the volume fraction $|Y_2|/|Y|$ of the inclusion in the periodic cell.
As illustrated in Figures~\ref{fig:scale}~(ab) and (ef), the change of the inclusion size leads to a combined effect, such that the horizontal axis of frequencies is linearly scaled while the vertical axis of mass matrix eigenvalues is mapped by an affine transformation.

To explain this phenomenon, let us consider the following affine mapping $\Fcal_\kappa:y_i \mapsto \kappa (y_i - \bar y_i) + \bar y_i$, $i = 1,\dots, d$, where $(\bar y_i) = \bar y \in Y$ is a given point and $\kappa \in \RR^+$.
We introduce new coordinates $z_i = \kappa (y_i - \bar y_i) + \bar y_i$ for $y \in Y_2$.
This transforms the inclusion $Y_2$ to $Z_2 = \Fcal_\kappa(Y_2)$, whereby $\ol{Z}\subset Y$ and $Z_1 := Y \setminus \ol{Z_2}$ is the transformed matrix part; 
obviously, there exists a homeomorphic mapping $\Gcal_\kappa:y \mapsto z$, such that $Z_1 := \Gcal_\kappa(Y_1)$, however, we do not need to know this mapping as far as the material in $Y_1$ is homogeneous.
Thus, the  decomposition of the periodic unit cell   transforms to $Y = Z_1 \cup \bar Z_2$.
By $\wtilde{A}$ we denote the corresponding quantity $A$ defined in sub-cells $Z_1$ and $Z_2$ obtained by transformations $\Gcal_\kappa$ and $\Fcal_\kappa$.
We shall now derive the transformed averaged density $\wtilde{\aver{\rho}}$. Using $ \kappa^d = \det{(\nabla_y\Fcal_\kappa)}$, thus $|Z_2| = \kappa^d |Y_2|$, we obtain
\begin{equation*}
\begin{split}
\wtilde{\aver{\rho}} & = \sum_{\beta=1,2} \rho^\beta \frac{|Z_\beta|}{|Y|} = 
\frac{\kappa^d}{|Y|}  \sum_{\beta=1,2} \rho^\beta + \frac{\rho^1}{|Y|}(1 - \kappa^d|Y_2|) - \rho^1\kappa^d |Y_1|\\
& = \kappa^d \aver{\rho} + \rho^1(1 - \kappa^d)\;,
\end{split}
\end{equation*}
where $\aver{\rho}$ is given by \eq{eq-36}$_2$. The transformed eigenvalue problem yields spectral elements $(\tilde \lam^r, \tilde \phibf^r)$ satisfying
\eq{eq-36a} transformed by $\Fcal_\kappa$ (note $\nabla_z = \kappa^{-1}\nabla_y$
\begin{equation*}
\begin{split}
\int_{Z_2} [\bar\Cop^2 \eeb^z(\tilde\phibf^r)]:\eeb^z(\vb) & = \tilde\lam^r \int_{Y_2} \rho^2 \tilde\phibf^r \cdot \vb, \quad \int_{Z_2} \rho^2 \tilde\phibf^r\cdot\tilde\phibf^s  = \delta_{rs}\;,\\
\Leftrightarrow & \\
\int_{Y_2}\kappa^{-2} [\Cop^2 \eeb^y(\tilde\phibf^r(z(y))]:\eeb^y(\tilde\vb)\kappa^d & = \tilde\lam^r \int_{Y_2} \rho^2 \tilde\phibf^r(z(y)) \cdot \tilde\vb\kappa^d, \quad \int_{Y_2} \rho^2 \tilde\phibf^r(z(y))\cdot\tilde\phibf^s(z(y)) \kappa^d= \delta_{rs}\;, 
\end{split}
\end{equation*}
for all $\vb \in \HOdb(Z_2)$ and all $\tilde\vb \in \HOdb(Y_2)$,
which reveals
\begin{equation*}
\begin{split}
\tilde\lam^s = \kappa^{-2}\lam^s,\quad
\tilde\phibf^s(z(y)) = \kappa^{-d/2} \phibf^s(y)\;, \quad s \in \Rcal_\theta\;,\\
\wtilde{\mb^s} := \int_{Z_2}\rho^2\tilde\phibf^s = \int_{Y_2}\rho^2\kappa^{-d/2}\phibf^s(y)\kappa^d = \kappa^{d/2} \phibf^s.
\end{split}
\end{equation*}
Hence, by invoking the definition  \eq{eq-37} of $\Mb(\om)$ transformed by $\Fcal_\kappa$ and using the observation listed in  Section~\ref{sec-RS2}, we can derive the following expression:
\begin{equation}\label{eq-R4}
\begin{split}
\wtilde{\Mb}(\om) & = \wtilde{\aver{\rho}}\Ib - \frac{1}{|Y|}
\sum_{r \in \Rcal_\theta} \frac{\om^2}{\om^2 - \tilde\lam^r}\wtilde{\mb^r}\otimes\wtilde{\mb^r} \\
& = \left(\kappa^d{\aver{\rho}}+ \rho^1(1 - \kappa^d)\right)\Ib - \frac{1}{|Y|}\sum_{r \in \Rcal_\theta} 
\frac{\kappa^2\om^2}{\kappa^2\om^2 - \lam^r}\kappa^{d/2}\mb^r\otimes \kappa^{d/2}\mb^r \\
& = \kappa^d {\Mb}(\kappa\om) + \rho^1(1 - \kappa^d)\Ib\;.
\end{split}
\end{equation}
It is worth to note that for $\kappa>1$, the frequency bands relevant to $\wtilde{\Mb}$ are decreased as well as the eigenvalues due to the negative  second term in \eq{eq-R4}.
This provokes an extension of the band gaps, see the discussion below and
Figure~\ref{fig:study-fraction} which illustrates this effect for the circular inclusion and  the variation of $\kappa \in \{0.5, 0.75, 1.0, 1.1, 1.2\}$.

\begin{table}
\centering
\begin{tabular}{c|ccc|c}
\hline
$\kappa$ & $\sqrt{\lam_2}$ & ${\ol{\om}}$ & $\sqrt{\lam_3}$ & ${\ol{\om} - \ul{\om}}$ \\
\hline
0.5 & $2.5021 \cdot 10^4$ & $2.5289 \cdot 10^4$ & $2.8011 \cdot 10^4$ & $268.1$ \\
0.75 & $1.6680 \cdot 10^4$ & $1.7123 \cdot 10^4$ & $1.8674 \cdot 10^4$ & $442.6$ \\
1.0 & $1.2510 \cdot 10^4$ & $1.3182 \cdot 10^4$ & $1.4005 \cdot 10^4$ & $672.1$ \\
1.1 & $1.1373 \cdot 10^4$ & $1.2163 \cdot 10^4$ & $1.2732 \cdot 10^4$ & $790.2$ \\
1.2 & $1.0425 \cdot 10^4$ & $1.1357 \cdot 10^4$ & $1.1671 \cdot 10^4$ & $931.2$ \\
\hline
\end{tabular}
\caption{Effect of volume fraction.
Notation: $\ul{\om} = \sqrt{\lam_2}$ is the lower bound, $\ol{\om} = {\om_2^1}$ is the upper bound of the 2nd weak band gap.
}\label{tab:effect_of_volume_fraction}
\end{table}

\subsection{Discussion of the rescaling and resizing effects}\label{sec:discussion-rescaling}
It has been shown in Section~\ref{sec-RS1}, how the structure size modifies the range of frequencies.
In this paragraph and in~\ApxA, we denote by $\ul{\om}$ and $\ol{\om}$ the lower and upper band gap bounds, respectively; the same notation is adhered in Tab.~\ref{tab:effect_of_volume_fraction}. 
In particular, the band gap $G^{\veps_0}$ of periodic structures characterized by the scale $\veps_0 \ll 1$ can be obtained easily by rescaling the band gap $G^1 = ]\ul{\om},\ol{\om}[$  computed for fictitious scale $\veps=1$ by factor $1/\veps^0$, so that $G^\veps = 1/\veps^0]\ul{\om},\ol{\om}[$.
This simple rule reveals limitations of exploiting the band gap effect in an audible frequency range (up to 20 kHz) for a given contrast of the two materials. 
On one hand, for the structure with $\veps = 0.01$, such that the real-sized cell $\veps Y$ is the square $1\times 1$ cm, the first band gap is in the range of 1000 kHz, see Table~\ref{tab:size_of_microstructure}.
On the other hand, for larger structures, say $\veps \approx 1$ for which  the band gaps are predicted in the range of 10 kHz, the model is not relevant, since the characteristic size of the heterogeneities is comparable with wave lengths\footnote{For  the composite made of the aluminium alloy matrix and the epoxy inclusions, the wave lengths of waves at $\om = 14$ kHz are at the range of $\approx 1$ m; in particular, with reference to Tab.~\ref{tab:size_of_microstructure}, for $\veps\in\{1, 0.1, 0.01\}$, the respective wavelengths are $\{2.89, 2.01, 1.52\}$m.}.

To allow for small structures with the characteristic size $\approx 1$cm and the band gaps in the audible spectrum, much softer material would need to be used to keep the resonant frequencies in the inclusion under the 1 kHz.
This can be achieved by means of resonators occupying $Y_2$ which themselves are inhomogeneous, being designed as a coated heavy particles; as an example, a hard sphere (in $\RR^d$) forming the central part coated by a very compliant, though light material, can serve the desired property of a low stiffness due to the coating and a large mass due to the central part. 

The second study reported in  Section~\ref{sec-RS2} demonstrates the quite natural effect of shifting the band gaps to lower frequency bands by increasing the volume fraction $\phi_2$. As another effect of such design modification,
the eigenvalues $\tilde\gamma(\om)$ of tensor $\wtilde{\Mb}(\om)$ become smaller (in comparison with $\gamma(\kappa\om)$) which leads to an extension of the band gaps due to the increased roots  $\wtilde{\om}^s$, $s = 1,\dots,d$ for which $\tilde\gamma^s(\wtilde{\om}^s) = 0$.

For completeness, we remark that the change in $\phi_2 = |Y_2|/|Y|$ can also be achieved by keeping the inclusion volume $|Y_2|$ unchanged while changing the cell size (note that $|Y|\not =1$).
In this case depicted by Figures~\ref{fig:scale}~(cd) and (ef), the change of the cell size keeps the resonance frequencies because they depend on the same inclusion shape and size. 
The eigenvalues of mass matrix are scaled with affine transformation,
however their qualitative behaviour remains the same as the shapes of corresponding curves are identical in (d) and (f);
in (f), the dash-and-dot line depict the initial horizontal axis from (b) or (d). Therefore, the size of band gaps heavily depends on volume fraction; importantly, its size can be easily predicted a posteriori, without recalculation of eigenvalues problem \eqref{eq-R1}.

The following remark is devoted to the \emph{choice of the unit cell} $Y$.

\begin{myremark}{rem-4}
The unit cell $Y = ]0,1[^d$  is the simplest and the most often used cell to generate a periodic structure, although $Y$ can be chosen more generally as a parallelepiped.
The assumption $|Y| = 1$ is frequently used, but not mandatory. In classical homogenization, in the effective properties do not depend on the size of $Y$, although the scale separation in the context of $\veps$ being small is understood in the sense $|Y| = O(1)$. In our situation, the effective mass $\Mb(\om)$
depends on the size of the microstructure, \ie on the scale, as discussed in Section~\ref{sec-RS1}, and on the volume fraction $\phi_2 = {|Y_2|}/{|Y|} = 1-\phi_1$, as discussed in Section~\ref{sec-RS2}. Contrary to this observation,
the elasticity $\Dop$ defined in 
\eq{eq-C*} does not depend on the scale of the structure, however, it depends on the mutual relationship between the shapes of the inclusion, $Y_2$, and their relative positions. Indeed,
while keeping the shape of $Y_2$ and the fraction $\phi_1$ constant, \ie $|Y|=1$,
  different effective elasticity of the homogenized medium is computed when changing the ratios of the 2 diagonals in the parallelepiped $Y$.
 By virtue of the periodicity, such arrangement with non-orthogonal faces of $Y$ corresponds to different relative placements of inclusions in the composite.
\end{myremark}

\section{Optimal design of the phononic structures}
\label{sec:opt}


We focus on the problem of maximization of one of the lowest band gaps by means of changing the shape of inclusions $Y_2$.
Due to the scaling properties presented in Section~\ref{sec-RS1}, the optimization can be related to the unit cell $Y$.
The required frequencies of band gaps can be obtained a~posteriori by choosing the appropriate microstructure size,
\ie scaling the unit cell by an appropriate factor,
see~\ApxA, to obtain a required frequency range.
 
  We assume the composite is made of two materials in the sense of Remark~\ref{rem-3}.
For a given $k \geq 1$, the objective function can be defined in terms of the band gaps $G_k^S$ or $G_k^W$:
\begin{equation*}
\begin{split}
\Phi_S & = |G_k^S|\;, \quad \Phi_W  = |G_k^S| + |G_k^W|\;.
\end{split}
\end{equation*}
Since the existence of $G_k^S$ is guaranteed for symmetric shapes of $Y_2$ only, cf. \cite{Avila2008MMS_BG}, whereas $|G_k^W|>0$ whenever $\lam^{k+1}>\lam^k$, we chose rather the function $\Phi_W$ to express the optimality criterion.
By virtue of the Remark~\ref{rem-1}(v) we shall assume the existence of $\om_k^1$
such that 
$\gamma^1(\om_k^1) = 0$  and define the objective function
\begin{equation}\label{eq-objfun2}
\begin{split}
\Phi_k & = \om_k^1 - \sqrt{\lam^k}\;.
\end{split}
\end{equation}
From now on we shall rather restrict to 2D problems, thus, we consider $d= 2$.
The optimization problem is presented in three steps: First we introduce the shape parametrisation of domain $Y_2$,
then we summarize the constraints which make the link between the design parameters and the state variables, finally we formulate the optimization problem.

 \ER{ In the present study, we have chosen to maximize the weak band gaps
(WBG) rather than the strong band gaps which, besides their obvious advantages, in the context of optimization bring more difficulties
 related to the non-smoothness of the objective function. Optimization
 of WBG is of interest since a larger band gap width can be
 achieved. Also in the WBG, the structure behaves as a polarization
 filter, which can be of interest for some applications.  Moreover,
 the WBG permits for larger anisotropy of the effective material
 elasticity.}

\subsection{Shape parametrization}

We are interested in variation of the shape of the inclusion $Y_2$ bounded by the interface  $\Gamma_{12} = \pd Y_2$. We assume existence of a design parametrization, such that 
\begin{equation}\label{eq-adm}
\begin{split}
& Y_2 \mbox{ is nonempty, simply connected Lipschitz domain};\\
& \Gamma_{12} \subset Y, \mbox{ thus, } \dist(\Gamma_{12},\pd Y) > 0.
\end{split}
\end{equation}

The shape of $Y_2$ can be defined by a smooth curve described by a finite number of design variables $\Valp$.
For this, we introduce a smooth mapping $\Sigma : \Valp \mapsto \Gamma_{12}(\Valp)$ 
In our work, $\Sigma$ is defined in terms of the cyclic spline parametrization of the closed curve $\Gamma_{12}$.
By $\{\Pb^i\}_{i=0}^{n-1}$, where $\Pb^i = (P_j^i)$ we denote the control polygon of a B-spline of order $p < n$,
where $n$ is the number of the control points $\Pb^i \in \RR^2$. We consider the uniform parametrization defined by the knot vector $\tb = \{t_i\}$ with $t_i = i$ for $i \in \ZZ$, such that any point $y(t)\in\Gamma_{12}$ is expressed by
\begin{align}\label{eq-Bspl0}
y_j(t) = \sum_{i=0}^{n-1} P_j^{i} \Bsc_{i,p}(t)\;,
\end{align}
where $\Bsc_{i,p}(t)$ are the B-spline basis functions.

The control points $\Pb^i$ are modified by the design variables $\Valp^i$. In particular, the modified interface $\Gamma_{12}(\Valp)$ is given according to \eq{eq-Bspl0} where  
$\Pb^i := \Xb^i + \Valp^i$ with the ``initial shape'', $\Gamma_{12}^0 = \Gamma_{12}(\bmi{0})$ defined by the polygon $\{\Xb^i\}$, so that  
\begin{align*}
\Gamma_{12}(\Valp) = \left\{ y_j(t) = \sum_{i=0}^{n-1} (X_j^{i} + \alpha_j^i)\Bsc_{i,p}(t)\;, \; j=1,2\quad \mbox{ for any } t\in [0,n[\right\}.
\end{align*}
Admissible designs of the inclusion $Y_2$ are generated by $\Valp^i\in \Acal=\{\Valp\in\xR^{2\times(n-1)}: \eq{eq-adm} \mbox{ hold}\}$.

By virtue of Remark~\ref{rem-3}, \ie assuming two constant materials  distributed in $Y_1$ and $Y_2$, respectively, the objective function depends only on the shape of $\Gamma_{12}$. However, to derive the sensitivity analysis formulas and to handle 
the numerical model associated with a computational finite element partitioning,  we need to introduce a parametrization of domain $Y$.
Let $Y_\beta(\Valp)$, $\beta=1,2$ be the domains shaped by $\Gamma_{12}(\Valp)$, recalling $\pd Y$ is fixed.
Further let $Y_\beta^0$ be the reference (initial) domains\footnote{Domains $Y_\beta^0$ are introduced in the context of FE mesh used for the numerical solutions.
The so-called re-meshing is accomplished using fields $\vec\Vcal^i = (\Vcal_j^i)$ computed by solving \eq{eq-Uvec2}.};
obviously $Y_1^0\cup Y_2^0 \cup \Gamma_{12}^0= Y$.
For a modified design represented by $\Valp$ we introduce a smooth mapping $\Fcal:(\Valp , Y_\beta^0)\mapsto Y_\beta(\Valp)$
such that
\begin{equation}\label{eq-Uvec}
\begin{split}
Y_\beta(\Valp) &= \{y + \vec\Ucal(y,\Valp), \mbox{ where } y \in Y_\beta^0\}\;, \\
\vec\Ucal(y,\Valp) & =  \sum_{i=0}^{n-1} \sum_{i=1}^{2}\alpha_j^i\vec\Vcal^{i(j)}(y)\;, \quad y \in Y_\beta^0\;,
\end{split}
\end{equation}
where $\Vcal_k^{i(j)}$, $k = 1,2$ are components of the so-called ``design velocity field'' associated with the optimization variable $\alpha_j^i$.
There are various approaches how to establish $\vec\Vcal^{i(j)}$.
A simple one is based on the following auxiliary elasticity problems imposed in $ Y_1^0\cup Y_2^0$, \ER{see \eg \cite{Rodenas2004}.}
For $i\in\{0,1,\dotsc,n-1\}$, being associated with the control points $\Pb^i$ of the B-spline, and $j\in\{1,2\}$ (the coordinate index),
find $\vec\Vcal^{i(j)}\in \HOdb(Y)$ satisfying {(in the distributional sense)}
\begin{equation}\label{eq-Uvec2}
\begin{split}
\nabla\cdot (\Aop \eeb(\vec\Vcal^{i(j)}) & = 0\;,\quad \mbox{ in } Y_1^0\cup Y_2^0\;,\\
\vec\Vcal^{i(j)} & = 0 \;,\quad \mbox{ on } \pd Y\;,\\
\vec\Vcal^{i(j)}(y(t)) & = \deltabf^{(j)} \Bsc_{i,p}(t)\quad \mbox{ for } y(t)\in\Gamma_{12}^0, 0\leq t<n\;,
\end{split}
\end{equation}
where $\deltabf^{(j)} = (\delta_k^{(j)}) = (\del_{kj})$ and $\Aop = (A_{ijkl})$ is the elasticity tensor of an artificial isotropic material.
It is worth to remark that the reference domains $Y_\beta^0$ can be updated during the optimization process, see Section~\ref{sec:numerical-examples}. \ER{In our numerical treatment, fields $\vec\Vcal^{i(j)}$ are defined at finite element mesh points, so that new fields are computed only after re-meshing.}

\subsection{Constraints}\label{sec-constr}
%
  Although the optimization criterion concerns with the band gap, thus, with the wave propagation phenomena, the material itself can be exposed independently to other types  of static loads.
On one hand, the size of the band gaps, being expressed by objective function 
\eqref{eq-objfun2}, can be easily enlarged by increasing the inclusion size,
see Section~\ref{sec:rescaling} about scaling or numerical results \cite{RoVoHe2014}.
On the other hand, the change in shape of the inclusions may lead to a very compliant material,  or even to a loss of material integrity; as an example in Section~\ref{sec-RS2} we demonstrated, how such effects can be provoked while desired increasing the bang-gap width by means of increasing the inclusion relative size. \newtext{To avoid possible collapse of the optimized material, it is natural to guarantee some minimum stiffness of the phononic structure.  For this we define the tensor $\Dop_\mathrm{min}$ which represents admissible effective elastic properties \eqref{eq:homog_elas}, and require
\begin{align}
\label{eq:const_elas}
[\Dop(\Valp)\Ve]:\Ve \geq [\Dop_\mathrm{min}\Ve]:\Ve > 0
\quad
\text{for any macroscopic strain }\Ve\in\set{R}^{d\times d}_{\mathrm{sym}},\quad |\eb|>0\;.
\end{align}
It can be seen that this condition constraints the smallest eigenvalue $\varsigma(\Valp)$ of the elasticity tensor $\Dop(\Valp)-\Dop_\mathrm{min}$. Therefore, \eq{eq:const_elas} can be formulated alternatively, as follows
\begin{align}
\label{eq:const_elast-R}
\varsigma(\Valp) \geq \varsigma^* \;,
\end{align}
where $\varsigma^* >0$ is a given value. 

In the numerical examples treated in Section~\ref{sec:numerical-examples},
 we simplify the constraint \eqref{eq:const_elas} by testing
only with selected strains. This may induce an anisotropy of the optimized material, since 
the optimal shapes have a predominant orientation,
see Sections~\ref{sec:optimization-with-constraint-on-elastic-diagonal}.
In contrast, if $\Dop_\mathrm{min}$ is an isotropic material, the formulation based on the eigenvalue constraint \eqref{eq:const_elast-R} does not introduce any additional preferred orientation and shape of the inclusions.}

\subsection{Setting the optimization problem}
\label{sec:setting-the-optimization-problem}
We shall deal with the optimization problem,
constituted by the objective function describing the band gap size \eqref{eq-objfun2} and by
the constraint \eqref{eq:const_elas}, recalling its possible variant \eq{eq:const_elast-R} with $\Dop_\mathrm{min} = \Iop$, as discussed above, 
\begin{subequations}
\label{eq:opt_problem}
\begin{align}\label{eq-OP0}
& \min_{\Valp \in \Acal} -\Phi_k(\Valp)
\\
\label{eq:opt_problem_con}
\mbox{ s.t. }& (\Dop(\Valp)-\Dop_\mathrm{min})\Ve:\Ve \geq 0\quad\forall\Ve\in\set{R}^{2\times 2}_{\mathrm{sym}}.
\end{align}
\end{subequations}

Problem \eq{eq-OP0} has a complicated structure because of 
the evaluation of the objective function value, as explained using Algorithm~\ref{alg:obj_fun}. It involves the design parameters $\alphabf$ and the state variables: the objective function depends on the spectrum $(\lam^r,\phibf^r) $, $r\in \Rcal$, whereas the elasticity constraint depends on the characteristic elastic response $\wb^{ij} \in \Hpdb(Y)$.

\begin{algorithm}[h]
\caption{Calculation of band gap size in continuous setting}
\label{alg:obj_fun}
\begin{algorithmic}[1]
\Require
{Control points $\Pb^i$ for $i=0,\dotsc,n-1$ of B-spline, index $k\in\set{N}$ of band gap}
\Procedure{band gap-size: $\Phi_k$}{$\Valp$}
\State{Set the domains $Y_1=Y_1(\Valp)$ and $Y_2=Y_2(\Valp)$}
\State{Calculate eigenvalue problem \eqref{eq-36a} for $(\lam^r_h,\phibf^r_h) \in \RR \times \HOdb(Y_2)$ with $r \in\set{N}$ 
}
\State{Calculate eigenmomentum $\mb^r = \int_{Y_2}  \phibf^r$ and average density $\aver{\rho} = \sum_{k=1,2}\int_{Y_k} \rho^k$.}
\State{Calculate the relevant spectrum with the index set $\Rcal\subset\set{N}$ in accordance with \eqref{eq-evS1}}
\State{Calculate $\omega_k^r\in \Lam_k$ using \eqref{eq-evM} such that $r$-th eigenvalue of $M(\omega_k^r)$ is zero
}
\EndProcedure{\textbf{: return } $\Phi_k(\Valp)=\omega_k^r(\Valp)-\sqrt{\lam^k(\Valp)}$}
\Comment{band gap size}
\end{algorithmic}
\end{algorithm}

For clarity,
we introduce  the following detailed formulation which brings explicitly relationships between the state variables involved in the problem on one hand, and the constraints, on the other hand:
\begin{equation}\label{eq-OP}
\begin{split}
& \min_{\alphabf \in \Acal} -\Phi_k(\om^1_k(\alphabf), \lam^k(\alphabf))\\
\mbox{ s.t. }&  
\left\{
\begin{array}{ll}
(i) \quad & (\lam^r,\phibf^r),\;r\in\Rcal \mbox{~ given by the eigenvalue problem ~} \eq{eq-36a}\;, \\
(ii) \quad & \om_k^1 \mbox{ solves } \eq{eq-evroot} 
\;,\\
& \mbox{ where } (\lam^r,\phibf^r) \mapsto \Mb(\cdot) \mbox{ is defined by } \eq{eq-37}
\;, \\
(iii) \quad & \mbox{ the elasticity constraint \eq{eq:const_elas} holds}\;,\\
&  \mbox{ where }(\wb^{ij}) \mapsto \Dop \mbox{ is defined by }\eq{eq-C*a},\eq{eq-C*}\;.
\end{array}\right.
\end{split}
\end{equation}

We now discuss several aspects of the optimization problem \eq{eq-OP}.  

\begin{enumerate}
\item The evaluation of elastic constraint \eqref{eq:const_elas} or \eqref{eq:opt_problem_con} requires calculation of effective elastic properties $\Dop(\Valp)$ from \eqref{eq:homog_elas}
and calculation of the smallest eigenvalue for $(\Dop(\Valp)-\Dop_\mathrm{min})$,
which has to be nonnegative.
\item Note that $\Valp \mapsto Y_2$, therefore  $\Valp \mapsto (\lam^r,\phibf^r) $, $r\in \Rcal$ and $\Valp \mapsto (\wb^{ij})$.
This dependence is explained in detail in Section~\ref{sec:sa}.
\item The sensitivity analysis
of the objective function $\Phi_k$ does not necessitate any adjoint problem to be solved, \ER{see \cite{Rohan2009-WCSMO,pzph-roma06} cf. \cite{Haug1986book}.}
In a case of eigenvalues with multiplicity higher than one, only a subgradient of $\Phi_k$ exists.
\item Although the stiffness constraint is associated with the state problem variable $(\wb^{ij})$, to compute the sensitivity,
there is no need to introduce an adjoint variable, \ER{see \eq{eq-S19} in \ref{apx-B}, cf. \cite{Rohan2006sa_piezoel}.}
\end{enumerate}

\section{Sensitivity analysis}\label{sec:sa}
In order to use a gradient-based optimization method,
the sensitivity analysis associated with the optimization problem \eq{eq:opt_problem} is needed.
Therein, the implicit dependence of $\hat\om_k(\alphabf), \lam^k(\alphabf)$ and $\Dop(\alphabf)$ on the optimization variables 
$(\alphabf)$ can be understood by virtue of the unfolded formulation \eq{eq-OP}.

The aim of this section is to derive the sensitivity formulas which are needed to compute the gradient $\nabla_{\alphabf} \Phi(\alphabf) = (\pd {\Phi}(\alphabf)/\pd \alpha_j^i)$ {of objective function $\Phi(\alphabf)$ computed according the Algorithm~\ref{alg:obj_fun}};
the gradient of the constraint involving elastic properties $\Dop(\Valp)$ is given separately in~\ApxB~ by formula \eq{eq-S19}.
We follow the standard approach based on the
concept of the shape and material derivatives; for all details on this issue we refer to \cite{Haug1986book} and
\cite{Haslinger1988book,Haslinger2003book}. To associate a locally perturbed shape with 
a shifted  ``material point'' position
defined for any $y \in Y$, the design velocity field $\vec\Vcal$ is introduced, such that
\begin{equation}\label{eq:ph4}
z_i(y,\tau) = y_i + \tau\Vcal_i(y)\;,\quad i=1,2\;,
\end{equation}
where $\tau$ is the ``time-like'' variable;
here we use the design velocity field $\vec\Vcal$ in much the same sense as the one introduced in \eq{eq-Uvec}, where it was associated with particular design variables $\alpha_j^i$. 

  Throughout the text below we shall use the notion of the
following derivatives:
\begin{center}
\begin{tabular}{lcl}
$\delta(\cdot)$ & \dots & total ({\it material}) derivative \\ 
$\delta_\tau(\cdot)$ & \dots & partial ({\it shape})
derivative \wrt $\tau$ in the context of \eq{eq:ph4}\;.
\end{tabular}
\end{center}
These derivatives are computed as the directional derivatives in the
direction of $\vec\Vcal(y)$, $y \in Y$, see \eg \cite{Haslinger2003book}
for the classical results.
In this paper we use $\vec\Vcal \in \HOdb(Y)$, so that $\pd Y$ is not being perturbed.

Below we derive formulas needed to compute $\delta \Phi$ for any $\vec\Vcal$.
Consequently, $\pd \Phi(\Valp)/\pd \alpha_j^i$ is evaluated using $\delta \Phi$ where one substitutes $\vec\Vcal = \vec\Vcal^{i(j)}$, see \eq{eq-Uvec}.

\subsection{Sensitivity of the spectrum}
The problem of shape sensitivity of eigenvalues and eigenvectors $(\lam^r,\phibf^r)$ for $r\in \Rcal$ is discussed
exhaustively e.g. in \cite{Haug1986book}.
Here we introduce the sensitivity for
the case of single, separated eigenvalues, i.e. $\lam^r\not= \lam^s$ for any $r,s \in \Rcal$.
In such a situation, all elements $(\lam^r,\phibf^r)$ for $r\in \Rcal$ are differentiable with respect to $\tau$.

\begin{myremark}{rem-7}\emph{(Multiple eigenvalues of $\Mb(\om)$)}
On the
other hand, a more complex treatment is required for situations with
eigenvalues of higher multiplicities; then only the directional shape
differentials exists. It should be emphasized that optimal solutions
may often be featured by multiple eigenvalues. In the context of our
problem, such situation would emerge for spherical inclusions or some other symmetric shapes provoking the strong band gaps occurrence, see \eg discussion in paper \cite{Avila2008MMS_BG}, cf. \cite{Taheri-Hassani2014}.
\end{myremark}

{For brevity in what follows we employ following notation
\begin{equation}\label{eq:notation_bilf}
\begin{split}
\aYi{\ub}{\vb}{2} & = \int_{Y_2} [\bar\Cop^2 \eeb^y(\ub)]:\eeb^y(\vb),\quad\quad
\rYi{\ub}{\vb}{2}  = \int_{Y_2} \rho^2 \ub \cdot \vb.
\end{split}
\end{equation}
In analogy we define $\aYi{\ub}{\vb}{1}$ using $Y_1$ and the elasticity tensor $\Cop^1$.}

The spectral problem \eq{eq-36a} can be rewritten using the notation introduced above,
\begin{equation}\label{eq-evS}
\phibf^{k} \in \HOdb(Y_2), \quad \aYi{\phibf^{k}}{\vb}{2} = {\lam^{k}} \rYi{\phibf^{k}}{\vb}{2}\quad
\forall \vb \in \HOdb(Y_2)\;, \quad\quad \varrho_{Y_2}({\phibf^{k}},\phibf^{j})= \delta_{kj}
\end{equation}
Differentiation of \eq{eq-evS} yields
\begin{equation}\label{eq:ph6}
\begin{split}
a_{Y_2}(\delta\phibf^\rr,\vb ) - \lam^r \varrho_{Y_2}(\delta\phibf^\rr,\vb )
= & \;\delta\lam^\rr \varrho_{Y_2}(\phibf^\rr,\vb ) 
 +\lam^\rr \delta_\tau\varrho_{Y_2}(\phibf^\rr,\vb )
 -\delta_\tau a_{Y_2}(\phibf^\rr,\vb )\;, 
\end{split}
\end{equation}
to be satisfied for all $ \vb \in \HOdb(Y_2)$.  Due to
$\delta\phibf^\rr \in \HOdb(Y_2)$, the \lhs of \eq{eq:ph6} vanishes
for $\vb = \phibf^\rr$ so that we have the sensitivity of the $r$-th
eigenvalue (recall the orthonormality $\rYi{\lam^\rr}{\lam^\kk}{2} = \delta_{rk}$):
\begin{equation}\label{eq:ph5}
\delta \lam^r = \delta_\tau a_{Y_2} (\phibf^\rr,\phibf^\rr )
-\lam^\rr \delta_\tau\varrho_{Y_2}(\phibf^\rr,\phibf^\rr )\;.
\end{equation}
The shape derivative $\delta\phibf^\rr$ can be projected to the
Hilbert space constituted by all eigenfunctions
$\{\phibf^\kk\}_{k}$, $k\geq 1$. For this the Fourier coefficients are needed:
\begin{equation}\label{eq:ph7a}
\xi_k^\rr = \varrho_{Y_2}(\delta\phibf^\rr,\,\phibf^\kk),\quad k = 1,2,\dots\;,\quad
\mbox{ so that }\delta\phibf^\rr = \sum_{k\geq 1}\xi_k^\rr \phibf^\kk\;.
\end{equation}
Now the l.h.s. in \eq{eq:ph6}
can be written in the form
\begin{equation}\label{eq:ph7}
\begin{split}
& \;a_{Y_2}(\delta\phibf^\rr,\vb ) - \lam^\rr \varrho_{Y_2}(\delta\phibf^\rr,\vb ) 
=\; \sum_{k \geq 1} (\lam^\kk - \lam^\rr) \varrho_{Y_2}(\phibf^\kk,\vb )\, \xi_k^\rr\;.
\end{split}
\end{equation}
Clearly \eq{eq:ph7} and also the first right hand side term in \eq{eq:ph6} vanish for $\vb = \phibf^\rr$ or for $k = r$. Therefore, from \eq{eq:ph6} and \eq{eq:ph7} we can 
obtain $\delta\phibf^\rr$ up to a multiple of $\phibf^\rr$. 
Using the Fourier coefficients $\xi_k^\rr$ and for any $\phibf^\ss \not= \phibf^\rr$
these two equalities yield
\begin{equation*}
\begin{split}
& \sum_{k \geq 1,\,k \not = r}
\xi_k^\rr(\lam^\kk - \lam^\rr) \varrho_{Y_2}(\phibf^\kk,\phibf^\ss ) = \lam^\rr \delta_\tau\varrho_{Y_2}(\phibf^\rr,\phibf^\ss ) -
\delta_\tau a_{Y_2}(\phibf^\rr,\phibf^\ss )
\end{split}
\end{equation*}
and thus we have for $s \not = r$
\begin{equation}\label{eq:ph6b}
\xi_s^\rr = \frac{
\lam^\rr \delta_\tau \varrho_{Y_2}(\phibf^\rr,\phibf^\ss )
-\delta_\tau a_{Y_2}(\phibf^\rr,\phibf^\ss )}{\lam^s - \lam^r}\;.
\end{equation}
In order to determine $\xi_r^\rr$, we differentiate the identity 
$\varrho_{Y_2}(\phibf^\rr,\phibf^\rr )=1$, thus, we get
\begin{equation}\label{eq:ph6c}
\xi_r^\rr = -\frac{1}{2}\delta_\tau \varrho_{Y_2}(\phibf^\rr,\phibf^\rr )\;.
\end{equation}
Now $\delta\phibf^\rr$ can be evaluated by \eq{eq:ph7a}$_2$.
The partial shape derivatives $\delta_\tau a_{Y_2}(\cdot,\cdot)$, $\delta_\tau \varrho_{Y_2}(\cdot,\cdot)$ are given in~\ApxB by expressions \eq{eq-S15}.

\subsection{Sensitivity of the homogenized mass tensor $\Mb$}
\label{sec:sa_mass}
Total variation of components $M_{ij}$ defined in \eq{eq-37} yields
\begin{equation}\label{eq:ph31}
\begin{split}
\delta M_{ij} = &\; \pd_{\omega} M_{ij}\,\delta\omega 
 +\sum_{r \in \Rcal} \left (\pd_{\lam^\rr}M_{ij}\,\delta\lam^\rr + 
\pd_{\mb^\rr} M_{ij}\cdot\delta \mb^\rr \right ) +
\delta_\tau \aver{\rho} \delta_{ij}\;,
\end{split}
\end{equation}
where $\pd_{\mb^\rr} = (\pd_{\mb^\rr}^i) = (\pd/\pd m_i^\rr)$.
Above we need the partial derivatives:
\begin{subequations}
\begin{align}
\pd_{\omega}M_{ij} &= \frac{2}{\om}(M_{ij}-\aver{\rho}\delta_{ij}) + 
\sum_{r \in {\Rcal}} \frac{2\omega^3}{(\omega^2 - \lam^\rr)^2}
m_i^\rr m_j^\rr \;,\quad\label{eq:ph32} \\
\pd_{\lam^\rr}M_{ij} &= -
 \frac{\omega^2}{(\omega^2 - \lam^\rr)^2}m_i^\rr m_j^\rr\;,\quad\label{eq:ph33} \\
\pd_{\mb^\rr}^k M_{ij}&= -
\frac{\omega^2}{\omega^2 - \lam^\rr}
( \delta_{ki} m_j^\rr +\delta_{kj} m_i^\rr )  \;.\quad\label{eq:ph34} 
\end{align}
\end{subequations}
In \eq{eq:ph31} we need the sensitivity of the eigenmomentum $\mb^r$.
The sensitivity $\delta \phibf^r$ is given by \eq{eq:ph7a}$_2$ involving the 
coefficients $\xi_s^r$, see \eq{eq:ph6b},\eq{eq:ph6c} which depend 
on the shape derivatives introduced in~\eq{eq-S15},~\ApxB. By virtue of \eq{eq-S15}$_2$
it is easy to get
\begin{equation}\label{eq-S20}
\begin{split}
\delta \mb^r & = \rho^2 \int_{Y_2}\left(\delta \phibf^r + \phibf^r \nabla\cdot\vec\Vcal
\right)
\;.
\end{split}
\end{equation}

The partial shape derivative $\pd_\tau \aver{\rho}$ is given by 
\eq{eq-S15}$_3$.

\subsection{Sensitivity of the upper band gap bound $\omega_k^r$}
The upper bound of the phononic gap is defined above in \eq{eq-evroot}.
We derive the sensitivity formula for any index $r$ such that the root $\om_k^r$ exists (in particular, we rely on the existence of $\om_k^1$) within the interval $\Lam_k$.
By virtue of \eq{eq-evM}, using \eq{eq:ph31} and assuming
$\gamma^r(\om_k^r)$ has multiplicity one
\footnote{This assumption is consistent with the isolated spectrum, $\lam^i \not= \lam^j$ for $i\not = j$.},
it is easy to find
\begin{equation*}
\pd \gamma^r(\om_k^r) = (\vb^r)^T \pd \Mb(\om_k^r) \vb^r\;.
\end{equation*}

Sensitivity of $\omega_k^r$ now follows from \eq{eq-evroot} on substituting from \eq{eq:ph31}. We get
\begin{equation*}
\begin{split}
0 & \stackrel{!}{=}  \pd \gamma^r(\Mb(\om_k^r)
 = (\vb^r)^T\left( \pd \Mb(\om_k^r) \right)\vb^r  \\
& = x_j^r \left( \pd_\om M_{ij}(\om_k^r) \,\delta\hat\om 
 +\sum_{r \in \Rcal} \left (\pd_{\lam_\rr}M_{ij}(\om_k^r)\,\delta\lam_\rr + 
\pd_{\mb^\rr} M_{ij}(\om_k^r)\cdot\delta \mb^\rr \right ) +
\delta_\tau\aver{\rho} \delta_{ij}(\om_k^r)
 \right)x_j^r
\;,
\end{split}
\end{equation*}
where all functions involved are evaluated at $\omega_k^r$, hence
\begin{equation*}
\begin{split}
 \delta{\omega_k^r} 
= & - \frac{1}{(\vb^r)^T \pd_{\om} \Mb(\om_k^r) \vb^r}
x_j^r \left(\sum_{r \in \Rcal} \left (\pd_{\lam_\rr}M_{ij}(\om_k^r)\,\delta\lam_\rr + 
\pd_{\mb^\rr} M_{ij}(\om_k^r)\cdot\delta \mb^\rr \right ) +
\delta_\tau \aver{\rho} \delta_{ij}(\om_k^r)
 \right)x_j^r
\;,
\end{split}
\end{equation*}
To evaluate $\delta{\omega_k^r}$, the expressions \eq{eq:ph5}, \eq{eq:ph32}, \eq{eq-S20} are employed. The partial shape derivatives $\delta_\tau \aver{\rho}$ is computed using  expression \eq{eq-S15}$_3$ given in~\ApxB.

\section{Numerical implementation and examples}
\label{sec:numerical-examples}
In this section, we give several illustrations of the optimization problems described in previous sections. In particular, Section~\ref{sec:discretization} is devoted to the discretisation and implementation issues. With reference to the theoretical parts of the paper, an emphasis is put on the algorithmic framework that was used to calculate and optimize band gaps. The initial structures and some parameter setting considered for the numerical optimization are presented in Section \ref{sec:optimization-problems}.
In Sections~\ref{sec:optimization-with-constraint-on-elastic-diagonal} and \ref{sec:optimization-with-constraint-on-elastic-eigenvalues} we report several numerical examples of the shape optimization with different initial layouts, and different objective and constraining criteria, as introduced in the previous section. 
It appears that complications associated with the band gap numerical optimization are highly influenced by the selected interval $[\sqrt{\lam^k},\sqrt{\lam^{k+1}}]$, $k=1,2,\dots$, where the band gap of interest is located. Optimization of the weak band gap size in the first interval $[\sqrt{\lam^1},\sqrt{\lam^2}]$ 
seems to bring about less difficulties contrary to the  optimization in the second interval. 
\newtext{In Sections~\ref{sec:weak-band-gap-in-the-first-interval} and \ref{sec:weak-band-gap-in-the-first-interval-eig} we report on the numerical solutions of the optimization of the band gaps located in the first interval, whereby two different elasticity constraints are prescribed. In analogy, optimization associated with the second interval is considered in Sections \ref{sec:weak-band-gap-in-the-second-interval} and \ref{sec:weak-band-gap-in-the-second-interval-eig}.
}

\subsection{Discretisation and implementation}
\label{sec:discretization}
Here, we briefly describe discretisation and implementation of equations leading to calculation and optimization of band gaps.
For discretisation,
Finite element method (FEM) based on Galerkin approximation was used with triangular elements and linear polynomials as basis functions.
Hence, all the function spaces such as $\HOdb(Y_1)$ are approximated by corresponding discrete ones $\set{H}^1_{0,h}(Y_{1,h})$ with discretisation parameter $h$ corresponding to a characteristic mesh size. 
The smooth B-spline inclusion boundary $\Gamma(\Valp)$ is approximated with a polygon $\Gamma_h(\Valp)$ depending on the FE partitioning of domains $Y_1$ and $Y_2$. Thus, we deal with approximating domains $Y_{1,h}(\Valp)$ or $Y_{2,h}(\Valp)$, however, still satisfying $Y_{1,h}(\Valp)\bigcup Y_{2,h}(\Valp)\bigcup \Gamma_h(\Valp)= Y$ and $Y_{1,h}(\Valp)\bigcap Y_{2,h}(\Valp)= \emptyset$. Although all variables, such as $\lambda^k_h, \varphi^k_h, \omega_{k,h}^1, \Mb_h(\omega)$ involved in the discretised optimization problems are mesh dependent, to simplify notation, we drop the discretisation parameter $h$ from the notations of all variables. Also to simplify the notation, by $\set{H}^1_{0}(Y_{1})$ we refer to the approximation spaces $\set{H}^1_{0,h}(Y_{1,h})$.

The numerical algorithms have been implemented using the following programming tools and software packages:
\begin{itemize}
	\item The band gap computation was implemented within
	\href{http://fenicsproject.org/}{FEniCS} project, which is a collection of free software with an extensive list of features for automated,
	efficient solution of differential equations, see \cite{Fenics_v15,AlnaesEtAl2012}.
	
	\item The computational domain of the periodic cell $Y$ was discretised with a finite element mesh using software package \href{http://geuz.org/gmsh/}{Gmsh} described in \cite{Geuzaine2009gmsh}.

	\item \newER{Optimization itself was calculated within an open-source software \href{http://www.pyopt.org/}{PyOpt},
		described in \cite{Perez2011pyOpt},
		which is a Python-based package for formulating and solving nonlinear constrained optimization problems in an efficient, 
		reusable and portable manner.
		As an optimization method, the Sequential Least SQuares Programming (SLSQP) algorithm was employed \cite{Kraft1988slsqp}, which allows to incorporate different types of constraints; we have experienced a good numerical behaviour for the particular type of considered problems.}
\end{itemize}

By virtue of the domain parameterization and the FE discretization, two variants  of the optimization problems introduced in section~\ref{sec:setting-the-optimization-problem} were considered with different constraints on the elastic properties of the homogenized structure, as discussed in section~\ref{sec-constr}. The first one arises form \eqref{eq:const_elas} where three selected strain modes $\Ve\in\set{R}^{2\times 2}_{\mathrm{sym}}$ are defined by the principal directions $\left(\begin{smallmatrix}1&0\\0&0\end{smallmatrix}\right)$, 
$\left(\begin{smallmatrix}0&0\\0&1\end{smallmatrix}\right)$,
and by shear mode $\left(\begin{smallmatrix}0&1\\1&0\end{smallmatrix}\right)$, to test the inequality. Numerical results for the following problem are reported in Section \ref{sec:optimization-with-constraint-on-elastic-diagonal}
\begin{subequations}
	\label{eq:opti_discrete}
	\begin{align}
	& \min_{\Valp \in \set{R}^{2\times (n-1)}} -\Phi_{k}(\Valp),
	&\text{the band gap size,}&
	\\
	\mbox{ s.t. }
	&|\alpha_i^j| \leq \alpha^\mathrm{max} = 0.05 \text{ for all }i,j,
	\label{eq:const_des_var}
	&\text{constraint on design variables},&
	\\
	\label{eq:const_elas_disc}
	\begin{split}
	&
	D_{1111}(\Valp) \geq 3D^\mathrm{min},
	\\
	&
	D_{2222}(\Valp) \geq 3D^\mathrm{min},
	\\
	&
	D_{1212}(\Valp) \geq D^\mathrm{min},
	\end{split}
	& \text{constraint on elastic properties,}&
	\end{align}
\end{subequations}
where $D^\mathrm{min} = 8$~GPa. \newtext{The second formulation arises from constraints on the  elasticity tensor eigenvalues  \eq{eq:const_elast-R}; the following problem is  studied in Section~\ref{sec:optimization-with-constraint-on-elastic-eigenvalues}
\begin{subequations}
	\label{eq:opti_discrete_eig}
	\begin{align}
	& \min_{\Valp \in \set{R}^{2\times (n-1)}} -\Phi_{k}(\Valp),
	&\text{the band gap size,}&
	\\
	\mbox{ s.t. }
	&|\alpha_i^j| \leq \alpha^\mathrm{max} = 0.05 \text{ for all }i,j,
	&\text{constraint on design variables},&
	\\
	&	\eigs_i [\Dop(\Valp)] \geq \varsigma^*
	\quad\text{for }i=1,2,3,
	&\text{constraint on elastic properties,}&
	\end{align}
\end{subequations}
where $\varsigma^*= 15\,\mathrm{GPa}$.}

{We shall now discuss few remarkable differences between continuous \eqref{eq:opt_problem} and discretized optimization problems \eqref{eq:opti_discrete} and \eqref{eq:opti_discrete_eig}.}
\begin{itemize}
	\item The evaluation of objective function $\Phi_{k}$ is provided by Galerkin approximation framework with discrete FEM spaces, compare Algorithm~\ref{alg:obj_fun} with Algorithm~\ref{alg:obj_fun_disc}.
	\item The major complexity arises from the need of computing the whole spectrum of the eigenvalue problem \eqref{eq:evp_disc} but also from the sensitivity for all eigenvectors.
	The computational costs are reduced by omitting non-relevant frequencies. For some threshold value $\theta$, a criterion can be based on the norm of eigenmomentum 
	$|\mb^r| > \theta$. \newtext{Here, since we calculate band gaps or eigenvalues of $\Mb(\omega)$ only in $]\sqrt{\lam^{k}},\sqrt{\lam^{k+1}}[$,
	we propose the following criterion modification leading to an index set of the relevant spectrum:
	\begin{align}
	\label{eq:spectrum_ind_set}
	R_\theta^k & = \{r\in\set{N}:r\leq\dim\set{H}_0^1(Y_2)\text{ and } \frac{w_k^r |\mb^r|^2}{\max_s w_k^s |\mb^s|^2} > \theta\}
	\;,\\
	\mbox{ where } w_k^r & := \bigl|\frac{\bar\om_k^2}{\bar\om_k^2-\lam^r}\bigr| \mbox{ with } \bar \om_k^2 := \frac{\lam^k+\lam^{k+1}}{2}\;. \nonumber
	\end{align}
	The weight $w_k^r$ is established by virtue of the term occurring in the mass tensor \eq{eq-37}; $w_k^r$  rapidly decreases for an increase of the distance $|\bar\om_k^2-\lam^r|$, see Figure~\ref{fig:eigenmomenta}.}
	
	\begin{figure}[ht] 
		\centering
		\subfigure[L-shaped inclusion]{
			\includegraphics[height=0.36\textwidth]{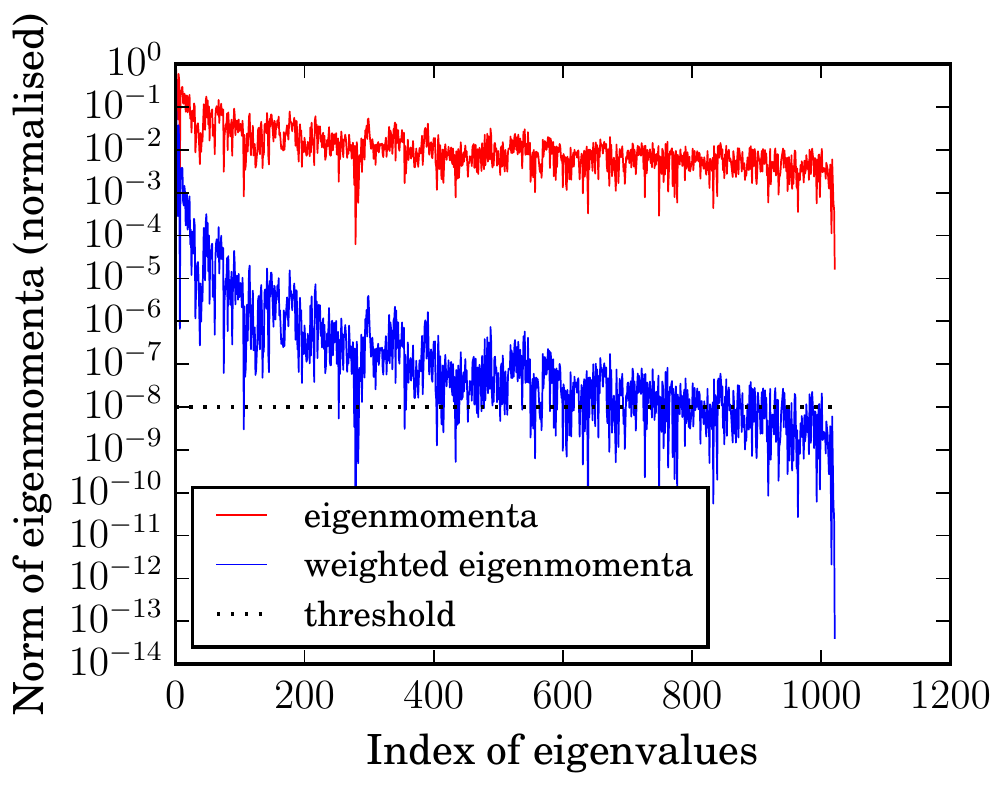}
		}
		\subfigure[Square inclusion]{
			\includegraphics[height=0.36\textwidth]{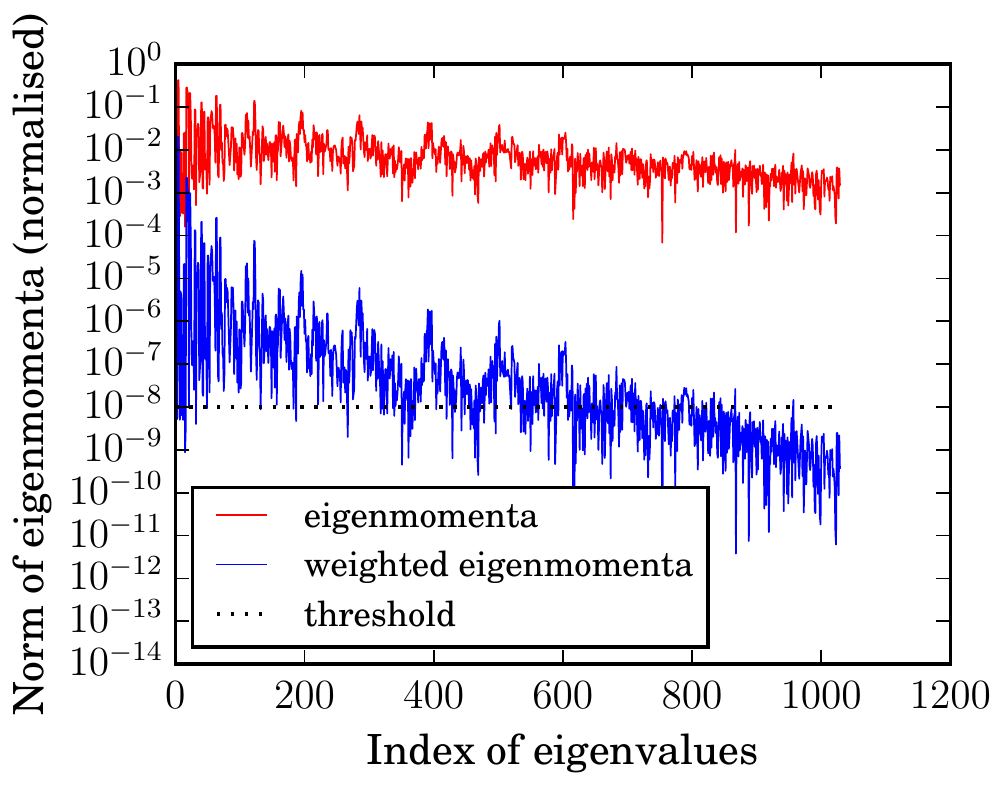}
		}
		\caption{\newtext{Normalized norm of eigenmomenta $\frac{|\mb^r|}{\max_s|\mb^s|}$ defined in \eqref{eq-36} and its comparison to the normalized norm of weighted eigenmomenta $\frac{w_k^r |\mb^r|^2}{\max_s w_k^s |\mb^s|^2}$ used in \eqref{eq:spectrum_ind_set}; the latter only was used to determine relevant eigenvalues using threshold $\theta=10^{-8}$ considered for computing the mass tensor; (a) 768 eigenvalues active (254 suppressed) of total 1022 (b) 637 eigenvalues active (393 suppressed)  of total 1230.}}
		\label{fig:eigenmomenta}
	\end{figure}
	
	\item As pointed out in Section~\ref{sec-constr}, the formulation \eq{eq:opti_discrete}
	imposes additional preferred anisotropy of the optimized designs.
	Since the selected test strains span the space of all symmetric matrices,
	the constraint still guarantees some minimal stiffness for any load, however, 
	it lacks the invariance with respect to rotations, in contrast with the more general constraint involved in \eq{eq:opti_discrete_eig}.

	\item The constraints on the design variables $\Valp\in\set{R}^{2\times(n-1)}$ is incorporated in \eqref{eq:const_des_var} to 
	protect the  domain $Y_2$ against undesired degeneration violating properties \eqref{eq-adm} and to keep the FE mesh quality.
	Despite these constraints, the optimization algorithm can still lead to distortion of the inclusion boundary $\pd Y_2$, or to intersection of the inclusion and cell boundaries. Therefore, to avoid such undesired effects, the quality of the mesh is further controlled, see the adaptive procedure presented in Algorithm~\ref{alg:overall}.
\end{itemize}

\begin{algorithm}[h]
	\caption{Calculation of weak band gap size in discrete setting}
	\label{alg:obj_fun_disc}
	\begin{algorithmic}[1]
		\Require
		{Control points $\Pb^i$ for $i=0,\dotsc,n-1$ of B-spline, index $k\in\set{N}$ of band gap, a corresponding FEM mesh, design velocities $\dV^{i(j)}$, threshold $\theta$ for \eqref{eq:spectrum_ind_set}}
		\Procedure{Objective-function}{FEM mesh, $\Valp$, $\dV^{i(j)}$, $\theta$}
		\Comment{band gap size}
		\State{Set $Y_1=Y_1(\Valp)$ and $Y_2=Y_2(\Valp)$ by updating the FEM mesh using $\Valp$ and $\dV^{i(j)}$}
		\State{Calculate eigenelements $(\lam^r,\phibf^r) \in \RR \times \set{H}^1_{0}(Y_2)$, $r = 1,2,\dots,\dim\set{H}_{0}^1(Y_2)$ of the Galerkin approximation to the eigenvalue problem \eqref{eq-36a}
			\begin{align}
			\label{eq:evp_disc}
			\int_{Y_2} [\Cop^\mater \eeb^y(\phibf^r)]:\eeb^y(\vb) &= \lam^r \int_{Y_2} \rho^2 \phibf^r \cdot \vb
			\quad \forall \vb \in \set{H}_0^1(Y_2),
			&
			\int_{Y_2} \rho^2 \phibf^r\cdot\phibf^s &= \delta_{rs}
			\end{align}
		}
		\State{Calculate eigenmomentum $\mb^r = \int_{Y_2}  \phibf^r$ and average density $\aver{\rho} = \sum_{s=1,2}\int_{Y_s} \rho^s$.}
		\State{Calculate the relevant spectrum using the index set $R_\theta^k$ in accordance with \eqref{eq:spectrum_ind_set}}
		\State{Calculate $\omega_k^1\in ]\sqrt{\lambda^k},\sqrt{\lambda^{k+1}}[$ using \eqref{eq-evM}, i.e. the smallest eigenvalue $\gamma^1(\omega_k^1)$ of
			approximated mass matrix $\Mb(\om) = \aver{\rho} \Ib - \frac{1}{|Y|}\sum\nolimits_{r \in R_\theta} \frac{\om^2}{\om^2 - \lam^r} \mb^r \otimes \mb^r$ which is zero
		}
		\EndProcedure{\textbf{: return }$\Phi_{k}(\Valp) = \omega_k^1(\Valp)-\sqrt{\lambda^k(\Valp)}$}
	\end{algorithmic}
\end{algorithm}

\begin{algorithm}[h]
	\caption{Overall optimization algorithm of band gap size}
	\label{alg:overall}
	\begin{algorithmic}[1]
		\Require{Control points $\Pb^i$ for $i=0,\dotsc,n-1$ of B-spline, index $k\in\set{N}$ of band gap, a discretization parameter $h$, threshold $\theta$ for \eqref{eq:spectrum_ind_set}}
		\Repeat
		\State{Generate a FEM mesh dependent on $\Pb^i$ and discretization parameter $h$}
		\State{Calculate design vel. $\dV^{i(j)}$ using Galerkin approximation of auxiliary problem \eqref{eq-Uvec2}}
		\State{Calculate $\Valp^i_\mathrm{opt}$ of discrete optimization problem \eqref{eq:opti_discrete} using SLSQP algorithm}
		\break
		\Comment{Use here Alg.~\ref{alg:obj_fun_disc},\ref{alg:grad_fun}, \ref{alg:elastic_grad} for evaluation of obj. function, constraint, and their gradients}
		\If{SLSQP converged and the quality of mesh is in tolerance}
		\State{$\Pb^i \gets \Pb^i+\Valp^i_\mathrm{opt}$}
		\Comment{Update of B-spline control points with an optimal solution}
		\Else
		\State{Stop or adjust parameters of the discrete optimization problem \eqref{eq:opti_discrete}}
		\EndIf
		\Until{The size of band gap increases}
		\\
		\Return{Admissible topology with highest band gap size}
	\end{algorithmic}
\end{algorithm}

\newER{The gradient-based optimization algorithm SLSQP is supplied by 
	the objective function value and by its gradient, both implemented according to Algorithm~\ref{alg:grad_fun}. Also the elastic constraints are evaluated and the associated gradients computed according to Algorithm~\ref{alg:elastic_grad} 
	which is presented for the two alternative constraint types involved in formulations \eq{eq:opti_discrete} and \eq{eq:opti_discrete_eig}.}

\begin{algorithm}[h]
	\caption{Calculation of gradient of band gap size w.r.t. design variables}
	\label{alg:grad_fun}
	\begin{algorithmic}[1]
		\Require
		{Requirements as in Algorithm~\ref{alg:obj_fun_disc} and its auxiliary results such as eigenelements $(\lam^r,\varphi^r)$ for $r\in\Rcal_\theta$, upper bound of band gap $\omega_k^1$, average density $\mean{\rho}$, eigenmomenta $\mb^r$}
		\Procedure{Grad-objective-function}{FEM mesh, $\Valp$, $\dV^{i(j)}$}
		\State{Set $Y_1=Y_1(\Valp)$ and $Y_2=Y_2(\Valp)$ by updating the FEM mesh using $\Valp$ and $\dV^{i(j)}$}
		\ForAll{design velocities $\Vcal^{i(j)}$}
		\State{Calculate sensitivity to eigenelements $(\lam^r,\varphi^r)$ for $r\in R_\theta$ using \eqref{eq:ph5} and \eqref{eq:ph7a}}
		\State{Calculate the sensitivity of mass matrix $\Mb(\omega_k^1)$ according to Section~\ref{sec:sa_mass}}
		\State{Calculate the sensitivity of eigenvalues of mass matrix $\Mb(\omega_k^1)$}
		\EndFor
		\EndProcedure{\textbf{: return }}$\nabla_{\Valp}\Phi_{k}(\Valp)=\nabla_{\Valp}\omega^1_k(\Valp) - \nabla_{\Valp} \lam^k(\Valp)$
	\end{algorithmic}
\end{algorithm}

\begin{algorithm}[h]
	\caption{Elastic constraint and its gradient}
	\label{alg:elastic_grad}
	\begin{algorithmic}[1]
		\Require
		{Requirements as in Algorithm~\ref{alg:obj_fun_disc}}
		\Procedure{Elastic-stiffness}{FEM mesh, $\Valp$, $\dV^{i(j)}$}
		\State{Set $Y_1=Y_1(\Valp)$ by updating the FEM mesh using $\Valp$ and $\dV^{i(j)}$}
		\State{Calculate corrector functions $\wb^{kl}\in\set{H}^1_\per(Y_1)$ for $k,l\in\{1,2\}$ with Galerkin approximation of \eqref{eq-C*a}}
		\State{Evaluate the effective stiffness tensor $\Dop$ from \eqref{eq-C*} and \newtext{compute its eigenpairs $(\varsigma^k,\eb^k)$, such that $\Dop\eb^k = \varsigma^k\eb^k$ for $k=1,2,3$}}
		\EndProcedure{\textbf{: return }} \newtext{$D_{1111}$, $D_{2222}$, $D_{1212}$ for \eqref{eq:opti_discrete} or $\varsigma^k$ with $k=1,2,3$ for \eqref{eq:opti_discrete_eig}}
		\Procedure{Grad-Elastic-stiffness}{FEM mesh, $\Valp$, $\Vcal^{i(j)}$, $\wb^{kl}$}
		\State{Set $Y_1=Y_1(\Valp)$ by updating the FEM mesh using $\Valp$ and $\dV^{i(j)}$}
		\ForAll{design variables $\Vcal^{i(j)}$}
		\State{Evaluate equation \eqref{eq-S19}.}
		\State{\newtext{Compute $\delta\varsigma^k = \delta D_{ijkl}e^k_{ij}e^k_{kl}$ for $k=1,2,3$.}}
		\EndFor
		\EndProcedure{\textbf{: return }}
		\newtext{$\delta D_{1111}$, $\delta D_{2222}$, $\delta D_{1212}$ for \eqref{eq:opti_discrete} or $\delta \varsigma^k$ with $k=1,2,3$ for \eqref{eq:opti_discrete_eig}}
	\end{algorithmic}
\end{algorithm}

\clearpage
\subsection{Initial parameters and layouts for optimization}
\label{sec:optimization-problems}

\begin{figure}[!hp] 
	\centering
	\subfigure[Mesh with inclusion]{
		\includegraphics[height=0.36\textwidth]{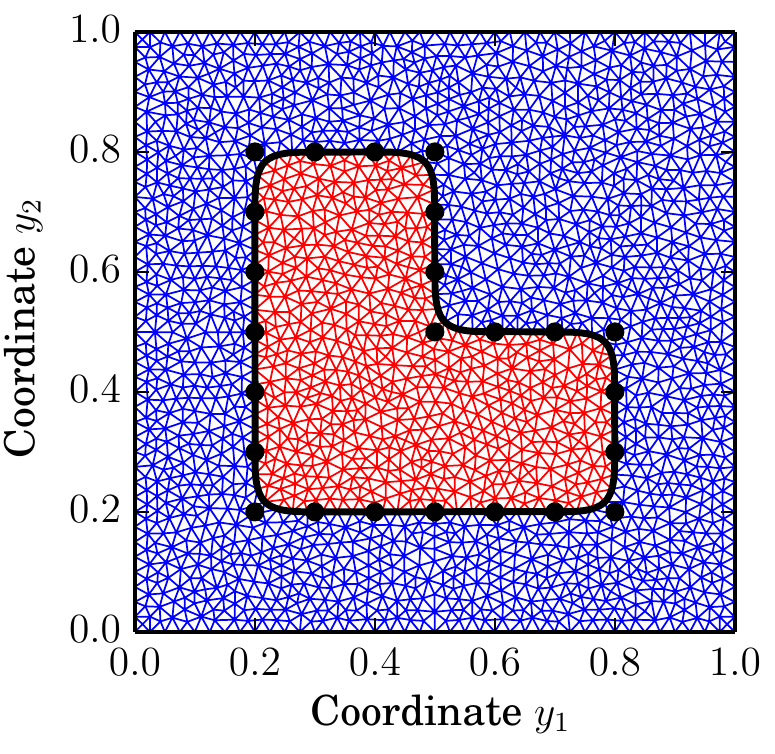}
	}
	\subfigure[band gaps and resonance frequencies]{
		\includegraphics[height=0.36\textwidth]{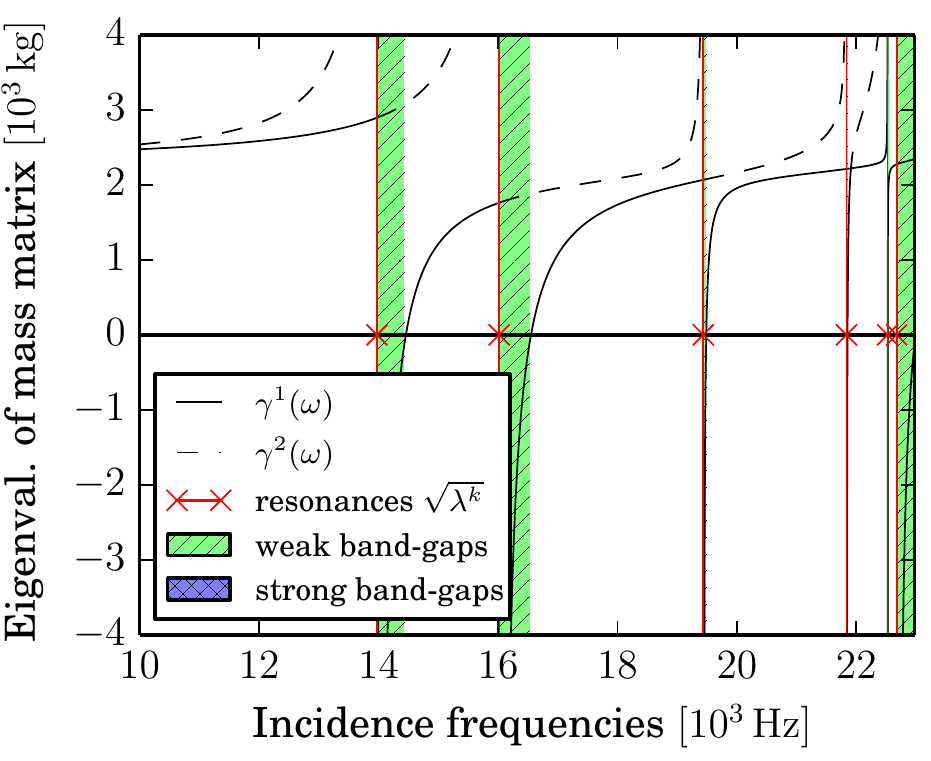}
	}
	\caption{The initial state; (a) the mesh with red L-shaped inclusion, blue matrix, and black-doted spline boundary with black control nodes;
		(b) eigenvalues of mass matrix for incidence frequencies with weak band gaps and no strong band gaps}
	\label{fig:initial_L_state}
\end{figure}

\begin{figure}[!htpb]
	\centering
	\subfigure[Mesh with inclusion]{
		\includegraphics[height=0.36\textwidth]{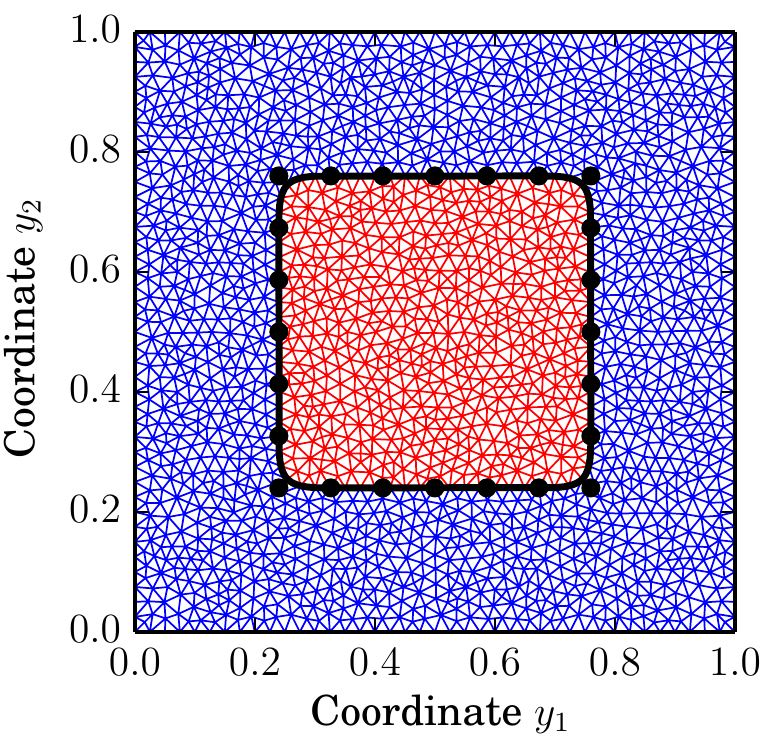}
	}
	\subfigure[band gaps and resonance frequencies]{
		\includegraphics[height=0.36\textwidth]{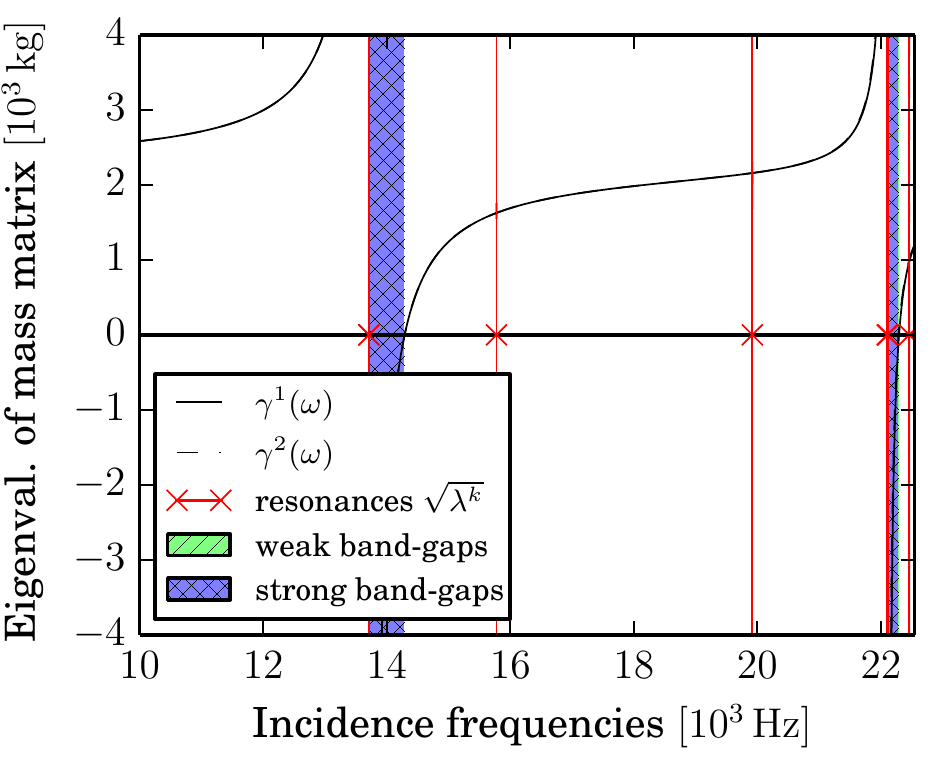}
	}
	\caption{The initial state; (a) the mesh with blue matrix, red inclusion, and black-doted spline boundary with black spline nodes; 
		(b)  eigenvalues of mass matrix for incidence frequencies with strong band gaps}
	\label{fig:initial_square}
\end{figure}

Here, we introduce and comment on problems and parameters that were used for numerical optimization,
as described in previous Section~\ref{sec:discretization}.
In Table~\ref{tab:material},
the material properties for both inclusion and matrix are presented together with an artificial elastic material used for the calculation of design velocities $\dV^{i(j)}$ in \eqref{eq-Uvec2}.
Two types of inclusion, L-shaped and square-shaped, were used for optimization, see Figures~\ref{fig:initial_L_state} and \ref{fig:initial_square} for both inclusion geometry with mesh and distribution of band gaps. The discretization parameters, such as number of mesh and spline nodes,
are summarized in Table~\ref{tab:discret_stat}.

\begin{table}[htbp]
	\centering
	\begin{tabular}{ll|lll}
		\hline
		material & region & lame coef. & shear modulus & density \\
		\hline
		aluminium & matrix ($Y_1$) & $58.98\cdot10^{9}\,\mathrm{Pa}$ & $26.81\cdot10^{9}\,\mathrm{Pa}$  & $2.799\cdot 10^3\,\mathrm{kg}/\mathrm{m}^3$ \\
		epoxy & inclusion ($Y_2$) & $1.798\cdot10^{9}\,\mathrm{Pa}$ & $1.48\cdot10^{9}\,\mathrm{Pa}$  & $1.142\cdot 10^3\,\mathrm{kg}/\mathrm{m}^3$ \\
		auxiliary for $\dV$ & cell ($Y$) & $\frac{1}{2}\,\mathrm{Pa}$ & $1\,\mathrm{Pa}$  & --- \\
		\hline
	\end{tabular}
	\caption{Material coefficients for optimization}\label{tab:material}
\end{table}

\ER{Due to the discussion reported in Section~\ref{sec:rescaling}, the optimization problems can be solved for assumed scale parameter $\veps_0 = 1$. Thus, we consider a unit cell with real materials distributed in both the matrix and the inclusion parts. The solutions of the particular optimization problems can be interpreted a~posteriori, in order to obtain the band gap within required frequencies.}

\begin{table}[h]
	\centering
	\begin{tabular}{ll|llll}
		\hline
		inclusion & Fig. & mesh vertices in $Y$ & spline points & $\alp^{\mathrm{max}}$ & $D^\mathrm{min}$ \\
		\hline
		L-shaped & \ref{fig:initial_L_state} & 2218 & 24 & 0.05 & 8 $\mathrm{GPa}$
		\\
		square & \ref{fig:initial_square} &  2153 & 24 & 0.05 & 8 $\mathrm{GPa}$
		\\
		\hline
	\end{tabular}
	\caption{Discretization parameters for initial layouts. 
	}\label{tab:discret_stat}
\end{table}

\subsection{Optimization with elasticity constraint by strain modes}
\label{sec:optimization-with-constraint-on-elastic-diagonal}

\ER{This section is dedicated to numerical optimization of weak band-gaps with constraint on the effective elastic stiffness relative to the selected strain modes, see formulation in~\eqref{eq:opti_discrete}.}

\subsubsection{Weak band gap in the first interval}
\label{sec:weak-band-gap-in-the-first-interval}

\begin{figure}[!htbp]
	\centering
	\subfigure[Mesh with inclusion]{
		\includegraphics[height=0.36\textwidth]{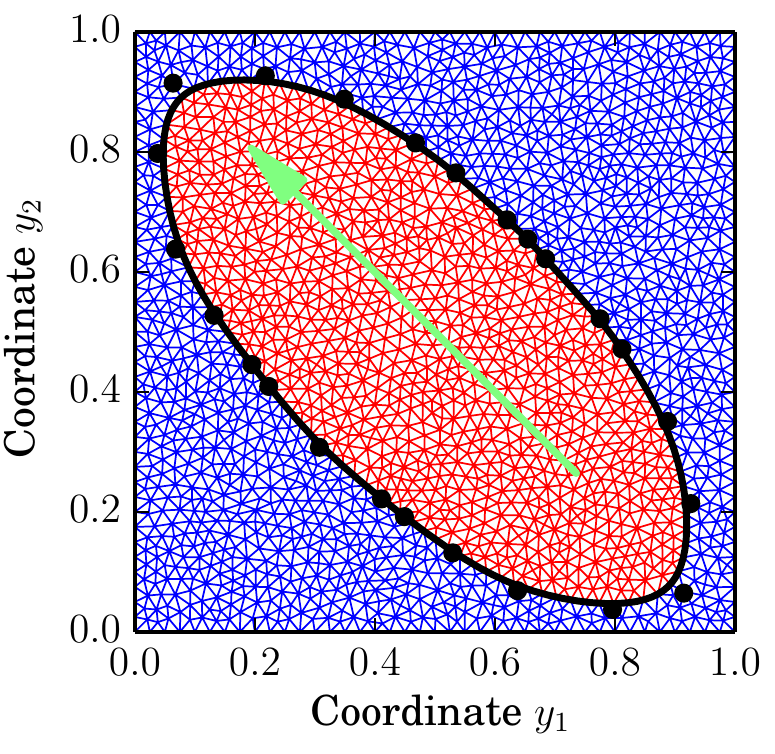}
	}
	\subfigure[band gaps and resonance frequencies]{
		\includegraphics[height=0.36\textwidth]{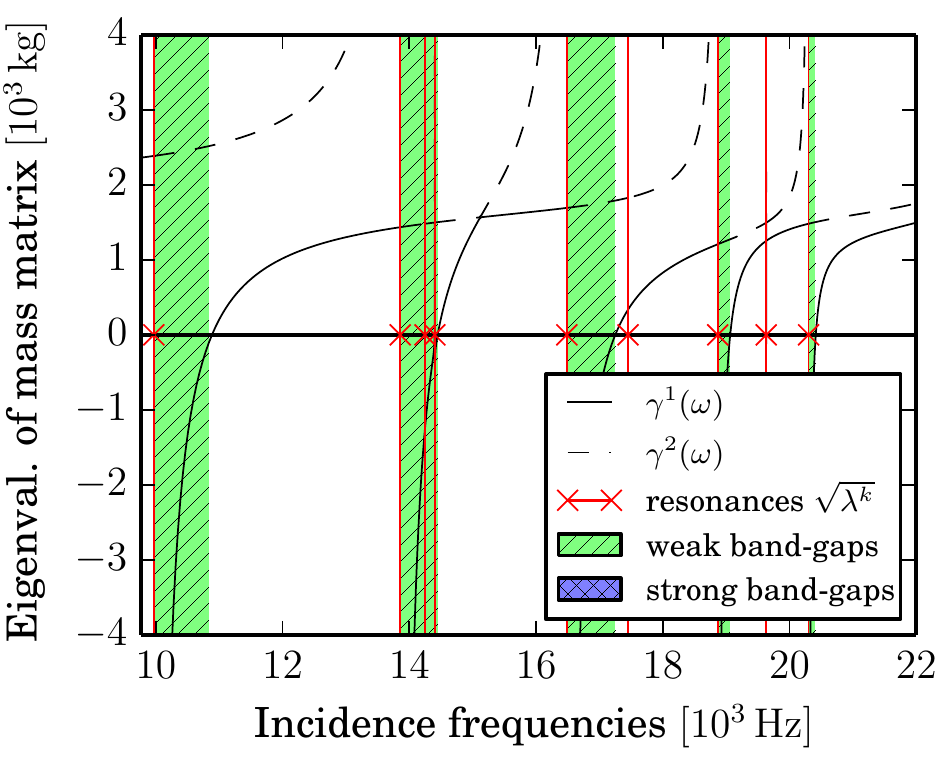}
	}
	\caption{\ER{The optimal layout for maximal weak band gap size in interval $[\sqrt{\lam^1}, \sqrt{\lam^2}]$; initial layout: L-shaped inclusion; 3 re-meshing used;
			(a) the mesh with blue matrix, red inclusion, and black boundary with spots of B-spline nodes, the green arrow shows the polarization of the suppressed waves;
			(b) the two eigenvalues of the mass tensor for incidence frequencies}
	}
	\label{fig:final_L_state}
\end{figure}

\begin{figure}[!htbp]
	\centering
	\subfigure[\footnotesize Before 1\textsuperscript{st} re-meshing]{
		\includegraphics[height=0.22\textwidth]{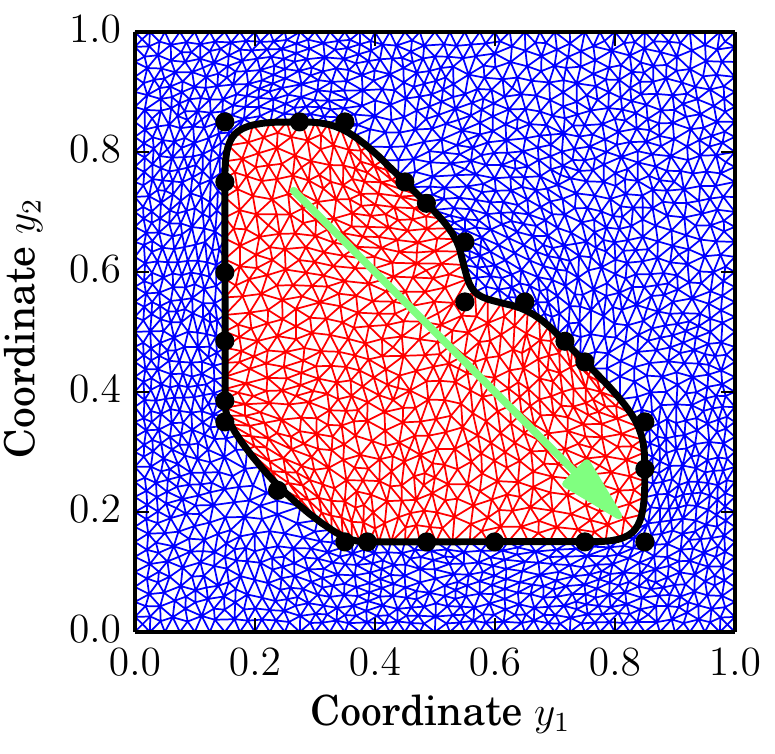}
	}
	\subfigure[\footnotesize Before 2\textsuperscript{nd} re-meshing]{
		\includegraphics[height=0.22\textwidth]{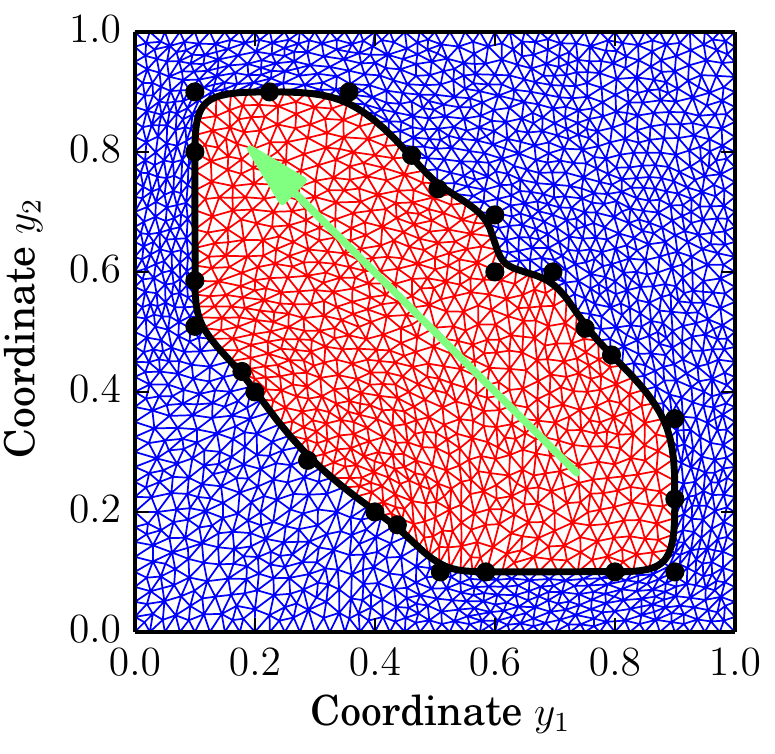}
	}
	\subfigure[\footnotesize Before 3\textsuperscript{rd} re-meshing]{
		\includegraphics[height=0.22\textwidth]{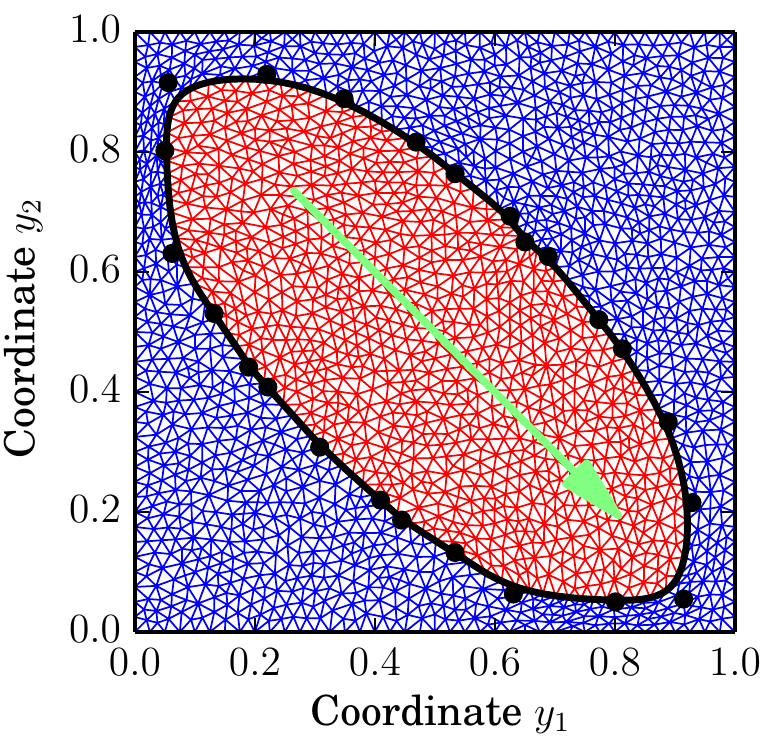}
	}
	\caption{The intermediate layouts obtained from SLSQP optimization leading to the optimal ellipse-like inclusion}
	\label{fig:p30_interstates}
\end{figure}

Optimization of initial L-shaped inclusion shown Figure~\ref{fig:initial_L_state}(a) leads to the optimal {layout},
presented in Figure~\ref{fig:final_L_state}(a)
which is characterized with an ellipse-like inclusion rotated by $45$ degrees.
The eigenvalues of the mass matrix, distribution of band gaps,
and resonance frequencies are \ER{depicted in Figures~\ref{fig:initial_L_state}(b),
	while in Figure~\ref{fig:final_L_state}(b) the band gap diagram for the optimal structure is displayed.} \newtext{The intermediate layouts are displayed in Figure~\ref{fig:p30_interstates}.}

The optimal design was obtained with three re-meshing corresponding to three restarts of the SLSQP algorithm. the weak band gap size was enlarged about $89.5\,\%$, other results are summarized in Table~\ref{tab:opti_first}.
The inclusion was enlarged, the volume fraction $|Y_2|/|Y|$ increased to  $64.7\,\%$.
\ER{However, there is also a significant decrease of effective elastic stiffness components $D_{1111}$ and $D_{2222}$. 
	The optimal shape is reached with all elastic inequality constraints 
	active, so that within the optimization algorithm they are satisfied  within the required tolerance.} 
The optimized band gap is shifted to lower frequencies in comparison with the band gap at the initial layout.

\begin{figure}[h]
	\centering
	\subfigure[Size of weak band gap in first interval]{
		\includegraphics[height=0.36\textwidth]{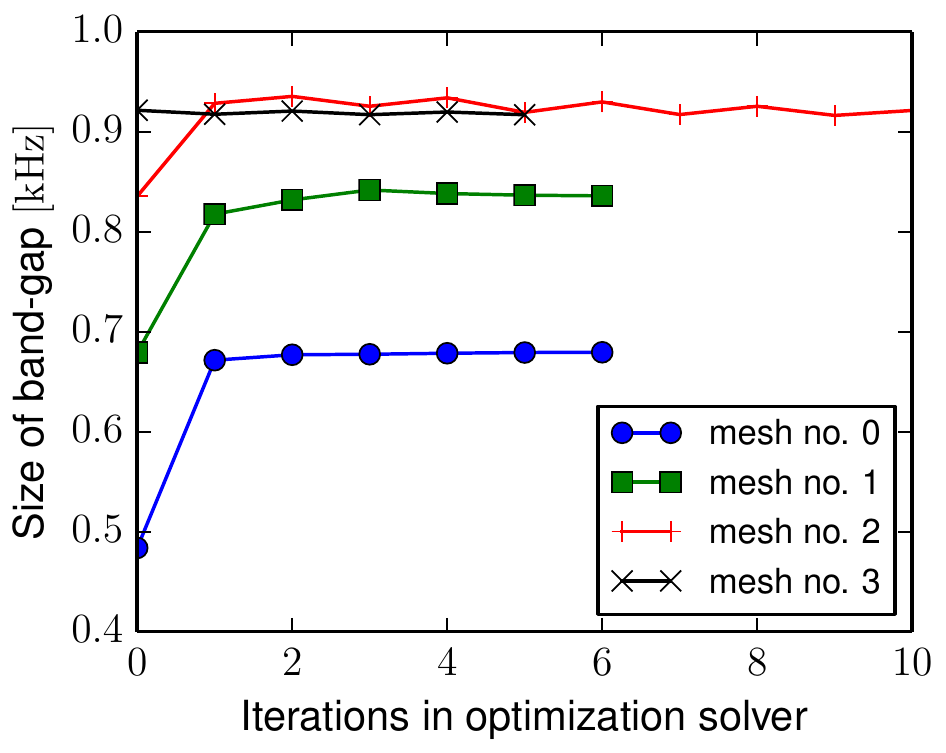}
	}
	\subfigure[Constraint on shear modulus]{
		\includegraphics[height=0.36\textwidth]{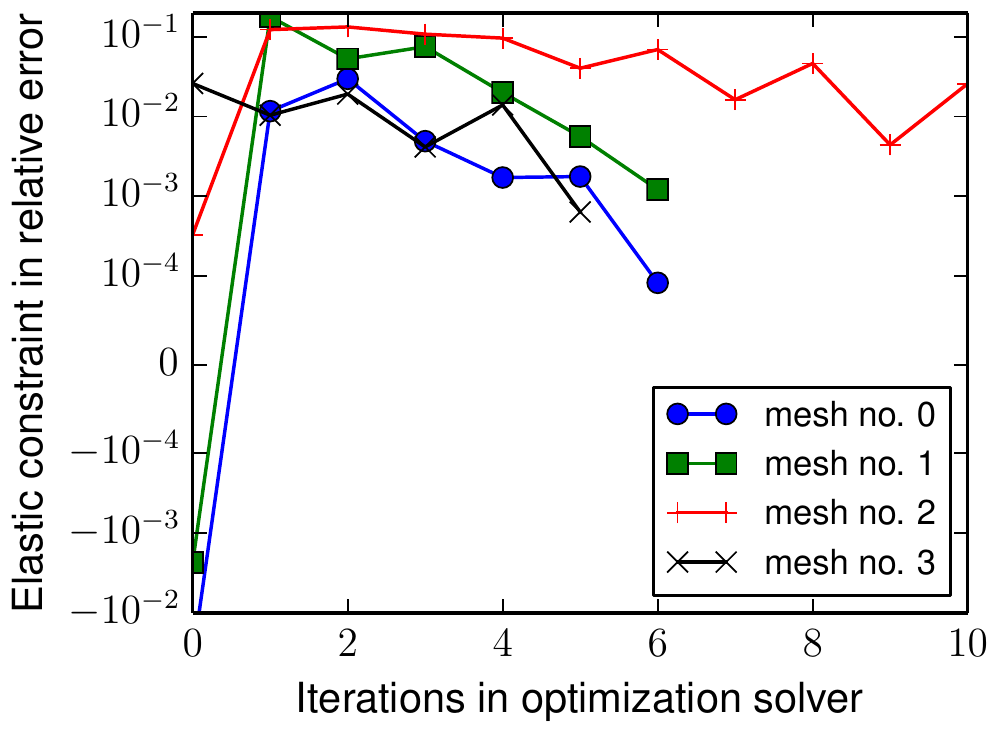}
	}
	\caption{The evolution of (a) objective function $\Phi_1(\Valp)$, i.e. size of band gap in first interval, during optimization and
		(b) elastic constraint in relative error
		$\frac{D^{\mathrm{min}}-D_{1212}(\Valp)}{D^{\mathrm{min}}}$,
		negative values denotes admissible area; the convergence tolerance in SLSQP algorithm corresponds to $10^{-3}$}
	\label{fig:evol_f_g_1}
\end{figure}

The evolution of band gap width during optimization is captured in Figure~\ref{fig:evol_f_g_1}(a) together with the  constraint on the shear elasticity $D_{1212}$ in Figure~\ref{fig:evol_f_g_1}(b),
which is the only inequality type constraint being active at all iterations of the optimization;
the other two constraints became active at the last iterations only.
At each of the three restarts of the SLSQP algorithm after re-meshing, the best improvement in the band gap width is provided at the first iteration.
\ER{Since the iterations of the optimization run leave the admissibility design area, the subsequent optimization merits to satisfy the admissibility constraints within the required tolerance,
	which is reached at the last iteration, except the third optimization run with mesh no. 2, which is stopped when reaching the maximal number of iterations.}
\newtext{Moreover, the conforming method in primal formulation, used here, leads to the overestimation of the homogenized stiffness \cite{Haslinger1995optimum,VoZeMa2014GarBounds}, which can cause leaving of admissible constraint area with mesh refinement. This can be avoided by using the dual formulation.}

\ER{The band gap optimization in the first interval $]\sqrt{\lam^1},\sqrt{\lam^2}[$, initiated with the square inclusion layout, is not directly possible because the first two eigenvalues coincide, \ie $\lam^1\approx\lam^2$. Nevertheless, optimization run starting with a slightly perturbed square inclusion (one control point of the B-spline was moved about $0.01$), This leads to the same ellipse-shaped inclusion, as the one obtained when initiating the optimization with the L-shaped domain.}

\begin{table}[h]
	\centering
	\begin{tabular}{ll|llllllll}
		\hline
		inclusion & Fig. & $\Phi_1$ & $\sqrt{\lam^1}$ & $\omega_1^1$ & $\sqrt{\lam^2}$ &
		$D_{1111}$ & $D_{2222}$ & $D_{1212}$ & $\frac{|Y_2|}{|Y|}$ \\
		\hline
		L-shape & \ref{fig:initial_L_state} & $0.484$ & $13.975$ & $14.459$ & $16.021$ & 
		$42.334$ & $42.335$ & $8.147$ & $0.266$
		\\
		optimal & \ref{fig:final_L_state} & $0.917$ & $9.969$ & $10.886$ & $13.857$ & 
		$23.998$ & $23.998$ & $7.997$ & $0.438$
		\\
		\hline
	\end{tabular}
	\caption{Optimization in first interval; the units are for band gap size $\Phi_1$ and resonances $\sqrt{\lam^k}$ in $[\mathrm{kHz}]$, for effective elasticity $\Dop$ in $[\mathrm{GPa}]$.
	}\label{tab:opti_first}
\end{table}

\begin{table}[h]
	\centering
	\begin{tabular}{ll|lllllllll}
		\hline
		inclusion & Fig. & $\Phi_2$ & 
		$\sqrt{\lam^2}$ & $\omega_2^1$ & $\sqrt{\lam^3}$ & $D_{1111}$ & $D_{2222}$ & $D_{1212}$ & $\frac{|Y_2|}{|Y|}$ \\
		\hline
		L-shaped & \ref{fig:initial_L_state} & $0.533$ & 
		$16.021$ & $16.554$ & $19.439$ & $42.334$ & $42.335$ & $8.147$ & $0.266$
		\\
		suboptimal & \ref{fig:fin_L2} & $0.884$ & 
		$15.559$ & $16.443$ & $17.359$ & $36.391$ & $36.205$ & $7.431$ & $0.321$
		\\
		\hline
		square & \ref{fig:initial_square} & $0.648$ & 
		$12.709$ & $13.357$ & $14.301$ & $45.346$ & $45.421$ & $7.996$ & $0.304$
		\\
		suboptimal & \ref{fig:fin_square2} & $0.782$ & 
		$15.301$ & $16.083$ & $16.107$ & $30.578$ & $29.220$ & $7.851$ & $0.327$
		\\
		\hline
	\end{tabular}
	\caption{Optimization in second interval; the units are for band gap size $\Phi_2$ and resonances $\sqrt{\lam^k}$ in $[\mathrm{kHz}]$, for effective elasticity $\Dop$ in $[\mathrm{GPa}]$.
	}\label{tab:opti_second}
\end{table}

\subsubsection{Weak band gap in the second interval}
\label{sec:weak-band-gap-in-the-second-interval}

\ER{The band gap optimization in the second interval $[\sqrt{\lam^2},\sqrt{\lam^3}]$  is more complicated than the one considered in the previous section, leading to an optimized design. Two examples illustrate the optimization initiated with the L-shape inclusion layout and with the square inclusion, see Table~\ref{tab:opti_second}.}

\begin{figure}[!htp]
	\centering
	\subfigure[Mesh with inclusion]{
		\includegraphics[height=0.36\textwidth]{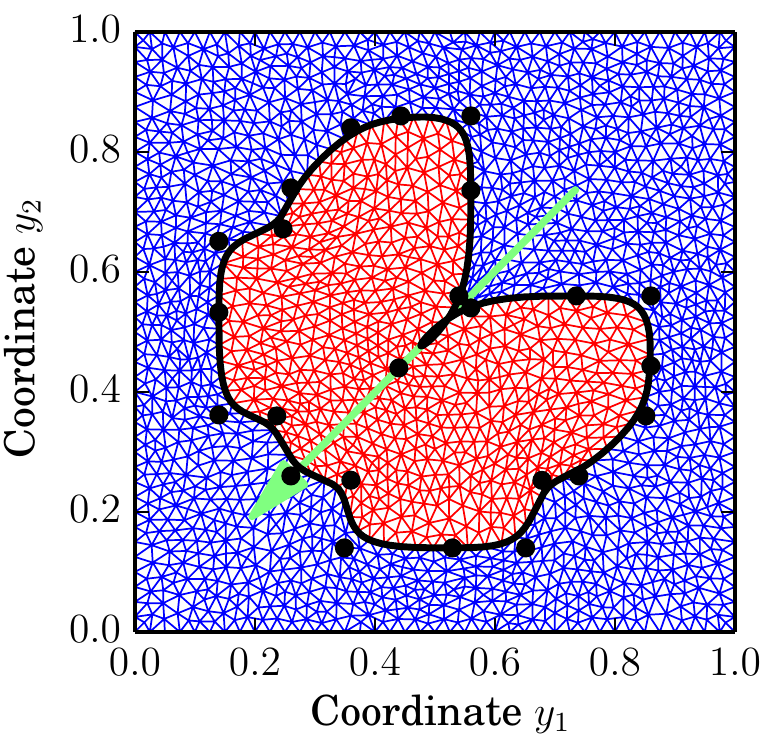}
	}
	\subfigure[band gaps and resonance frequencies]{
		\includegraphics[height=0.36\textwidth]{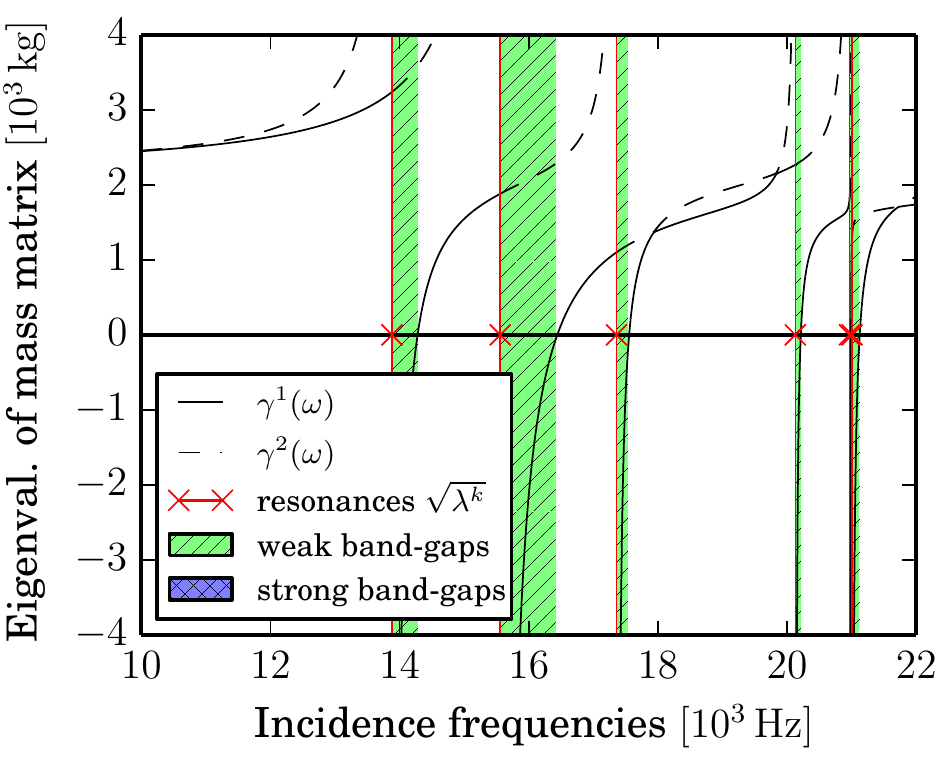}
	}
	\caption{\newtext{The sub-optimal layout for maximal weak band gap size in interval $[\sqrt{\lam^2}, \sqrt{\lam^3}]$; initial layout: L-shaped inclusion; 1 re-meshing used; (a) the mesh with blue matrix, red inclusion, and black boundary with spots of B-spline nodes, the green arrow shows the polarization of the suppressed waves;
			(b) the two eigenvalues of the mass tensor for incidence frequencies}
	}
	\label{fig:fin_L2}
\end{figure}

\ER{Optimization starting with the L-shape inclusion depicted in Figure~\ref{fig:initial_L_state}(a) leads to a suboptimal layout depicted in Figure~\ref{fig:fin_L2},
	which is obtained after one re-meshing, but using constraints on maximal movement of spline nodes. After re-meshing, the second optimization run leads to a serious mesh collapse related to a topology defect in the central area, where the inclusion shape becomes concave.
	Here, the \ER{potential} reformulation to topology optimization might be of a great importance, because spline parametrisation can barely lead to more complicated inclusion topologies.}

\begin{figure}[!htpb]
	\centering
	\subfigure[Mesh with inclusion]{
		\includegraphics[height=0.36\textwidth]{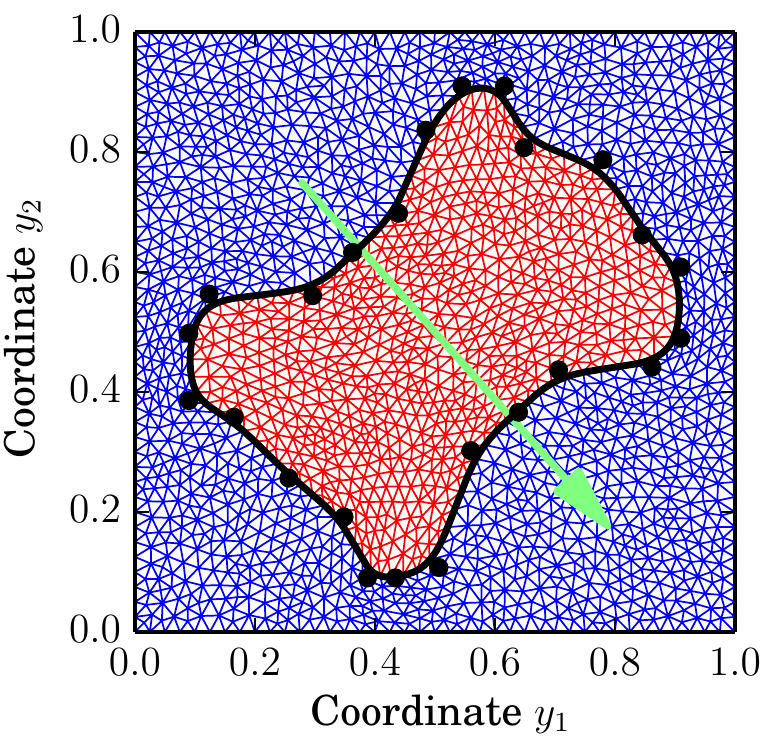}
	}
	\subfigure[Band-gaps and resonance frequencies]{
		\includegraphics[height=0.36\textwidth]{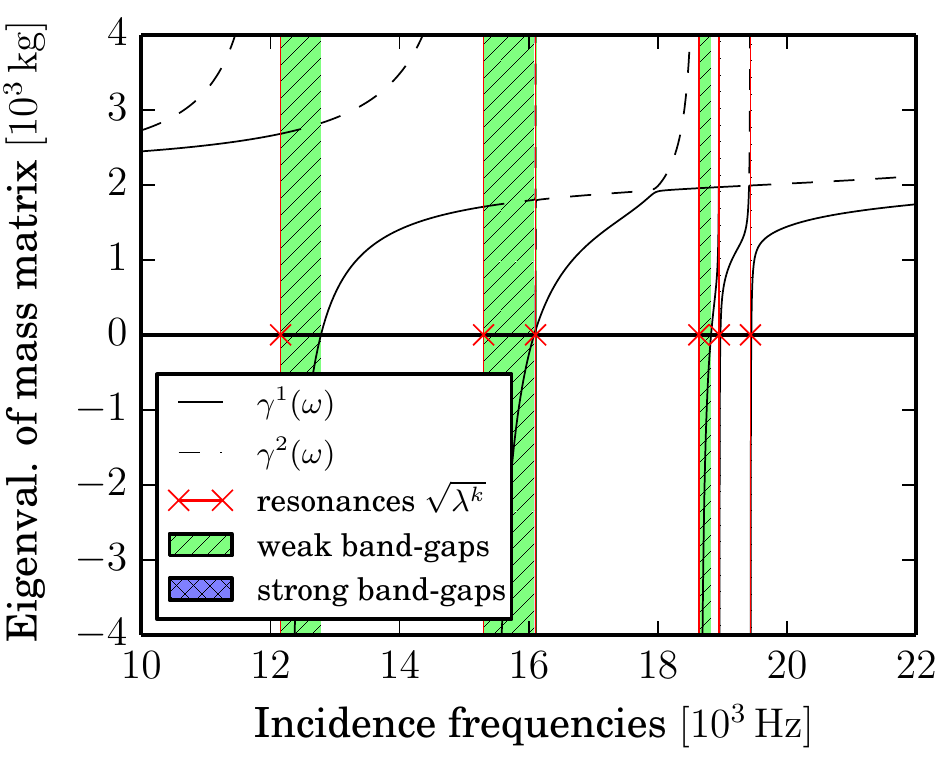}
	}
	\caption{\newtext{The sub-optimal layout for maximal weak band gap size in the second interval $[\sqrt{\lam^2}, \sqrt{\lam^3}]$; initial layout: square-shaped inclusion; 4 re-meshing used;
			(a) the mesh with blue matrix, red inclusion, and black boundary with spots of B-spline nodes, the green arrow shows the polarization of the suppressed waves;
			(b) the two eigenvalues of the mass tensor for incidence frequencies}
	}
	\label{fig:fin_square2}
\end{figure}

\begin{figure}[!htbp]
	\centering
	\subfigure[\footnotesize Before 1\textsuperscript{st} re-meshing]{
		\includegraphics[height=0.22\textwidth]{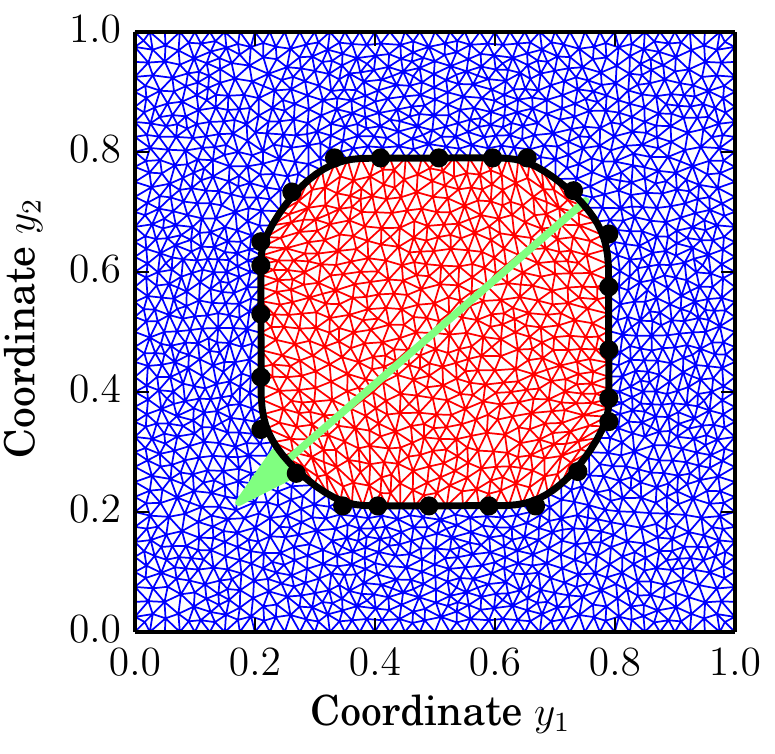}
	}
	\subfigure[\footnotesize Before 2\textsuperscript{nd} re-meshing]{
		\includegraphics[height=0.22\textwidth]{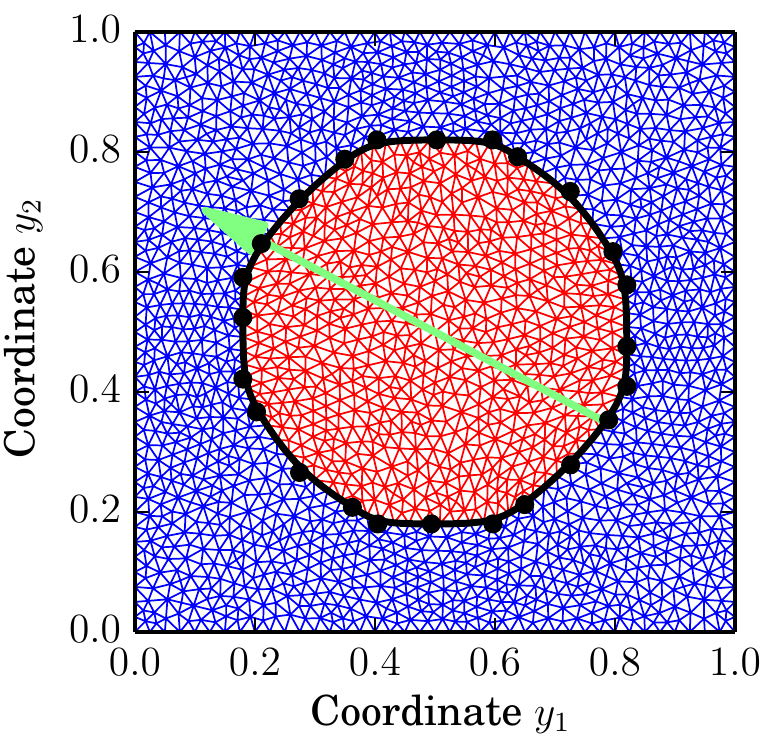}
	}
	\subfigure[\footnotesize Before 3\textsuperscript{rd} re-meshing]{
		\includegraphics[height=0.22\textwidth]{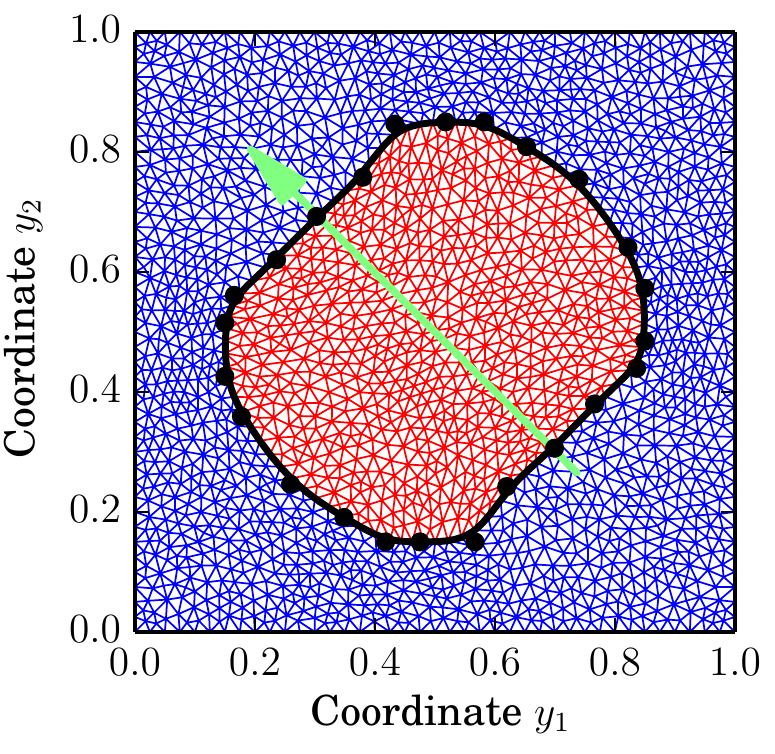}
	}
	\subfigure[\footnotesize Before 4\textsuperscript{th} re-meshing]{
		\includegraphics[height=0.22\textwidth]{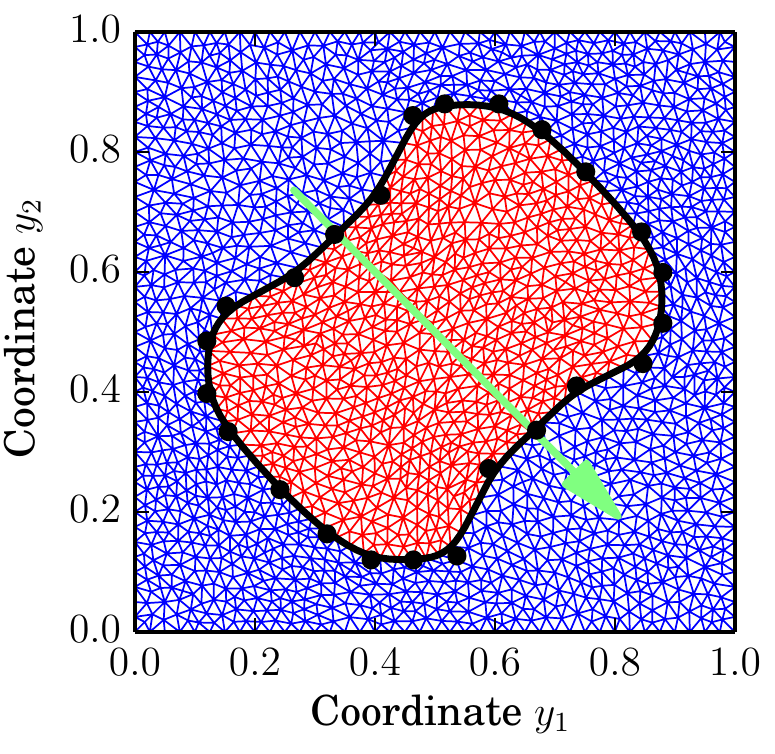}
	}
	\caption{The intermediate layouts obtained from SLSQP optimization leading to the sub-optimal inclusion depicted in Figure~\ref{fig:fin_square2}}
	\label{fig:p20_interstates}
\end{figure}

For the square-shape inclusion considered as the initial layout for which the strong band gaps occur, see Figure~\ref{fig:initial_square},
a different topology and numerical difficulties are observed during optimization, which leads to the suboptimal inclusion shape depicted in Figure~\ref{fig:fin_square2}\newtext{; the intermediate layouts are displayed in Figure~\ref{fig:p20_interstates}.}

\ER{The optimization was stopped after three remeshing, because  the upper bound of the weak band gap in second interval reaches the 3rd eigenvalue border. 
	Further optimization with the same objective function leads to numerical difficulties, since the maximal eigenvalue of mass matrix goes to infinity at the resonance frequency $\sqrt{\lam^3}$.}
\newtext{As a remedy, an additional constraint on a size of band gap with respect to the size of the corresponding interval $\sqrt{\lam^{k+1}}-\sqrt{\lam^k}$ could be combined with an internal barrier-type optimization algorithms such as \cite{Wachter2006}, which avoid movement from the admissible area.}

\clearpage
\subsection{Optimization with constraint on elastic eigenvalues}\label{sec:optimization-with-constraint-on-elastic-eigenvalues}

\newtext{
	This section is dedicated to numerical optimization of weak band-gaps with constraint on the eigenvalues of the effective elastic stiffness, see the formulation in~\eqref{eq:opti_discrete_eig}.
}

\subsubsection{Weak band gap in the first interval}\label{sec:weak-band-gap-in-the-first-interval-eig}

\begin{figure}[!htpb]
	\centering
	\subfigure[Mesh with inclusion]{
		\includegraphics[height=0.36\textwidth]{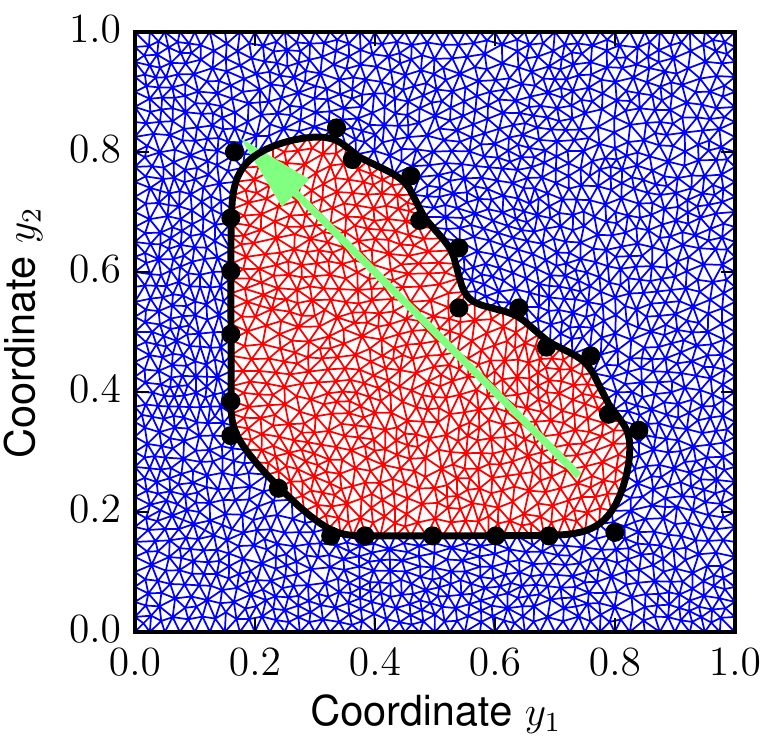}
	}
	\subfigure[Band-gaps and resonance frequencies]{
		\includegraphics[height=0.36\textwidth]{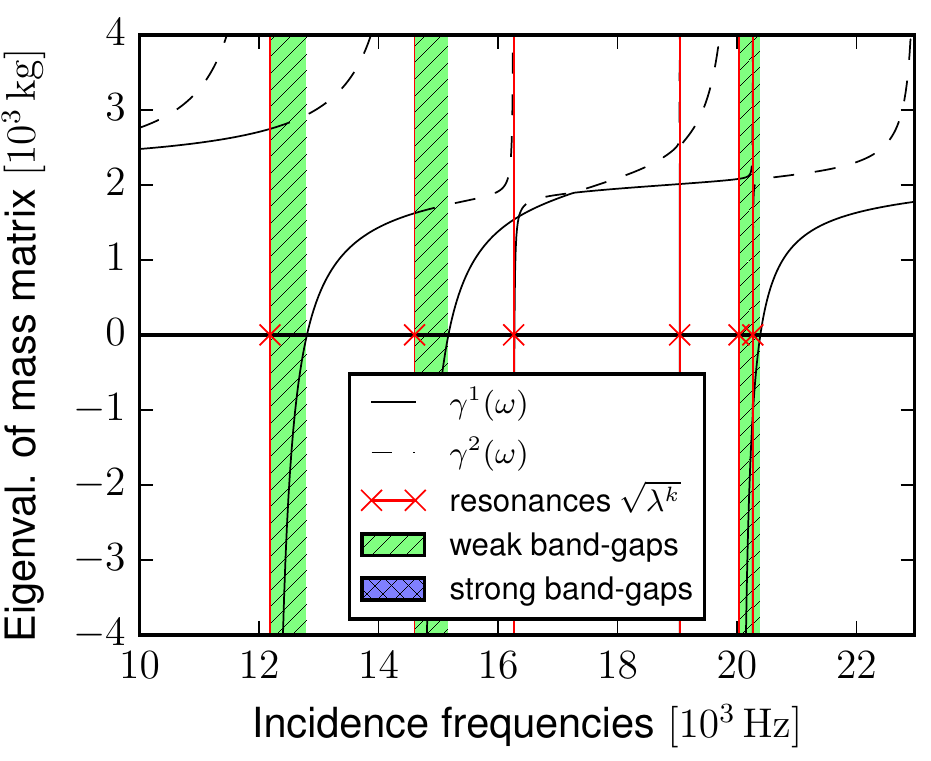}
	}
	\subfigure[Mesh with inclusion]{
		\includegraphics[height=0.36\textwidth]{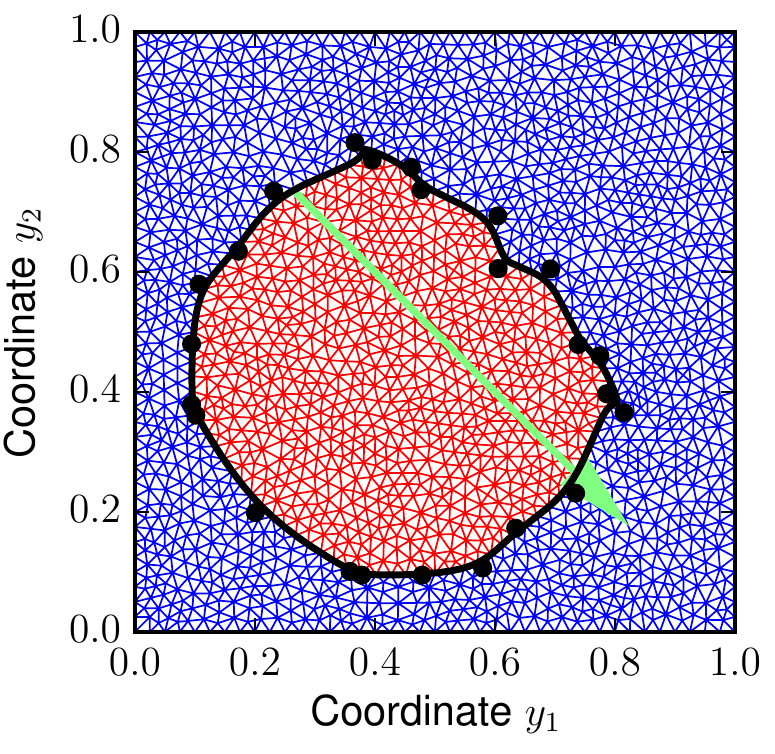}
	}
	\subfigure[Band-gaps and resonance frequencies]{
		\includegraphics[height=0.36\textwidth]{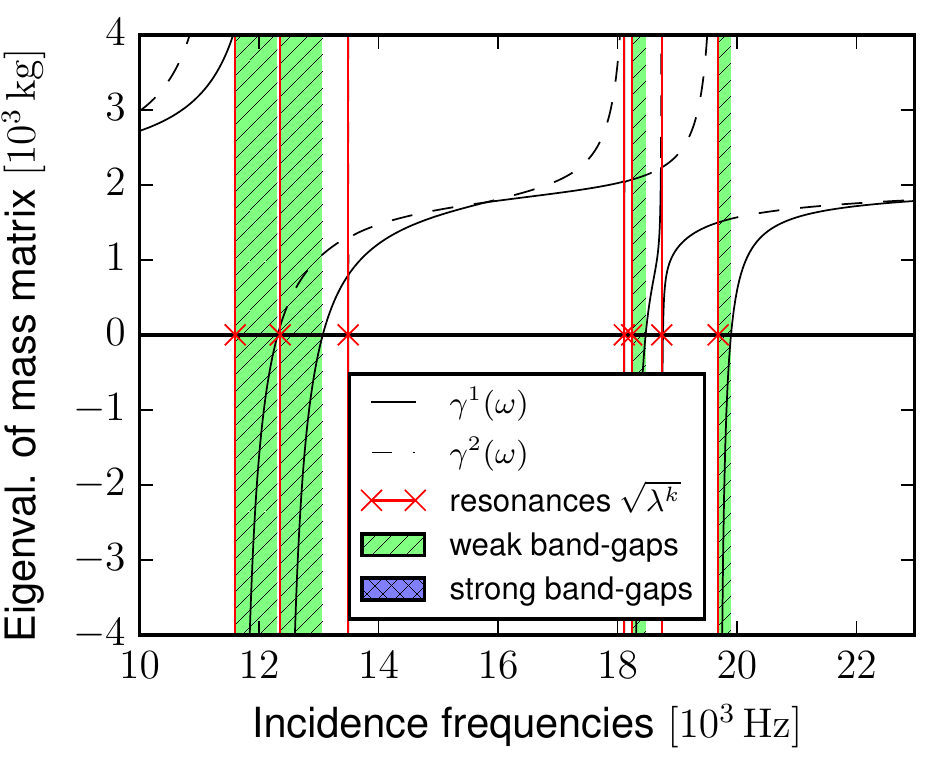}
	}
	\caption{\newtext{The sub-optimal layout for maximal weak band gap size in interval $[\sqrt{\lam^1}, \sqrt{\lam^2}]$; initial layout: L-shaped inclusion; 1 re-meshing used for (ab), 5 re-meshing for (cd); (ac) the mesh with blue matrix, red inclusion, and black boundary with spots of B-spline nodes, the green arrow shows the polarization of the suppressed waves;
			(bd) the two eigenvalues of the mass tensor for incidence frequencies}
	}
	\label{fig:opti_eig_L_first}
\end{figure}

\newtext{The optimization initiated with the L-shaped inclusion, see Figure~\ref{fig:initial_L_state}, leads to sub-optimal shape depicted in Figure~\ref{fig:opti_eig_L_first}. In analogy with the example described in the preceding section, the optimization algorithm is terminated when the upper bound of the weak band-gap in the first interval reaches the resonance value $\sqrt{\lam^2}$. We note that five restarts (after re-meshing) of the optimization procedure were used. 
	
	As already mentioned in Section~\ref{sec:weak-band-gap-in-the-first-interval},
	the optimization initiated with the square inclusion is not directly possible, because the first two eigenvalues coincide resulting in no band-gap in the first interval.
	However, a very slight perturbation of the square shape causes mutual separation of the two eigenvalues and, thus, enable to run the optimization algorithm. It then leads to a sub-optimal shape similar to the optimal shape in Figure~\ref{fig:opti_eig_sq_second} with connected weak band-gap spanning the first two intervals between $\sqrt{\lam^1}$ and $\sqrt{\lam^3}$. Because of the reasons described in the preceding section, the optimization is also terminated when the weak band-gap fills the whole first interval.
}

\clearpage
\subsubsection{Weak band gap in the second interval}
\label{sec:weak-band-gap-in-the-second-interval-eig}
\begin{figure}[!htpb]
	\centering
	\subfigure[Mesh with inclusion]{
		\includegraphics[height=0.36\textwidth]{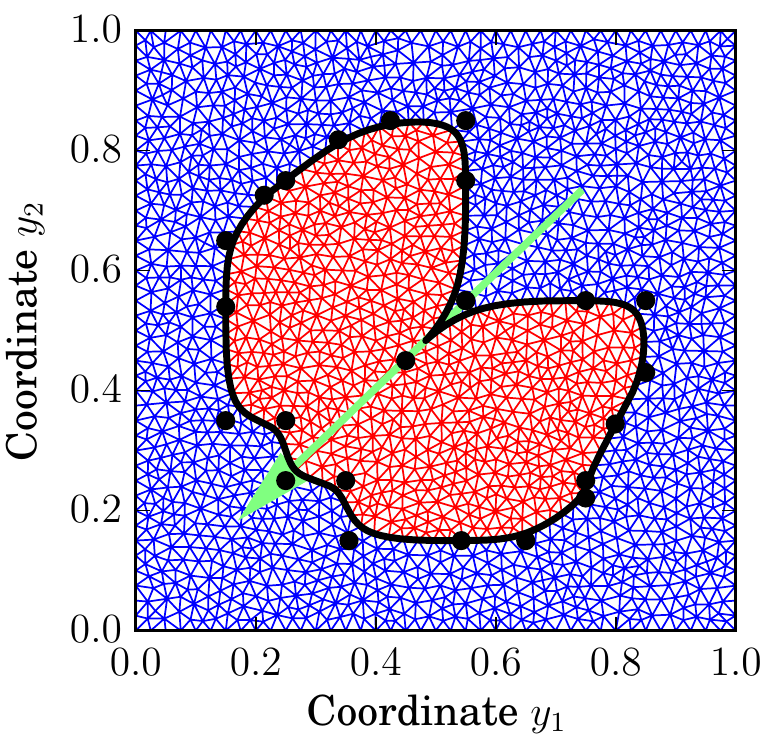}
	}
	\subfigure[Band-gaps and resonance frequencies]{
		\includegraphics[height=0.36\textwidth]{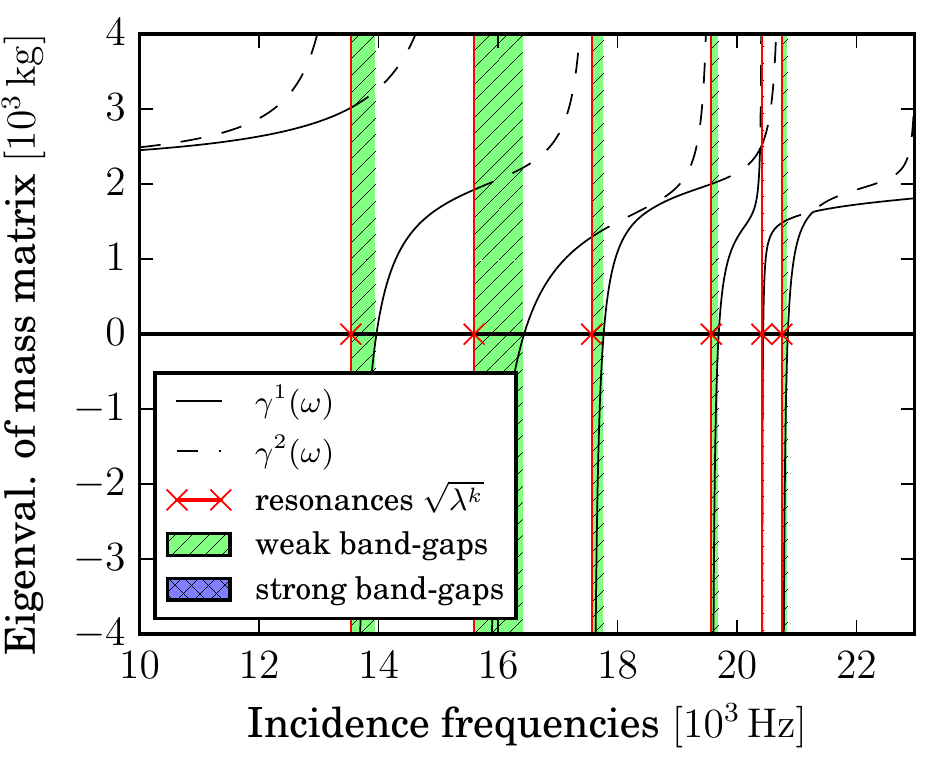}
	}
	\caption{The sub-optimal layout for maximal weak band gap size in interval $[\sqrt{\lam^2}, \sqrt{\lam^3}]$; initial layout: L-shaped inclusion; 1 re-meshing used; (a) the mesh with blue matrix, red inclusion, and black boundary with spots of B-spline nodes, the green arrow shows the polarization of the suppressed waves;
		(b) the two eigenvalues of the mass tensor for incidence frequencies}
	\label{fig:opti_eig_L_second}
\end{figure}

\newtext{
	In the second interval, the optimization run initiated with the L-shaped domain leads to a sub-optimal shape depicted in Figure~\ref{fig:opti_eig_L_second}. The optimization is terminated after one restart (re-meshing) of the optimization solver because the inclusion shape becomes very distorted.
}

\begin{figure}[!htpb]
	\centering
	\subfigure[Mesh with inclusion]{
		\includegraphics[height=0.36\textwidth]{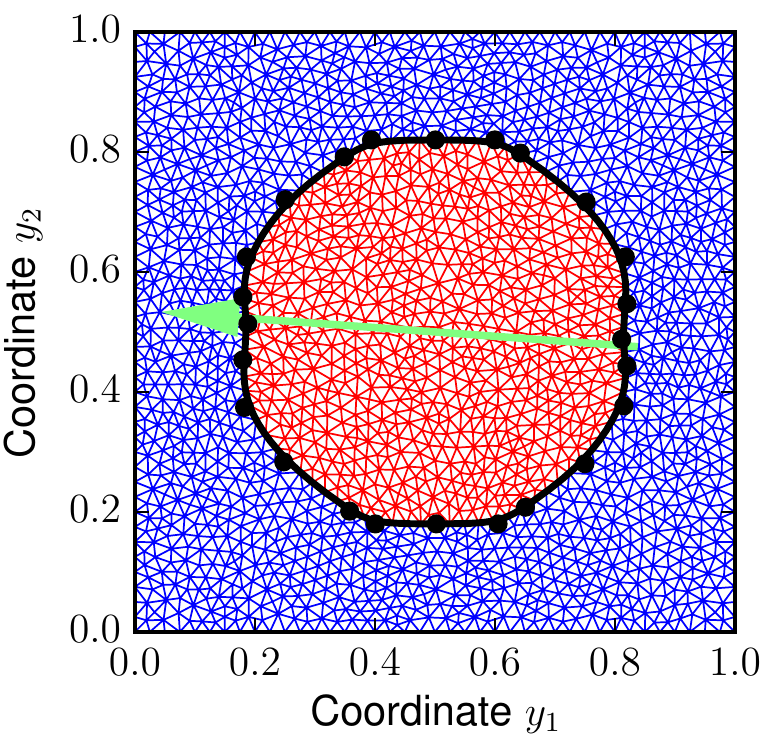}
	}
	\subfigure[Band-gaps and resonance frequencies]{
		\includegraphics[height=0.36\textwidth]{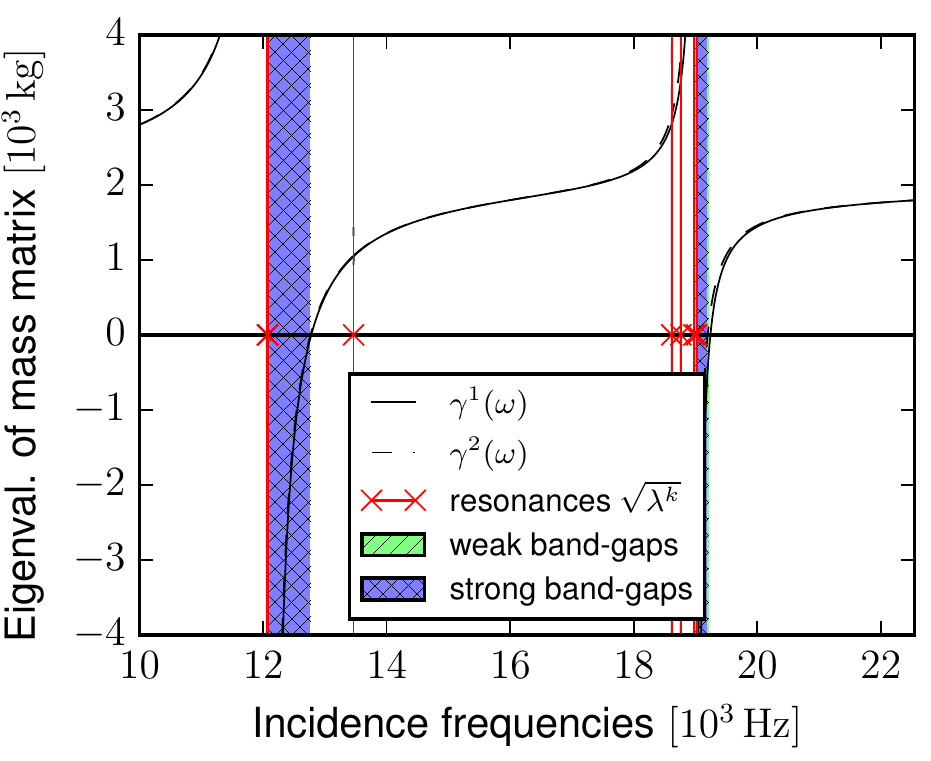}
	}
	\subfigure[Mesh with inclusion]{
		\includegraphics[height=0.36\textwidth]{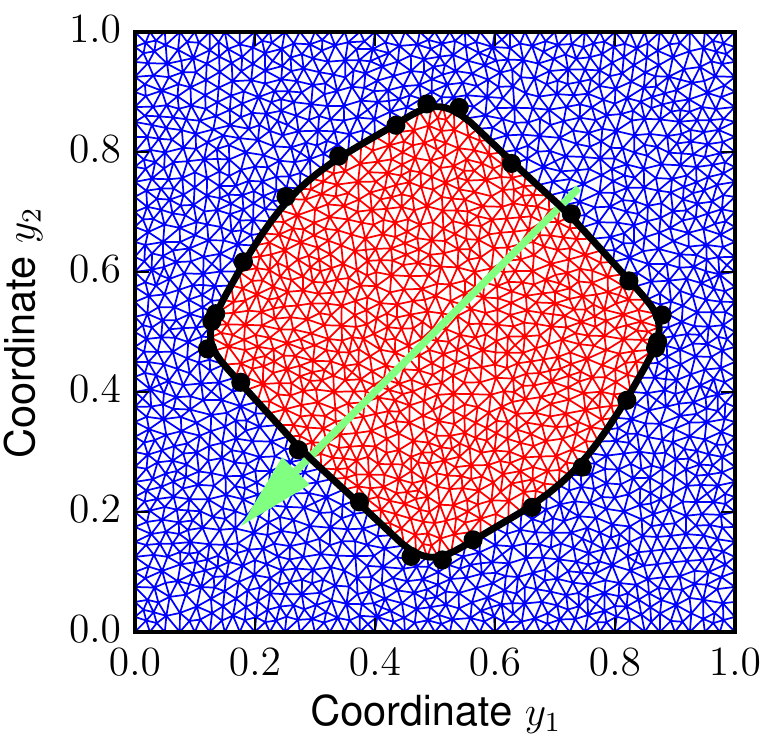}
	}
	\subfigure[Band-gaps and resonance frequencies]{
		\includegraphics[height=0.36\textwidth]{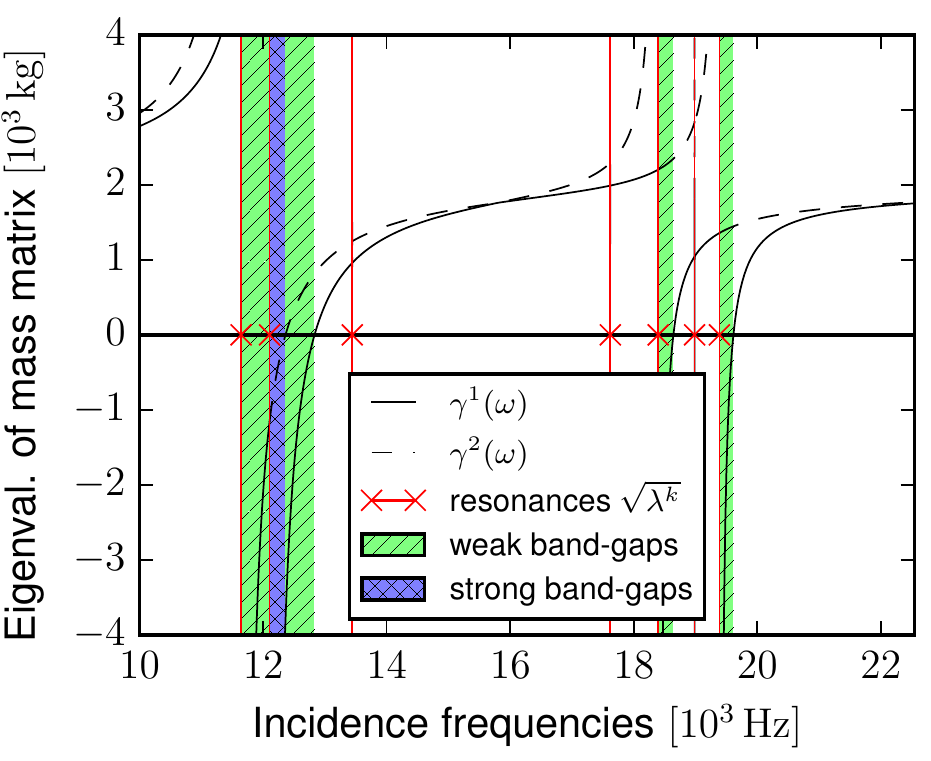}
	}
	\caption{\newtext{The optimal layout for maximal weak band gap size in interval $[\sqrt{\lam^2}, \sqrt{\lam^3}]$; initial layout: square-shaped inclusion; 2 re-meshing used for (ab), 5 re-meshing for (cd); (ac) the mesh with blue matrix, red inclusion, and black boundary with spots of B-spline nodes, the green arrow shows the polarization of the suppressed waves;
			(bd) the two eigenvalues of the mass tensor for incidence frequencies}
	}
	\label{fig:opti_eig_sq_second}
\end{figure}

\newtext{
	On the other hand, the optimization in the second interval, being initiated with the square inclusion,  leads to an optimal diamond-like shape depicted in Figure~\ref{fig:opti_eig_sq_second} together with an intermediate circle-like shape of the inclusion. The optimal shape is obtained after five restarts (re-meshing) of the optimization algorithm.
}

\clearpage

\section{Conclusion}
We presented a methodology for
the optimization of 2D phononic crystals with the main objective to maximize acoustic band-gaps.
The band gap identification is based on the model of wave propagation in homogenized periodic strongly heterogeneous two-component materials which are featured by a high-contrast in the elastic coefficients.
This modelling approach explained in \cite{Avila2008MMS_BG} and validated in \cite{RohanSeifrt2009} using the Bloch-Floquet theory, cf. \cite{Auriault1985,Auriault2012},
enables to determine the band gap bounds from the analysis of the homogenized mass (or the anisotropic effective density)
tensor which depends on the frequency.
Recall that the ``metamaterial'' property of the model is relevant for situations when the wave length is significantly larger that the periodicity gauge of the phononic crystal which, thus,
confines the range of frequencies considered depending on the given material components and the crystal size.

\newtext{In the paper, the optimization problem was formulated with the objective function describing the weak band-gap size and with constraints on the effective material elasticity, see Section~\ref{sec:opt}. The shape of the inclusions occupied by the weaker material was described by cyclic B-spline curves which, by the consequence, prevent the inclusion boundary oscillation associated with the spatial finite element discretisation.}
The most relevant results are now summarized:

\begin{itemize}
	
	\item The size-effect of elastic wave propagation is described for a change of a microstructure size and for a change of inclusion size only, see Section~\ref{sec:rescaling}. 
	As a consequence, the behaviour of the model can be described only with an inclusion shape on a unit cell. For a specific chosen microstructure size, the results can be interpreted.

	\item For numerical optimization,
	the employed gradient-based optimization algorithm SLSQP \ER{ proved good performance, although it does not enforce the design admissibility constraints in intermediate iterations.}
	The complexity of the optimization problem is driven by the calculation of the whole spectral problem \eqref{eq-36a},
	and the sensitivity analysis of corresponding eigenvectors. 

	\item \newtext{The optimized layout of the phononic crystal depends on the type of elasticity constraint, for which we have considered two variants.
		One optimization problem has led to an optimal layout, see  Figures~\ref{fig:final_L_state}, \ref{fig:opti_eig_sq_second}, some have terminated due to mesh distortion, see Figures~\ref{fig:fin_L2}, \ref{fig:opti_eig_L_second}, and some has led to sub-optimal layouts with termination due to closeness of upper-band gap bound with the next eigenvalue \ref{fig:fin_square2}, \ref{fig:opti_eig_L_first}.}
	

	\item \ER{The band gap width maximization is restricted by the interval given by two consecutive resonance frequencies associated with the effective mass tensor. To allow for connecting band gaps spanning more such intervals will require much more complex optimal problem formulation involving additional constraints related to the resonance frequencies.}
	
\end{itemize}

As a difference to the classical approach which relies on the Bloch-Floquet theory, there is no need to  search for the band gaps in the first Brillouin zone, in contrast with usual approach \cite{Qian2011,Sigmund2003bandgaps}.
In this context, our approach can distinguish strong band gaps, such that there are no propagating modes,
or the weak band gaps which admit propagation of waves of certain polarizations only;
this classification is independent of the direction of the wave propagation, as proved in \cite{RohanSeifrt2009}.



For the objective function expressing a selected band gap length, the sensitivity analysis formulas have been derived so that gradient based method can be used to solve the optimization problem.
To avoid all difficulties arising in a case of nondifferentiable objective function,
we confined to maximization of the weak band gaps which restrict the wave propagation only partially,
for some polarizations (however, independently on the direction of the wave propagation).

To handle also the strong band gaps, typically admitted by symmetric shapes of inclusions
featured by resonant frequencies with higher multiplicity, 
the sensitivity analysis complicates and the nonsmooth optimization tools, like the bundle-type methods working with the notion of subdifferentials,
must be resorted for.
The sensitivity analysis of such a nonsmooth problem has been treated in\cite{Rohan2009-WCSMO} for piezo-phononic structures.
In the context of desired strong band gaps maximization, to avoid  hurdles arising with multiple eigenvalues, ad~hoc symmetric 
shapes could be considered on reduced geometries with symmetry boundary conditions, \cf \cite{Taheri-Hassani2014}.


\section*{Acknowledgement}
E. Rohan and J. Heczko have been supported by the Czech Science Foundation through project  No.~P101/12/2315 and in part 
by the project LO 1506 of the Czech Ministry of Education, Youth and Sports
and J.~Vond\v{r}ejc partially by the project EXLIZ -- CZ.1.07/2.3.00/30.0013, which is co-financed by the European Social Fund and the state budget of the Czech Republic.


\appendix
\section{Transformation formulas for the size effect}\label{apx-A}
We consider a given scale $\veps_0$, such that  $\Cop^{\veps_0} = \Cop^\mater$ is the realistic value, see Remark~\ref{rem-5}.
Let $\veps_1$ be another scale of the structure
and define $a = \frac{\veps_1}{\veps_0}$. Further, by subscript $\brex{k}$ we label the quantities associated with the microstructure characterized by $\veps_k$, $k = 1,2$. The following observations are straightforward:
\begin{list}{}{} 
\item (i) the eigenfunctions are identical, $\phibf_\brex{1}^r = \phibf_\brex{0}^r$ for $r = 1,2,\dots$.
\item (ii) By virtue of  \eq{eq-36a}, $\lam_\brex{1}^r$ computed for $\bar\Cop_\brex{1}:=\veps_\brex{1}^{-2}\Cop^\mater = \bar\Cop_\brex{0} \left(\frac{\veps_0}{\veps_1}\right)^2 = a^{-2}\bar\Cop_\brex{0}$ is related to $\lam_\brex{0}^r$,
as follows: $\lam_\brex{1}^r = \left(\frac{\veps_0}{\veps_1}\right)^{-2}\lam_\brex{0}^r = a^{-2}\lam_\brex{0}^r$.

\item (iii) Since $\frac{\om^2}{\om^2 - \lam_\brex{1}^r} = \frac{a^2\om^2}{a^2\om^2 - \lam_\brex{0}^r}$,
the two effective mass tensors are equivalent for rescaled frequencies: it holds that
$\Mb_\brex{1}(\om) = \Mb_\brex{0}(a\om)$.

\item (iv) Due to the  above observation (iii), the band gaps distributions are scaled by $a$.
Let $G_\brex{0} = ]\ul\om,\ol\om[$, then
$G_\brex{1} = ]a^{-1}\ul\om,a^{-1}\ol\om[$, thus, $|G_\brex{1}| = a^{-1}(\ol\om-\ul\om) =  a^{-1}|G_\brex{0}|$.

\end{list}

\section{Shape sensitivity -- auxiliary results}\label{apx-B}
By virtue of the shape sensitivity based on the domain parametrization
\cite{Haslinger1988book}{, and using notation \eqref{eq:notation_bilf}}, the
following formula hold,
\begin{equation}\label{eq-S15}
\begin{split}
\delta_\tau \aYi{\ub}{\vb}{\beta} &= \intY_{Y_\beta} 
C_{irks}^\beta\left(\delta_{rj}\delta_{sl} \nabla_y\cdot \vec\Vcal - \delta_{jr}\pd_s^y \Vcal_l - \delta_{ls}\pd_r^y \Vcal_j\right)
e_{kl}^y(\ub) e_{ij}^y(\vb)\;,\quad \beta = 1,2\;,\\
\delta_\tau \rYi{\ub}{\vb}{2} & = \intY_{Y_2} \rho_2\ub\cdot\vb  \nabla_y\cdot \vec\Vcal\;,\\
\delta_\tau \aver{\rho} & = \rho^1\intY_{Y_1} \nabla_y\cdot \vec\Vcal
+\rho^2\intY_{Y_2} \nabla_y\cdot \vec\Vcal\;,
\end{split}
\end{equation}
where $\intYsmall_{Y_\beta} = |Y|^{-1}\int_{Y_\beta}$.

\paragraph{Sensitivity of the effective elasticity $\Dop$}
We can now differentiate the expressions for $\Dop$ defined in \eqref{eq-C*}, to obtain $\delta\Dop$. This yields
\begin{equation*}
\begin{split}
\delta D_{ijkl}|Y| & = \delta_\tau
\aYi{\wb^{ij}+\Pibf^{ij}}{\wb^{kl}+\Pibf^{kl}}{1} 
+ \aYi{\delta_\tau\Pibf^{ij}}{\wb^{kl}+\Pibf^{kl}}{1} 
+\aYi{\wb^{ij}+\Pibf^{ij}}{\delta_\tau \Pibf^{kl}}{1} \\
& \quad + \aYi{\delta\wb^{ij}}{\wb^{kl}+\Pibf^{kl}}{1} 
+\aYi{\wb^{ij}+\Pibf^{ij}}{\delta \wb^{kl}}{1}\;,
\end{split}
\end{equation*}
where $\delta_\tau \Pi_k^{ij} = \Vcal_j \delta_{ik}$, see \eq{eq-C*}. Note that 
$\delta_t|Y| =0$ due to the design parametrization.  
The last two integrals involving the undesired differentials 
$\delta \wb^{kl},\delta \wb^{ij}$ cancel due to \eqref{eq-C*a}
evaluated for $\vb = \delta \wb^{kl}$, or  $\vb = \delta \wb^{ij}$.
Hence, using \eq{eq-S15}$_1$, we get the following explicit formula:
\begin{equation}\label{eq-S19}
\begin{split}
\delta D_{mnpq}  &= \intY_{Y_\beta} 
C_{irks}^\beta\left(\delta_{rj}\delta_{sl} \nabla_y\cdot \vec\Vcal - \delta_{jr}\pd_s^y \Vcal_l - \delta_{ls}\pd_r^y \Vcal_j\right)
e_{kl}^y(\wb^{mn}+\Pibf^{mn}) e_{ij}^y(\wb^{pq}+\Pibf^{pq}) \\
& \quad + \intY_{Y_\beta} \left(
C_{pjkl}^\beta e_{kl}^y(\wb^{mn}+\Pibf^{mn}) \pd_j^y\Vcal_q
+ C_{ijml}^\beta  \pd_l^y\Vcal_n e_{ij}^y(\wb^{pq}+\Pibf^{pq}) \right)
\;.
\end{split}
\end{equation}




\end{document}